\documentclass{article}
\pdfoutput=1

\usepackage{jheppub}
\usepackage{dashrule}
\usepackage[T1]{fontenc} 
\usepackage{braket}
\usepackage{float}
\usepackage{amsmath,amssymb}
\usepackage{graphicx}
\usepackage{textcomp}
\usepackage{mathtools}
\usepackage{makecell}

\usepackage{eucal}
\usepackage{mathrsfs}
\usepackage{tikz}
\usepackage{pdflscape}

\newcommand{\JPb}{\multj_{\bar{P}}}
\newcommand{\JFb}{\multj_{\bar{F}}}

\newcommand{\gO}{G^{\text{Out}}_\nn}
\newcommand{\gs}{G^{\text{Out}*}_\nn}
\newcommand{\gI}{G^{\text{In}}_\nn}
\newcommand{\pO}{\pi^{\text{Out}}_\nn}
\newcommand{\ps}{\pi^{\text{Out}*}_\nn}

\newcommand{\phN}{\varphi_{_\nn}}
\newcommand{\piN}{\pi_{\nn}}

\newcommand{\mathinhead}[2]{\texorpdfstring{$\boldsymbol{#1}$}{#2}}

\newcommand{\del}{\partial}

\newcommand{\w}{\omega}
\newcommand{\Om}{\Omega}

\newcommand{\nn}{\mathcal{N}}
\newcommand{\Dp}{D_+}
\newcommand{\Dm}{D_-}

\newcommand{\Khout}{\widehat{K}^{\text{Out}}}

\newcommand{\ct}{\mathscr{C}}

\newcommand{\SCIP}{S_{\text{CIP}}}
\newcommand{\kO}{K_{\text{Out}}}
\newcommand{\kI}{K_{\text{In}}}
\newcommand{\khO}{\widehat{K}_{\text{Out}}}
\newcommand{\khI}{\widehat{K}_{\text{In}}}
\newcommand{\multj}{\mathcal{J}}
\newcommand{\spL}{\mathbb{L}}

\title{Influence Phase of a dS Observer I : Scalar Exchange }
\author[a]{R. Loganayagam,}
\author[a]{Omkar Shetye.}

\emailAdd{nayagam@icts.res.in}
\emailAdd{omkar.shetye@icts.res.in}

\affiliation[a]{ International Centre for Theoretical Sciences (ICTS-TIFR),
	Tata Institute of Fundamental Research,
	Shivakote, Hesaraghatta,
	Bangalore 560089, India.}

\abstract{Inspired by real-time computations in AdS black holes, we propose a method to obtain the influence phase of a cosmological observer by calculating the on-shell action on a doubled spacetime geometry. The influence phase is the effective action for an open system: for a dS static patch observer coupled to a scalar field it incorporates the radiation reaction due to the bulk fields and their dS Hawking radiation. For a general extended source in dS, we describe how to account for finite size effects. In the long-time limit, we get a Markovian open quantum system susceptible to cosmological fluctuations, whereas the short-time limit reproduces the worldline theory of flat-space radiation reaction. We also present a fully covariantised form for the cubic corrections to the radiation reaction in even spacetime dimensions, including Hubble contributions, and find an intriguing recursive structure across dimensions.} 

\begin{document}
	\maketitle
	\section{Introduction}

    Over the last few decades, many independent lines of evidence have converged on the fact that our universe has a positive cosmological constant\cite{Carroll:2000fy,Peebles:2002gy,Frieman:2008sn,Weinberg:2013agg}. This has presented a difficult conundrum for those who want to think about the relation between gravity and quantum mechanics\cite{Witten:2001kn,Bousso_2004}. Among the most fruitful ideas coming out of research in quantum gravity has been holography, i.e. the statement that a gravitational theory is equivalent to a quantum system living on its boundary. However, spacetimes with positive cosmological constant do not have any time-like boundaries for a dual quantum system to live in. Thus, it seems, that gravity in such spacetimes cannot have a holographic dual theory (or at least we cannot have a dual which is a conventional quantum dynamical system). 

    One attempt to overcome this obstacle is as follows\cite{Anninos:2011af,Nakayama:2011qh}: imagine a lone observer probing such a spacetime. The worldline of such an observer can then be thought of as a time-like boundary where a possible holographic description might reside. This is the idea of \emph{solipsistic holography}, which posits that a  quantum system\footnote{Perhaps a large $N$ matrix model as in BFSS duality\cite{Banks:1996vh} (See \cite{Taylor:2001vb,Ydri:2017ncg} for a review).} living on such a worldline encodes the  quantum theory of gravity that describes the universe. To claim that the information about the entire universe can be gleaned from a single worldline within it might seem speculative: but it is pertinent to remember that all existing knowledge about our universe can be traced ultimately to measurements around the earth. Thus,  we might want to assess the viability of such a proposal by examining it further.\footnote{Another alternative approach is to focus not on the observables but rather on meta-observables like the global wave-function of the universe\cite{Strominger:2001pn,Maldacena:2002vr,Marolf:2008hg,Anninos:2011ui,Anninos:2017eib,Hogervorst:2021uvp,Chakraborty:2023yed,Loparco:2023rug}. As has been emphasised in these references, this approach is especially suited to model the physics of inflation, with us serving as meta-observers to some extent. How the spacetime dynamics gets encoded in  the proposed dual is not yet entirely  clear, though much progress has been achieved over the past few years\cite{Arkani-Hamed:2018kmz, Sleight:2019hfp, Goodhew:2020hob, Baumann:2022jpr}.} 
     
	Any object in a dynamical spacetime influences and is influenced by its surroundings. In this sense, any gravitational observer should be thought of as an open quantum system constantly interacting with the rest of the universe. On the quantum mechanical side, the emergence of the open system is due to integrating out observer's internal degrees of freedom. This is the cosmological analogue of the fluid-gravity correspondence\cite{Hubeny:2011hd}. In the AdS/CFT context,
    on the gravity side, fluid dynamics emerges by integrating out the physics in radial direction, whereas on the gauge theory side, it is a consequence of coarse-graining quarks and gluons. There is, by now, non-trivial evidence supporting this statement, including precise matching of anomalous effects on both sides\cite{Jensen:2013kka,Jensen:2013rga}. In a similar vein, one might ask how we could go about checking the cosmological version of this statement.

    The central challenge in answering this question is  twofold: first, to derive an open quantum system on the worldline from the ambient dynamics. As we shall see, a precise definition of this first step already involves some work.\footnote{Systematic description of observers in the middle of a  spacetime (as opposed to asymptotic observers) is well-known to be a hard problem. Some of the approaches to the AdS version of this question, starting from the CFT side, can be found in \cite{Hamilton:2006fh,Papadodimas:2012aq,Maxfield:2017rkn,Jafferis:2020ora}. It would be interesting to extend these ideas to take into account the open nature of the observer, as we do here. } More precisely, what we need is a cosmological analogue of GKPW prescription in AdS/CFT that will allow us to derive the open system for an observer. This note is aimed at addressing this issue.
    
    The second step would be to construct a dual unitary quantum system that, after integrating out appropriate degrees of freedom, leads to the same open theory as gravity. This might be a hard undertaking: after all, even in the fluid-gravity correspondence, to derive the fluid dynamics from a strongly coupled gauge theory is practically impossible. But, since we are dealing with a quantum mechanical system here, there is reason for hope. One immediate goal would be to check whether the putative open quantum system derived in the first step shows the right structural features to admit a solipsistic interpretation. We will postpone further thoughts on this issue to the discussion section.

	Let us return now to the issue of constructing the open system on the gravitational side.
    Imagine a universe described by a dynamical spacetime along with a variety of fields living on it. A local observer in such a theory may be modelled as a source for these fields: a source that emits/radiates as well as a source that absorbs/detects. Any autonomous motion of the observer is then accompanied by an outgoing radiation and an associated radiation reaction. This results in a dissipation of the observer's energy, and we seek an open quantum system should describe this physics. The open quantum theory on the world line should also describe the influence of the incoming radiation from the rest of the universe. As we shall elaborate on later,
	this incoming radiation also includes the Hawking radiation from the Hubble horizon.\footnote{See \cite{Chandrasekaran:2022cip} for an analysis of dS observer from the point of view of von Neumann algebras. It would be interesting to link such an analysis to the ideas discussed in this note, e.g., one may ask how the physics of radiation reaction is encoded within von Neumann algebras. Another algebraic statement of potential interest is the `time-like tube theorem'\cite{borchers1961vollstandigkeit,araki1963generalization,Strohmaier:2023opz,Witten:2023qsv}, but, again, it is unclear to us how  such formal statements relate to the description of dS observer as an open system. }
 
    Quite independently of such holographic quests, the worldline open quantum theory under question shows up in a variety of concrete physical questions. As an example, worldline EFT has emerged as a useful way to organise the post-Newtonian expansion of a binary system radiating gravitational waves\cite{Goldberger:2004jt,Goldberger:2007hy,Porto:2016pyg,Levi:2018nxp,Barack:2018yvs}. The basic idea in such approaches is to systematically integrate out the short-distance gravitational physics that binds the binaries to get an effective theory that describes the inspiral process. Due to the radiation of gravitational waves, ultimately such a binary is also an open system of the type described above. These ideas can be generalised into a cosmological setting where, for example, a worldline EFT which takes also the expansion of the universe into account might be useful in studying the dynamics of galactic formation, cooling and mergers.\footnote{See \cite{Porto:2013qua} for the role played by worldline methods in the effective field theory(EFT) of large scale structure(LSS).}  A motivation of this note is to describe an approach that might help us systematically derive such an EFT.

	We will conclude this preamble with a broad outline of what follows: in section \S\ref{sec:OQScosmo}, we begin by describing the basic geometric set-up used in deriving the open quantum mechanics associated with the cosmological observer. The prescription we propose is inspired by the recent developments in real-time AdS/CFT\cite{Skenderis:2008dg,Skenderis:2008dh,Glorioso:2018mmw,deBoer:2018qqm,Chakrabarty:2019aeu,Jana:2020vyx,Chakrabarty:2020ohe,Ghosh:2020lel,  He:2022jnc, Loganayagam:2022zmq} that have led to systematic derivation of open quantum systems by integrating out a thermal holographic CFT bath. The essential idea here is a real-time version of Gibbons-Hawking procedure\cite{Gibbons:1976ue}: one proposes an appropriate semi-classical geometry only containing the relevant region (BH exterior for Gibbons-Hawking, dS static patch in the current problem), and computes the path integral in a saddle-point approximation by evaluating on-shell action. We will argue that such a prescription leads to an answer which correctly encodes both the radiation reaction and Hawking radiation from the Hubble horizon.
    
    The problem of cosmological observer exhibits broad structural similarities to the AdS case, which we exploit. But we also find significant differences: for one, much of the standard holographic machinery (e.g. GKPW prescription, counter-term procedure) available on the AdS side is simply absent. We outline a regularisation procedure that gives finite answers in \S\ref{sec:OQScosmo}, relegating the details to appendices.  
	
	In section \S\ref{sec:CIP} we use our prescription to derive the open effective action/influence phase for an observer coupled to a class of generalised free scalar fields. Next, we examine the flat space limit of our influence phase in \S\ref{sec:flatRR} demonstrating how the already known expressions of the flat space radiation reaction are reproduced in this limit. We also compute the leading cosmological corrections to the Abraham-Lorentz-Dirac radiation-reaction force. In the penultimate section \S\ref{sec:interactions}, we sketch how interactions can be incorporated into our formalism. We conclude with a summary and a discussion of further directions in \S\ref{sec:conc}.
	
	To enable readability, we confine ourselves to describing the basic physical ideas as well as the central results in the main sections. Much of the relevant technical details are presented in appendices.	The first appendix \S\ref{app:FlatMult} is a review of multipole expansion in flat spacetime, along with a description of the multipole expansion in terms of symmetric trace-free(STF) tensors. Much of this is standard material just cast into a notation convenient for our purposes. In the next appendix \S\ref{app:DesSc}, we review the outgoing scalar solutions in dS and describe a counterterm procedure to deal with point-like sources placed at the centre of the static patch. In appendix \S\ref{app:Inf}, we show that the counterterm procedure extends to the most general scalar configurations and describe how to deal with extended sources. The discussion in these two appendices culminates in an effective action describing arbitrary scalar sources in the dS static patch. The next appendix \S\ref{app:RadReact} specialises to point-like observers in arbitrary motion in $dS_{d+1}$ with $d$ odd: we show that our effective action evaluated for such sources re-assembles into a  generally covariant  radiation-reaction force with Hubble corrections.

	\section{The cosmological influence phase \mathinhead{\SCIP}{SCIP}}\label{sec:OQScosmo}
	Our goal is to describe the experience of an observer in an expanding spacetime. This, in turn, will help us in understanding the spacetime itself. In particular, we want to ask how to construct the open quantum system that describes the cosmological observer. In its full generality, this is a difficult problem, but we can start with a simple model for the observer. We can think of the observer as a single worldline undergoing absorption and emission processes. So the observer is privy to 3 kinds of data:
	\begin{itemize}
		\item Outgoing radiation: Emission data along with the outgoing propagator tells us the field values at a later time in the spacetime.
		\item Incoming radiation: The fields in the past can be reconstructed by using an incoming propagator given the absorption data. 
		\item Fluctuations: The observer will also be sensitive to \emph{cosmic noise}, which shows up in the absorption data. 
	\end{itemize} 
\begin{figure}
    \centering

\tikzset{every picture/.style={line width=0.75pt}} 

\begin{tikzpicture}[x=0.75pt,y=0.75pt,yscale=-1,xscale=1]

\draw    (138.5,60) -- (138.5,197.5)(135.5,60) -- (135.5,197.5) ;
\draw    (137,128.75) .. controls (134.64,128.75) and (133.46,127.57) .. (133.46,125.21) .. controls (133.46,122.86) and (132.28,121.68) .. (129.93,121.68) .. controls (127.57,121.68) and (126.39,120.5) .. (126.39,118.14) .. controls (126.39,115.79) and (125.21,114.61) .. (122.86,114.61) .. controls (120.5,114.61) and (119.32,113.43) .. (119.32,111.07) .. controls (119.32,108.72) and (118.14,107.54) .. (115.79,107.54) .. controls (113.43,107.54) and (112.25,106.36) .. (112.25,104) .. controls (112.25,101.65) and (111.07,100.47) .. (108.72,100.47) .. controls (106.36,100.47) and (105.18,99.29) .. (105.18,96.93) .. controls (105.18,94.57) and (104,93.39) .. (101.64,93.39) .. controls (99.29,93.39) and (98.11,92.21) .. (98.11,89.86) .. controls (98.11,87.5) and (96.93,86.32) .. (94.57,86.32) .. controls (92.22,86.32) and (91.04,85.14) .. (91.04,82.79) .. controls (91.04,80.43) and (89.86,79.25) .. (87.5,79.25) .. controls (85.15,79.25) and (83.97,78.07) .. (83.97,75.72) .. controls (83.97,73.36) and (82.79,72.18) .. (80.43,72.18) .. controls (78.08,72.18) and (76.9,71) .. (76.9,68.65) .. controls (76.9,66.29) and (75.72,65.11) .. (73.36,65.11) -- (70.75,62.5) -- (70.75,62.5) ;
\draw    (263.5,59.5) -- (263.5,197.5)(260.5,59.5) -- (260.5,197.5) ;
\draw    (199,191.5) .. controls (199,189.14) and (200.18,187.96) .. (202.54,187.96) .. controls (204.89,187.96) and (206.07,186.78) .. (206.07,184.43) .. controls (206.07,182.07) and (207.25,180.89) .. (209.61,180.89) .. controls (211.96,180.89) and (213.14,179.71) .. (213.14,177.36) .. controls (213.14,175) and (214.32,173.82) .. (216.68,173.82) .. controls (219.03,173.82) and (220.21,172.64) .. (220.21,170.29) .. controls (220.21,167.93) and (221.39,166.75) .. (223.75,166.75) .. controls (226.1,166.75) and (227.28,165.57) .. (227.28,163.22) .. controls (227.28,160.86) and (228.46,159.68) .. (230.82,159.68) .. controls (233.18,159.68) and (234.36,158.5) .. (234.36,156.14) .. controls (234.36,153.79) and (235.54,152.61) .. (237.89,152.61) .. controls (240.25,152.61) and (241.43,151.43) .. (241.43,149.07) .. controls (241.43,146.72) and (242.61,145.54) .. (244.96,145.54) .. controls (247.32,145.54) and (248.5,144.36) .. (248.5,142) .. controls (248.5,139.65) and (249.68,138.47) .. (252.03,138.47) .. controls (254.39,138.47) and (255.57,137.29) .. (255.57,134.93) .. controls (255.57,132.58) and (256.75,131.4) .. (259.1,131.4) -- (262,128.5) -- (262,128.5) ;
\draw  [fill={rgb, 255:red, 0; green, 0; blue, 0 }  ,fill opacity=1 ] (102.96,102.78) -- (98.53,89.75) -- (111.97,92.73) -- (103,93.75) -- cycle ;
\draw  [fill={rgb, 255:red, 0; green, 0; blue, 0 }  ,fill opacity=1 ] (221.23,156.59) -- (234.94,155.35) -- (228.88,167.71) -- (230,158.75) -- cycle ;
\draw   (65.75,62.5) .. controls (65.75,61.12) and (66.87,60) .. (68.25,60) .. controls (69.63,60) and (70.75,61.12) .. (70.75,62.5) .. controls (70.75,63.88) and (69.63,65) .. (68.25,65) .. controls (66.87,65) and (65.75,63.88) .. (65.75,62.5) -- cycle ;
\draw   (196.5,189) .. controls (196.5,187.62) and (197.62,186.5) .. (199,186.5) .. controls (200.38,186.5) and (201.5,187.62) .. (201.5,189) .. controls (201.5,190.38) and (200.38,191.5) .. (199,191.5) .. controls (197.62,191.5) and (196.5,190.38) .. (196.5,189) -- cycle ;
\draw    (374.5,60.5) -- (374.5,198.5)(371.5,60.5) -- (371.5,198.5) ;
\draw    (310,192.5) .. controls (310,190.14) and (311.18,188.96) .. (313.54,188.96) .. controls (315.89,188.96) and (317.07,187.78) .. (317.07,185.43) .. controls (317.07,183.07) and (318.25,181.89) .. (320.61,181.89) .. controls (322.96,181.89) and (324.14,180.71) .. (324.14,178.36) .. controls (324.14,176) and (325.32,174.82) .. (327.68,174.82) .. controls (330.03,174.82) and (331.21,173.64) .. (331.21,171.29) .. controls (331.21,168.93) and (332.39,167.75) .. (334.75,167.75) .. controls (337.1,167.75) and (338.28,166.57) .. (338.28,164.22) .. controls (338.28,161.86) and (339.46,160.68) .. (341.82,160.68) .. controls (344.18,160.68) and (345.36,159.5) .. (345.36,157.14) .. controls (345.36,154.79) and (346.54,153.61) .. (348.89,153.61) .. controls (351.25,153.61) and (352.43,152.43) .. (352.43,150.07) .. controls (352.43,147.72) and (353.61,146.54) .. (355.96,146.54) .. controls (358.32,146.54) and (359.5,145.36) .. (359.5,143) .. controls (359.5,140.65) and (360.68,139.47) .. (363.03,139.47) .. controls (365.39,139.47) and (366.57,138.29) .. (366.57,135.93) .. controls (366.57,133.58) and (367.75,132.4) .. (370.1,132.4) -- (373,129.5) -- (373,129.5) ;
\draw   (337.17,164.23) -- (341.5,161) (351.6,153.46) -- (347.27,156.69) (336.57,154.39) -- (346.43,167.61) (342.34,150.08) -- (352.2,163.31) ;
\draw   (307.5,195) .. controls (307.5,193.62) and (308.62,192.5) .. (310,192.5) .. controls (311.38,192.5) and (312.5,193.62) .. (312.5,195) .. controls (312.5,196.38) and (311.38,197.5) .. (310,197.5) .. controls (308.62,197.5) and (307.5,196.38) .. (307.5,195) -- cycle ;

\draw (74,209) node [anchor=north west][inner sep=0.75pt]   [align=left] {Emission};
\draw (242,206) node [anchor=north west][inner sep=0.75pt]   [align=left] {{\small Absorption}};
\draw (186,107) node [anchor=north west][inner sep=0.75pt]  [font=\small] [align=left] {{\small incoming }\\{\small propagator}};
\draw (309,124) node [anchor=north west][inner sep=0.75pt]   [align=left] {{\small noise}};
\draw (49,108) node [anchor=north west][inner sep=0.75pt]  [font=\small] [align=left] {outgoing\\{\small propagator}};

\end{tikzpicture}
    \caption{A cosmological observer can access 3 kinds of data: radiation due to its own emissions, incoming radiation from sources in the environment and noise.}
    \label{fig:enter-label}
\end{figure}
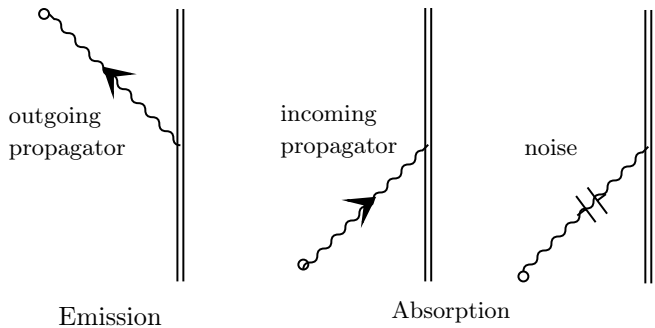
	This is reminiscent of the motion of a Brownian particle in Langevin theory. A pollen grain in water is sensitive not only to coarse-grained currents in the water (analogous to the incoming radiation) but also to fluctuations arising from the motion of water molecules. Finally, the motion of the Brownian particle can influence the dynamics of water as well (analogous to outgoing radiation). 
	
	The dynamics of such open quantum systems can be derived by the path integral prescription of Feynman and Vernon\cite{FEYNMAN1963118} describing the density matrix evolution. According to the authors of \cite{FEYNMAN1963118}, the effective description of the open system can be derived starting from two non-interacting copies of each of the system as well as the environment (describing the combined density matrix). Integrating out two copies of the environment then induces new interactions between the copies of the system, resulting in a non-unitary evolution of the system state. These terms constitute the \emph{influence phase}, which encodes completely the effect of the environment on the system. Applying this insight to the question at hand, we conclude that all cosmological effects on an observer(the system) are succinctly summarised in a \emph{cosmological influence phase} $\SCIP$.
	
	What does $\SCIP$ depend on? It should depend on how effective the observer is at emitting/absorbing radiation of a given frequency $\w$ and a given multipole type $\spL$. Say we have two sets of functions $\multj_A(\w,\spL)$ and $\multj_D(\w,\spL)$ characterising the emission/absorption efficiency of the observer. From the Feynman-Vernon viewpoint, $\multj_A(\w,\spL)$ and $\multj_D(\w,\spL)$ have the following interpretation: to begin with, we have two copies of the observer (left/right), each probing their copy of the universe via their respective multipole moments $\multj_L(\w,\spL)$ and $\multj_R(\w,\spL)$ respectively. The influence phase, which results from integrating out the universe, then depends on the average \[
 \multj_A(\w,\spL)\equiv\frac{1}{2}[\multj_R(\w,\spL)+\multj_L(\w,\spL)]\ ,\] as well as the difference \[
 \multj_D(\w,\spL)\equiv \multj_R(\w,\spL)-\multj_L(\w,\spL) \] of these two multipole moments. The fact that the average/difference sources characterise its emissive/absorptive properties is a well-known feature of the Feynman-Vernon formalism\cite{Schwinger:1960qe,Keldysh:1964ud,kamenev_2011,Sieberer:2015svu}: this fact can ultimately be traced to the past/future boundary conditions on the two copies imposed within this formalism. To conclude, the cosmology as seen by an observer with multipole moments $\multj_A(\w,\spL)$ and $\multj_D(\w,\spL)$ is encoded in a single influence functional  $\SCIP\left[\multj_A(\w,\spL),\multj_D(\w,\spL)\right]$. In terms of the Schwinger-Keldysh path integral of quantum gravity, we can write
     \begin{equation}
	e^{i\SCIP} \equiv \int [d\varphi_R][d\varphi_L]\  e^{iS_g[\varphi_R,\multj_R]-iS_g[\varphi_L,\multj_L]} \ ,
	\end{equation}
    where $\varphi_{L,R}$ denote the bra/ket copy of the bulk quantum fields in cosmology (including the spacetime metric) and $S_g[\varphi,\multj]$ is the full gravitational action in the background of an observer with multipole moments $\multj$. The above path integral should then be interpreted in a wilsonian sense: we want to integrate out the fast modes of quantum gravitational theory, while freezing the slow degrees of freedom of the observer, and obtain an effective action which describes the open dynamics of such an observer.

	The cosmological influence phase $\SCIP$  is a direct observable. Given an expanding universe, assuming we have a sufficiently long-lived observer with arbitrary multipole moments in some region,  the force on an observer due to radiation reaction as well as radiation reception can directly be measured.
    This force serves to determine all terms in the `effective action' $\SCIP$ that encodes the influence of the ambient universe. All the \emph{real} observables of astrophysics and cosmology, e.g. the sky maps at different frequencies, can be incorporated this way into the absorptive part of $\SCIP$. 
	
	From this viewpoint, all cosmological calculations should, in principle, be recast in terms of $\SCIP$ to connect them with observations. This is already implicit in the existing approaches to cosmology: for example, the final step in CMB power spectrum computation is to expand it in spherical harmonics centred around us. Phrasing observables in terms of  $\SCIP$ makes explicit this observer-dependence (which is probably essential for defining observables \emph{within} a quantum spacetime). Talking in terms of a single functional $\SCIP$ may also be convenient for effective field theory (EFT) based approaches to cosmology based on direct observables (e.g. those based on classifying sources in the red-shift space\cite{Senatore:2014vja,Schmittfull:2020trd,Philcox:2022frc}). More ambitiously, one may conceive of a  bootstrap program based on the cosmological influence phase that complements existing proposals for cosmological bootstrap\cite{Arkani-Hamed:2018kmz,Sleight:2019hfp,Sleight:2021plv,Baumann:2022jpr,Hogervorst:2021uvp,Loparco:2023rug}.

	What are the general principles that constrain $\SCIP$? First of all, when $\multj_D(\w,\spL)$ is set to zero, $\SCIP$ should vanish. This statement arises from the microscopic unitarity of the environment: if the two copies of the observer in Feynman-Vernon formalism introduce identical perturbations into the environment, their effect cancels out of all correlators\cite{kamenev_2011}. From the viewpoint of the observer, the above condition is equivalent to the conservation of the observer density matrix's trace. Apart from this, there are also constraints on $\SCIP$ coming from causality. For example, causality implies that the coefficient of $\multj_D^{*}(\w,\spL)\multj_A(\w,\spL)$ is analytic in the upper half plane of complex $\w$ : this coefficient is the retarded correlator on the worldline of the observer\cite{Schwinger:1960qe,Keldysh:1964ud,Caldeira:1982iu,kamenev_2011,breuer2002theory,Sieberer:2015svu}. A similar statement holds for the coefficients of any term of the form $\multj_D^{*}(\w,\spL)\prod_k\multj_A(\w_k,\spL_k)$.
	
	Evaluation of the influence phase requires us to know the real-time or Schwinger-Keldysh(SK) propagators of the environment. It is unclear how to perform such computations in generic cosmological spacetimes, especially if gravity is also to be quantised. We will show that for an observer in dS, this computation can be geometrised roughly akin to recent implementations of SK path integrals in case of AdS black holes\cite{Skenderis:2008dg,Skenderis:2008dh,Glorioso:2018mmw,deBoer:2018qqm,Chakrabarty:2019aeu,Jana:2020vyx,Chakrabarty:2020ohe,Ghosh:2020lel,  He:2022jnc, Loganayagam:2022zmq}. The hope then is that one can later generalise it beyond dS to incorporate full FRW cosmology. 
 
    Specifically, in the case of dS, we conjecture that the computation of cosmological influence phase $\SCIP$ is dominated by a geometric saddle point built out of two copies of the static patch stitched together at their horizons. We will call this doubled geometry, the dS Schwinger-Keldysh(dS-SK) spacetime. In the rest of this subsection, we will describe this geometry in more detail before moving to the evidence for our conjecture in the subsequent sections.
	 
\begin{figure}[H]
	\centering

\tikzset{every picture/.style={line width=0.75pt}} 

\begin{tikzpicture}[x=0.75pt,y=0.75pt,yscale=-1,xscale=1]

\draw   (410,66.5) -- (573.5,66.5) -- (573.5,230) -- (410,230) -- cycle ;
\draw  [dash pattern={on 4.5pt off 4.5pt}]  (410,66.5) -- (573.5,230) ;
\draw  [dash pattern={on 4.5pt off 4.5pt}]  (573.5,66.5) -- (410,230) ;
\draw  [draw opacity=0][fill={rgb, 255:red, 10; green, 247; blue, 31 }  ,fill opacity=1 ] (492.01,147.98) -- (573.53,230) -- (573.5,66.5) -- cycle ;
\draw    (573.5,66.5) .. controls (565,70.5) and (546,175.5) .. (573.53,230) ;
\draw [color={rgb, 255:red, 0; green, 0; blue, 0 }  ,draw opacity=1 ]   (573.5,66.5) .. controls (543,98.5) and (541,202.5) .. (573.5,230) ;
\draw    (573.53,230) .. controls (563,194.5) and (563,101.5) .. (573.5,66.5) ;
\draw   (102,66.5) -- (265.5,66.5) -- (265.5,230) -- (102,230) -- cycle ;
\draw  [dash pattern={on 4.5pt off 4.5pt}]  (102,66.5) -- (265.5,230) ;
\draw  [dash pattern={on 4.5pt off 4.5pt}]  (265.5,66.5) -- (102,230) ;
\draw  [draw opacity=0][fill={rgb, 255:red, 10; green, 247; blue, 31 }  ,fill opacity=1 ] (184.01,147.98) -- (265.53,230) -- (265.5,66.5) -- cycle ;
\draw  [fill={rgb, 255:red, 230; green, 139; blue, 0 }  ,fill opacity=1 ] (259.5,125) .. controls (261.5,111) and (265,79) .. (265.5,66.5) .. controls (266,54) and (266.06,239) .. (265.53,230) .. controls (265,221) and (262,196) .. (260,182) .. controls (258,168) and (257.5,139) .. (259.5,125) -- cycle ;
\draw    (300,90.33) .. controls (284.16,75.48) and (222.26,87.1) .. (259.83,110.29) ;
\draw [shift={(261,111)}, rotate = 210.61] [color={rgb, 255:red, 0; green, 0; blue, 0 }  ][line width=0.75]    (10.93,-4.9) .. controls (6.95,-2.3) and (3.31,-0.67) .. (0,0) .. controls (3.31,0.67) and (6.95,2.3) .. (10.93,4.9)   ;
\draw [color={rgb, 255:red, 0; green, 0; blue, 255 }  ,draw opacity=1 ]   (203.5,129) -- (266,191.5) ;
\draw [color={rgb, 255:red, 0; green, 0; blue, 255 }  ,draw opacity=1 ]   (221.5,110) -- (266,154.5) ;
\draw [color={rgb, 255:red, 0; green, 0; blue, 255 }  ,draw opacity=1 ]   (237.5,94) -- (265,121.5) ;
\draw [color={rgb, 255:red, 0; green, 0; blue, 255 }  ,draw opacity=1 ]   (254.5,78) -- (266,89.5) ;
\draw [color={rgb, 255:red, 0; green, 0; blue, 255 }  ,draw opacity=1 ]   (511.5,130) -- (574,192.5) ;
\draw [color={rgb, 255:red, 0; green, 0; blue, 255 }  ,draw opacity=1 ]   (529.5,111) -- (574,155.5) ;
\draw [color={rgb, 255:red, 0; green, 0; blue, 255 }  ,draw opacity=1 ]   (545.5,95) -- (573,122.5) ;
\draw [color={rgb, 255:red, 0; green, 0; blue, 255 }  ,draw opacity=1 ]   (562.5,79) -- (574,90.5) ;

\draw (477,40.4) node [anchor=north west][inner sep=0.75pt]    {$\mathscr{I}^{+}$};
\draw (479,235.4) node [anchor=north west][inner sep=0.75pt]    {$\mathscr{I}^{-}$};
\draw (362,126) node [anchor=north west][inner sep=0.75pt]   [align=left] {North\\Pole};
\draw (581.77,136.98) node [anchor=north west][inner sep=0.75pt]   [align=left] {South\\Pole};
\draw (169,40.4) node [anchor=north west][inner sep=0.75pt]    {$\mathscr{I}^{+}$};
\draw (171,235.4) node [anchor=north west][inner sep=0.75pt]    {$\mathscr{I}^{-}$};
\draw (54,126) node [anchor=north west][inner sep=0.75pt]   [align=left] {North\\Pole};
\draw (275.77,134.98) node [anchor=north west][inner sep=0.75pt]   [align=left] {South\\Pole};
\draw (282.63,90.58) node [anchor=north west][inner sep=0.75pt]   [align=left] {$\displaystyle r=r_{c}$};

\end{tikzpicture}

\caption{Penrose diagrams of dS with the static patch of the south pole observer shown in green. Constant $u$ slices are shown in blue. \textbf{Left :} a localised observer at the south pole whose worldline is thickened to a world-tube(orange) of radius $r_c$. \textbf{Right :} an extended observer modelled as a sequence of spherical shells of radius $r_i$ with $i=1,\ldots, N$.}\label{fig:dSPen}
\end{figure}
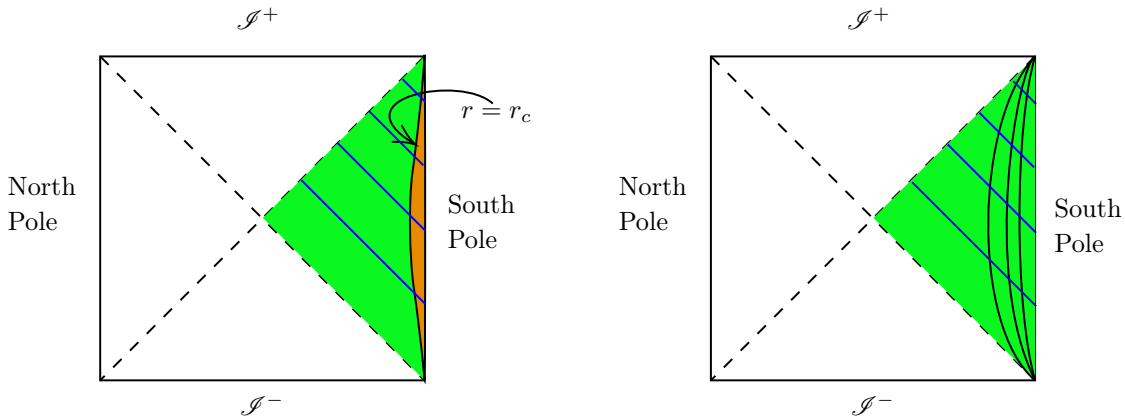

   Let us begin by setting up the basic notation required: consider a  $(d+1)$-dimensional dS spacetime dS$_{d+1}$ whose Penrose diagram is shown in Fig.\ref{fig:dSPen}. A horizontal slice (i.e., a constant time slice) in this diagram denotes the prime-meridian on a spatial sphere $S^d$, with the two ends denoting the poles. Each point in the horizontal slice corresponds to a sphere $S^{d-1}$, which shrinks to a point near the poles. The first example we will consider is a co-moving dS observer whom we place at the south pole. Our focus will be on the static patch of such an observer, i.e., the patch between the past and future cosmological horizons of the observer. We will later describe a more general class of observers spread arbitrarily over this static patch, modelled as a sequence of spherical shells around south pole (Fig.\ref{fig:dSPen}).  
	
	We will find it convenient to work with \emph{outgoing} Eddington-Finkelstein coordinates on the static patch. The metric in this coordinate system takes the form
	\begin{equation}
		ds^2=-(1-r^2H^2)\ du^2-2dudr+r^2d\Om_{d-1}^2\ .
	\end{equation} 
	Here $H$ is the Hubble constant of dS spacetime, $r$ is the radial distance from the observer, $u$ denotes the outgoing time labelling the outgoing waves and $d\Om_{d-1}^2$ is the line element on a unit $S^{d-1}$ sphere. The south-pole observer sitting at $r=0$ sees a future horizon at $r=1/H$ where the outgoing coordinates are well-behaved. In most of what follows, we will set $H=1$ for convenience and restore it later when we examine the flat space (i.e. $H\to 0$) limit.  
    We now turn to the model of the observer: Conceptually, the simplest model is that of a point particle with specified multipole moments sitting at $r=0$.	However, such a model needs to be regulated with appropriate counter-terms to allow the computation of radiation reaction effects.
    To this end, we will take the observer to be a small sphere of radius $r_c$ and thicken its worldline into a time-like `world-tube'. The point particle limit then corresponds to taking $r_c\to 0$
    limit \emph{after} the addition of counter-terms: both the Green functions and required counter-terms can be determined exactly for a dS observer coupled to generalised free scalar fields. The radius $r_c$ then acts like a UV regulator for the problem.
    
    Apart from the formal requirements of regularisation, we are also interested in the actual problem of an extended observer of a finite size. In such a case, there are no divergences. Nevertheless, finite counter-terms are needed to renormalise the bare parameters into the physically measured properties of the observer. As mentioned before, a simple model of the extended observer is a sequence of spherical shells of radius $r_i$ with $i=1,\ldots, N$:  their complement in the static patch is then the rest of the universe to be integrated out. We can define multipole moments for such extended observers and still write down a cosmological influence phase as a function of those multipole moments. The locality in radial direction gets obscured in such a description:  this is however natural in the solipsistic viewpoint where radial locality is an approximate/emergent property of the dual quantum mechanics.

    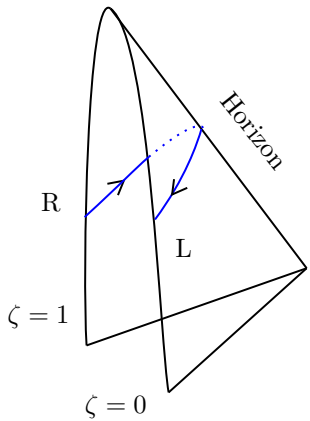
\begin{figure}[H]
\centering

\tikzset{every picture/.style={line width=0.75pt}} 

\begin{tikzpicture}[x=0.75pt,y=0.75pt,yscale=-1,xscale=1]

\draw    (315.84,42.81) .. controls (304.16,33.61) and (301.59,176.33) .. (303.78,212.86) ;
\draw    (315.84,42.81) .. controls (335.97,57.16) and (340.84,212.54) .. (345.17,236.57) ;
\draw    (414.79,173.88) -- (345.17,236.57) ;
\draw    (303.78,212.86) -- (414.79,173.88) ;
\draw    (315.84,42.81) -- (414.79,173.88) ;
\draw [color={rgb, 255:red, 0; green, 0; blue, 255 }  ,draw opacity=1 ]   (361.74,103.56) .. controls (359.65,110.82) and (353.52,128.79) .. (338.31,149.29) ;
\draw   (314.27,131.27) -- (322.57,129.69) -- (319.61,136.98) ;
\draw   (354.56,133.31) -- (346.93,136.67) -- (347.91,128.93) ;
\draw [color={rgb, 255:red, 0; green, 0; blue, 255 }  ,draw opacity=1 ]   (302.78,148.09) .. controls (321.32,131.21) and (327.35,124.41) .. (335.2,117.67) ;
\draw [color={rgb, 255:red, 0; green, 0; blue, 255 }  ,draw opacity=1 ] [dash pattern={on 0.84pt off 2.51pt}]  (335.2,117.67) .. controls (365.64,92.18) and (363.61,106.45) .. (361.74,103.56) ;

\draw (347.22,157.7) node [anchor=north west][inner sep=0.75pt]   [align=left] {L};
\draw (279.33,134.37) node [anchor=north west][inner sep=0.75pt]   [align=left] {R};
\draw (378.09,81.48) node [anchor=north west][inner sep=0.75pt]  [rotate=-51.47] [align=left] {Horizon};
\draw (301.41,236) node [anchor=north west][inner sep=0.75pt]   [align=left] {$\displaystyle \zeta =0$};
\draw (262.41,191) node [anchor=north west][inner sep=0.75pt]   [align=left] {$\displaystyle \zeta =1$};

\end{tikzpicture}
\caption{The two sheeted complex dS-SK geometry can be thought of as two static patches smoothly connected at the future horizon. The radial contour along an outgoing Eddington-Finkelstein slice (i.e., a constant $u$ slice) is shown in blue. The radial contour has an outgoing R branch and an incoming L branch.}
\label{fig:ds-sk}
\end{figure}

	\begin{figure}[H]\label{fig:SKcontour}
		\centering
	\begin{tikzpicture}[x=0.75pt,y=0.75pt,yscale=-1,xscale=1]

\draw    (108,153) -- (108,123) ;
\draw [shift={(108,121)}, rotate = 90] [color={rgb, 255:red, 0; green, 0; blue, 0}  ][line width=0.75]    (10.93,-3.29) .. controls (6.95,-1.4) and (3.31,-0.3) .. (0,0) .. controls (3.31,0.3) and (6.95,1.4) .. (10.93,3.29)   ;
\draw    (108,153) -- (140,153) ;
\draw [shift={(142,153)}, rotate = 180] [color={rgb, 255:red, 0; green, 0; blue, 0 }  ][line width=0.75]    (10.93,-3.29) .. controls (6.95,-1.4) and (3.31,-0.3) .. (0,0) .. controls (3.31,0.3) and (6.95,1.4) .. (10.93,3.29)   ;
\draw  [color={rgb, 255:red, 155; green, 25; blue, 25 }  ,draw opacity=1 ] (282,140) .. controls (283.55,141.02) and (285.03,142) .. (286.75,142) .. controls (288.47,142) and (289.95,141.02) .. (291.5,140) .. controls (293.05,138.98) and (294.53,138) .. (296.25,138) .. controls (297.97,138) and (299.45,138.98) .. (301,140) .. controls (302.55,141.02) and (304.03,142) .. (305.75,142) .. controls (307.47,142) and (308.95,141.02) .. (310.5,140) .. controls (312.05,138.98) and (313.53,138) .. (315.25,138) .. controls (316.97,138) and (318.45,138.98) .. (320,140) .. controls (321.55,141.02) and (323.03,142) .. (324.75,142) .. controls (326.47,142) and (327.95,141.02) .. (329.5,140) .. controls (331.05,138.98) and (332.53,138) .. (334.25,138) .. controls (335.97,138) and (337.45,138.98) .. (339,140) .. controls (340.55,141.02) and (342.03,142) .. (343.75,142) .. controls (345.47,142) and (346.95,141.02) .. (348.5,140) .. controls (350.05,138.98) and (351.53,138) .. (353.25,138) .. controls (354.97,138) and (356.45,138.98) .. (358,140) .. controls (359.55,141.02) and (361.03,142) .. (362.75,142) .. controls (364.47,142) and (365.95,141.02) .. (367.5,140) .. controls (369.05,138.98) and (370.53,138) .. (372.25,138) .. controls (373.97,138) and (375.45,138.98) .. (377,140) .. controls (378.55,141.02) and (380.03,142) .. (381.75,142) .. controls (383.47,142) and (384.95,141.02) .. (386.5,140) .. controls (388.05,138.98) and (389.53,138) .. (391.25,138) .. controls (392.97,138) and (394.45,138.98) .. (396,140) .. controls (397.55,141.02) and (399.03,142) .. (400.75,142) .. controls (402.47,142) and (403.95,141.02) .. (405.5,140) .. controls (407.05,138.98) and (408.53,138) .. (410.25,138) .. controls (411.97,138) and (413.45,138.98) .. (415,140) .. controls (416.55,141.02) and (418.03,142) .. (419.75,142) .. controls (421.47,142) and (422.95,141.02) .. (424.5,140) .. controls (426.05,138.98) and (427.53,138) .. (429.25,138) .. controls (430.97,138) and (432.45,138.98) .. (434,140) .. controls (435.55,141.02) and (437.03,142) .. (438.75,142) .. controls (440.47,142) and (441.95,141.02) .. (443.5,140) .. controls (445.05,138.98) and (446.53,138) .. (448.25,138) .. controls (449.97,138) and (451.45,138.98) .. (453,140) .. controls (454.55,141.02) and (456.03,142) .. (457.75,142) .. controls (459.47,142) and (460.95,141.02) .. (462.5,140) .. controls (464.05,138.98) and (465.53,138) .. (467.25,138) .. controls (468.97,138) and (470.45,138.98) .. (472,140) .. controls (473.55,141.02) and (475.03,142) .. (476.75,142) .. controls (478.47,142) and (479.95,141.02) .. (481.5,140) .. controls (483.05,138.98) and (484.53,138) .. (486.25,138) .. controls (487.97,138) and (489.45,138.98) .. (491,140) .. controls (492.55,141.02) and (494.03,142) .. (495.75,142) .. controls (497.47,142) and (498.95,141.02) .. (500.5,140) .. controls (502.05,138.98) and (503.53,138) .. (505.25,138) .. controls (506.97,138) and (508.45,138.98) .. (510,140) .. controls (511.55,141.02) and (513.03,142) .. (514.75,142) .. controls (516.47,142) and (517.95,141.02) .. (519.5,140) .. controls (521.05,138.98) and (522.53,138) .. (524.25,138) .. controls (525.97,138) and (527.45,138.98) .. (529,140) .. controls (530.55,141.02) and (532.03,142) .. (533.75,142) .. controls (535.47,142) and (536.95,141.02) .. (538.5,140) .. controls (540.05,138.98) and (541.53,138) .. (543.25,138) .. controls (544.97,138) and (546.45,138.98) .. (548,140) .. controls (549.55,141.02) and (551.03,142) .. (552.75,142) .. controls (554.47,142) and (555.95,141.02) .. (557.5,140) .. controls (559.05,138.98) and (560.53,138) .. (562.25,138) .. controls (563.97,138) and (565.45,138.98) .. (567,140) ;
\draw  [color={rgb, 255:red, 0; green, 0; blue, 255}  ]  (279,159.02) -- (535.51,159.02) ;
\draw   [color={rgb, 255:red, 0; green, 0; blue, 255}  ]  (279,120) -- (534,120) ;
\draw  [color={rgb, 255:red, 0; green, 0; blue, 255}  ] [draw opacity=0] (534,120) .. controls (540.57,108.6) and (553.1,100.91) .. (567.47,100.91) .. controls (588.66,100.91) and (605.83,117.65) .. (605.83,138.31) .. controls (605.83,158.96) and (588.66,175.71) .. (567.47,175.71) .. controls (554.13,175.71) and (542.39,169.08) .. (535.51,159.02) -- (567.47,138.31) -- cycle ; \draw  [color={rgb, 255:red, 0; green, 0; blue, 255}  ]  (534,120) .. controls (540.57,108.6) and (553.1,100.91) .. (567.47,100.91) .. controls (588.66,100.91) and (605.83,117.65) .. (605.83,138.31) .. controls (605.83,158.96) and (588.66,175.71) .. (567.47,175.71) .. controls (554.13,175.71) and (542.39,169.08) .. (535.51,159.02) ;  
\draw  [fill={rgb, 255:red, 15; green, 1; blue, 1 }  ,fill opacity=1 ] (564.47,141.31) .. controls (564.47,139.65) and (565.81,138.31) .. (567.47,138.31) .. controls (569.12,138.31) and (570.47,139.65) .. (570.47,141.31) .. controls (570.47,142.96) and (569.12,144.31) .. (567.47,144.31) .. controls (565.81,144.31) and (564.47,142.96) .. (564.47,141.31) -- cycle ;
\draw   (403,165) .. controls (400,161.67) and (397,159.67) .. (394,159) .. controls (397,158.33) and (400,156.33) .. (403,153) ;
\draw   (389,114) .. controls (392,117.33) and (395,119.33) .. (398,120) .. controls (395,120.67) and (392,122.67) .. (389,126) ;
\draw   (611.5,132.5) .. controls (608.17,135.5) and (606.17,138.5) .. (605.5,141.5) .. controls (604.83,138.5) and (602.83,135.5) .. (599.5,132.5) ;

\draw (153,141) node [anchor=north west][inner sep=0.75pt]   [align=left] {Re($\displaystyle r$)};
\draw (103,96) node [anchor=north west][inner sep=0.75pt]   [align=left] {Im($\displaystyle r$)};
\draw (252,166) node [anchor=north west][inner sep=0.75pt]   [align=left] {$\displaystyle \zeta ( 0-i\epsilon ) =0$};
\draw (247,100) node [anchor=north west][inner sep=0.75pt]   [align=left] {$\displaystyle \zeta ( 0+i\epsilon ) =1$};
\draw (541,118) node [anchor=north west][inner sep=0.75pt]   [align=left] {$\displaystyle r=H^{-1}$};
\draw (422.83,100) node [anchor=north west][inner sep=0.75pt]   [align=left] {R};
\draw (424.83,165) node [anchor=north west][inner sep=0.75pt]   [align=left] {L};

\end{tikzpicture}

	\caption{The branch cut structure of $\zeta(r)$ in the complex $r$ plane at fixed $u$: branch-cut shown as a wiggly line. We also show the \emph{clockwise} dS-SK radial contour running from $\zeta=1$ to $\zeta=0$ (the blue curve in this figure and in Fig.\ref{fig:ds-sk}). The $\text{Im}\ r>0$ branch is the time-ordered/right branch, whereas the $\text{Im}\ r<0$ branch is the anti-time-ordered/left branch. }
	\end{figure}
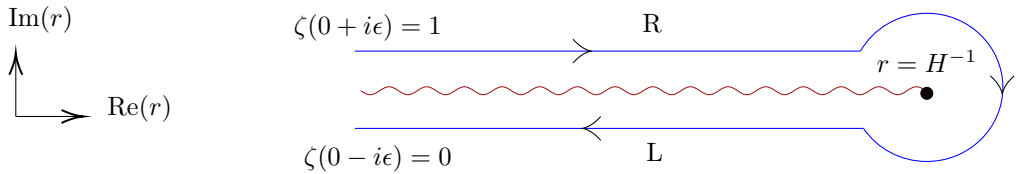

	We now turn to our conjecture for the dS-SK geometry, i.e., the semi-classical saddle point that dominates the quantum gravity path integral for $\SCIP$. What we seek is a real-time analogue of the Gibbons-Hawking-construction\cite{Gibbons:1977mu} as well as gr-SK construction in AdS\cite{Skenderis:2008dg,Skenderis:2008dh,Glorioso:2018mmw,deBoer:2018qqm,Chakrabarty:2019aeu,Ghosh:2020lel}, which would compute for us the cosmological influence phase. Here is the geometry we propose: take two copies of the static patch and stitch them together smoothly at the future horizon (Fig.\ref{fig:ds-sk}). To parametrise this geometry, we complexify the radial coordinate and think of dS-SK as a co-dimension one contour in the complex $r$ plane (Fig.\ref{fig:SKcontour}). To make this precise, let us define a \emph{mock tortoise coordinate} $\zeta$ as follows:
	\begin{equation}
		\zeta(r)=\frac{1}{i\pi}\int\limits^{0-i\epsilon}_r\frac{dr'}{1-r'^2}=\frac{1}{2\pi i}\ln\left(\frac{1-r}{1+r}\right)\label{eq:zeta-def}
	\end{equation}
	This integral has logarithmic branch points at $r=\pm 1$ and we choose its branch-cut to be over the interval $r\in[-1,1]$ on the real line. As shown in Fig.\ref{fig:SKcontour}, our normalisation is such that, if we begin from $0+i\epsilon$ (i.e., just above the midpoint of the branch-cut) and then go clockwise around the branch cut to $0-i\epsilon$ (i.e., just above the midpoint of the branch-cut), we pick up a discontinuity in $\zeta$ equal to negative unity. The choice of the overall constant in \eqref{eq:zeta-def} is such that the real part is 1 on the $R$ static patch (the $r+i\epsilon$ contour), and the real part falls to zero as we move clockwise and turn to traverse the $L$ boundary(the $r-i\epsilon$ contour).\footnote{ The reader should note the use of clockwise contours in the complex $r$ plane for dS, in contrast to the counter-clockwise contours used in the AdS black-brane case. This fact means that we need to be careful to add appropriate minus signs whenever we use the residue theorem, but this inconvenience seems unavoidable given the standard time orientations of the Schwinger-Keldysh contour.}

    We are now ready to state our prescription:
    \begin{equation}\label{eq:dSSKPresc}
       \textbf{Cosmological influence phase} = \textbf{On-shell gravitational action of the dS-SK geometry}\ .
    \end{equation}
    To be clear, on both sides of this equality, we treat observer(s) as prescribed sources, viz., we take it off-shell by freezing its dynamics. Both sides can then be thought of as functionals of the observer multipole moments that emit/detect fields. In the dual quantum mechanics, these multipole moments should be thought of as the `slow macroscopic degrees of freedom' whose influence phase is computed by integrating out the `fast microscopic degrees of freedom'. The solipsistic holography would then imply that we can replace the LHS in the above equality with such an influence phase computed in the dual quantum mechanics. The above statement can then be thought of as giving a  GKPW-like prescription\cite{Gubser:1998bc,Witten:1998qj} for solipsistic holography. The primary aim of this note is to exhibit simple example systems where we can show that the above prescription yields sensible answers.

    Before we turn to examples, we would like to comment on an interesting philosophical point: In this geometric picture, the cosmology reduces entirely to the static patch accessible to the observer, bypassing questions about the rest of the universe (or multi-verse as the case may be). We think of this focus on actual observables as a desirable feature of our proposal, in contrast to traditional descriptions of quantum gravity in dS spacetime phrased in terms of global questions. In the AdS black-brane case, gravitational Schwinger-Keldysh geometry (and its Gibbons-Hawking predecessor) divorces the phenomenology of the exterior from speculations about singularity and BH interior. In a similar vein, our geometric proposal aims at isolating the physics of the static patch from speculations about super-horizon modes, side-stepping the measure problem in cosmology. Our saddle point geometry can be thought of as a way to implement the causal-diamond-based cosmological measures ala Bousso\cite{Bousso:2006ev,Bousso:2006xc}.

	\section{Computing \mathinhead{\SCIP}{SCIP} from on-shell effective action}\label{sec:CIP}

    In the following sections, we will evaluate the on-shell action on the dS-SK geometry described above and show that we get meaningful semi-classical results for the cosmological influence phase $\SCIP$. We will do this in three parts: First, in this section, we will describe a class of systems where observers act as sources for scalar fields. We will describe how the on-shell action can be computed for these systems to yield  $\SCIP$. Next, in Sec\S\ref{sec:flatRR}, we will argue how $\SCIP$ does indeed capture the physics of radiation reaction for a moving dS observer. Finally, in Sec\S\ref{sec:interactions}, we will describe how field interactions could be taken into account.

    Let us begin by examining the mode decomposition in dS-SK geometry. Outgoing modes of frequency $\w$ in the static patch have the form:
	\begin{equation}
		\mathfrak{f}(r,\w,\ell)\ \mathscr{Y}_{\spL}(\Om)\ e^{-i\w u}.
	\end{equation}
	where $\mathfrak{f}(r,\w,\ell)$ is an analytic function of $r$ in the region $0<r\leq 1$: since we are working in outgoing Eddington-Finkelstein coordinates, analyticity near $r=1$ is equivalent to the outgoing boundary condition. Here, the field is decomposed into spherical harmonics $\mathscr{Y}_{\spL}(\Om)$ on $\mathbb{S}^{d-1}$ with labels $\spL\equiv\left\{\ell,\vec{m}\right\} $. The spherical harmonics with label $\ell$ are eigenfunctions of the sphere Laplacian with eigenvalue $-\ell(\ell+d-2)$. Given the analyticity of $\mathfrak{f}(r,\w,\ell)$, the outgoing modes can be analytically continued to the complex $r$ plane without any branch cuts. Consequently, on the dS-SK geometry, the outgoing modes become modes which are \emph{identical} in the right/left branches of the static patch. 
 
     The incoming modes are readily found by time reversing the outgoing mode. Time-reversal isometry of dS-SK geometry is implemented by taking $u\to 2\pi i \zeta-u$ and $\w\to-\w$. We then get an incoming mode of the form
	\begin{equation*}
		\mathfrak{f}(r,-\w,\ell)\ \mathscr{Y}_{\spL}(\Om)\ e^{-2\pi\w\zeta-i\w u}=\mathfrak{f}^\ast(r,\w,\ell)\ \mathscr{Y}_{\spL}(\Om)\ e^{-2\pi\w\zeta-i\w u}.
	\end{equation*} 
    To get the last equality, we have assumed $\mathfrak{f}(r,\w,\ell)$ to be a Fourier transform of a real function. The reader should note here the presence of the non-analytic factor $e^{-2\pi\w\zeta}$, thus resulting in a branch cut for the incoming mode. The incoming mode hence picks up a factor of $e^{2\pi\w}$ if the argument crosses the branch cut from above ($\zeta=1$) to below $(\zeta=0)$, i.e., as we move from right to the left static patch. As we will see below, this is indeed the appropriate  Boltzmann factor for the static patch, encoded automatically in the incoming modes.

    Consider a free scalar field theory on dS$_{d+1}$. Let $\gO(r,\w,\ell)$ denote the radial part of the \emph{outgoing} boundary-to-bulk Green function, i.e., the outgoing field created by a unit point source placed at the south pole. Here, and in what follows, we use the subscript $\nn$ to denote the exponent that characterises the near origin behaviour of the scalar field. More precisely, we define $\gO(r,\w,\ell)$ as the solution of an appropriate radial ODE that obeys the following boundary conditions: at the worldline, we impose a Dirichlet condition
    \begin{equation}\label{eq:PhiBCMain}
	\begin{split}
		\lim_{r\to 0}r^{\nu+\frac{\nn-1}{2}}\gO(r,\w,\ell) &=1\ ,
	\end{split}
\end{equation}
    where we have defined $\nu\equiv \ell+\frac{d}{2}-1$ and taken the  behaviour of the Green function to be $r^{-\nu-\frac{\nn-1}{2}}$ near the source. As an example, for a massless minimal scalar field, we have the fall-off $r^{-(\ell+d-2)}$ corresponding to $\nn=d-1$. 
    
    Apart from the above condition imposed at the origin, we impose analyticity/outgoing boundary conditions at the dS horizon $(r=1)$. Note however that this is \emph{not} the appropriate solution on the dS-SK geometry: its boundaries are not the worldline $+$ dS horizon but rather the right/left worldlines. It is then more natural to impose a \emph{double} Dirichlet boundary condition. To this end, we begin with the most general linear combination of outgoing/incoming modes for the radial part
    \begin{equation}\label{eq:phN_FPbasis}
	\begin{split}
		\phN(\zeta,\w,\spL)=-\gO(r,\w,\ell)\JFb(\w,\spL)+e^{2\pi\w(1-\zeta)}\gs(r,\w,\ell)\JPb(\w,\spL)\ .
	\end{split}
\end{equation}
Here the subscripts $F$ and $P$ denote the sources that radiate to the future and detectors that absorb from the past respectively. We use $\zeta$ to indicate the radial argument of $\phN$ to emphasise that this general linear combination takes two different values in the two branches of dS-SK geometry.

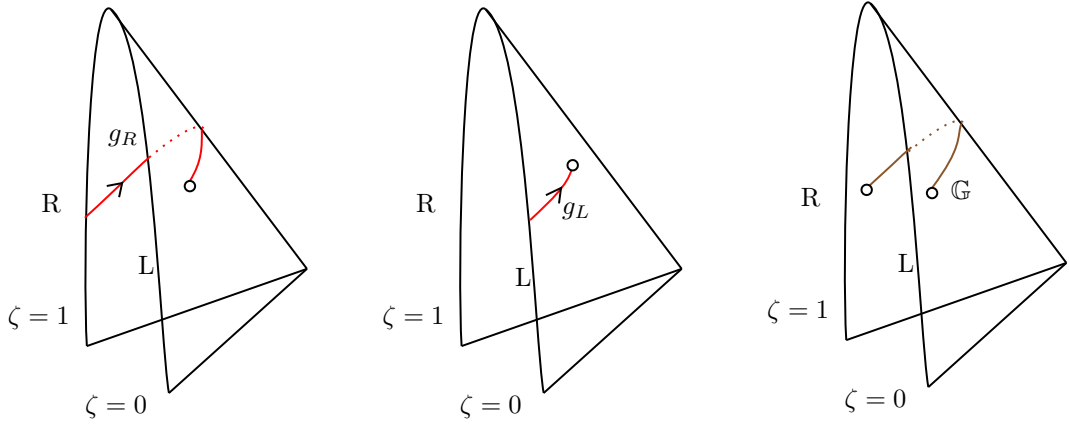
\begin{figure}
    \centering

\tikzset{every picture/.style={line width=0.75pt}} 

\begin{tikzpicture}[x=0.75pt,y=0.75pt,yscale=-1,xscale=1]

\draw    (146.84,32.81) .. controls (135.16,23.61) and (132.59,166.33) .. (134.78,202.86) ;
\draw    (146.84,32.81) .. controls (166.97,47.16) and (171.84,202.54) .. (176.17,226.57) ;
\draw    (245.79,163.88) -- (176.17,226.57) ;
\draw    (134.78,202.86) -- (245.79,163.88) ;
\draw    (146.84,32.81) -- (245.79,163.88) ;
\draw [color={rgb, 255:red, 255; green, 0; blue, 0 }  ,draw opacity=1 ]   (192.74,93.56) .. controls (192.9,102.5) and (192.9,109.5) .. (186.75,119.5) ;
\draw [color={rgb, 255:red, 255; green, 0; blue, 0 }  ,draw opacity=1 ]   (133.78,138.09) .. controls (152.32,121.21) and (158.35,114.41) .. (166.2,107.67) ;
\draw [color={rgb, 255:red, 255; green, 0; blue, 0 }  ,draw opacity=1 ] [dash pattern={on 0.84pt off 2.51pt}]  (166.2,107.67) .. controls (196.64,82.18) and (194.61,96.45) .. (192.74,93.56) ;
\draw    (335.84,32.81) .. controls (324.16,23.61) and (321.59,166.33) .. (323.78,202.86) ;
\draw    (335.84,32.81) .. controls (355.97,47.16) and (360.84,202.54) .. (365.17,226.57) ;
\draw    (434.79,163.88) -- (365.17,226.57) ;
\draw    (323.78,202.86) -- (434.79,163.88) ;
\draw    (335.84,32.81) -- (434.79,163.88) ;
\draw [color={rgb, 255:red, 255; green, 6; blue, 6 }  ,draw opacity=1 ]   (379.13,114.13) .. controls (377.03,121.39) and (369.9,128.5) .. (358.31,139.29) ;
\draw   (366.29,126.69) -- (373.91,123.05) -- (372.9,130.85) ;
\draw    (529.84,29.81) .. controls (518.16,20.61) and (515.59,163.33) .. (517.78,199.86) ;
\draw    (529.84,29.81) .. controls (549.97,44.16) and (554.84,199.54) .. (559.17,223.57) ;
\draw    (628.79,160.88) -- (559.17,223.57) ;
\draw    (517.78,199.86) -- (628.79,160.88) ;
\draw    (529.84,29.81) -- (628.79,160.88) ;
\draw [color={rgb, 255:red, 139; green, 87; blue, 42 }  ,draw opacity=1 ]   (575.74,90.56) .. controls (573.65,97.82) and (576.28,102.5) .. (561.07,123) ;
\draw [color={rgb, 255:red, 139; green, 87; blue, 42 }  ,draw opacity=1 ]   (529.52,121.81) .. controls (548.05,104.93) and (542.81,109.22) .. (550.65,102.48) ;
\draw [color={rgb, 255:red, 139; green, 87; blue, 42 }  ,draw opacity=1 ] [dash pattern={on 0.84pt off 2.51pt}]  (549.2,104.67) .. controls (579.64,79.18) and (577.61,93.45) .. (575.74,90.56) ;
\draw   (189.38,122.13) .. controls (189.38,120.68) and (188.2,119.5) .. (186.75,119.5) .. controls (185.3,119.5) and (184.13,120.68) .. (184.13,122.13) .. controls (184.13,123.57) and (185.3,124.75) .. (186.75,124.75) .. controls (188.2,124.75) and (189.38,123.57) .. (189.38,122.13) -- cycle ;
\draw   (382.34,112.27) .. controls (382.71,110.86) and (381.88,109.43) .. (380.48,109.05) .. controls (379.08,108.68) and (377.64,109.51) .. (377.27,110.91) .. controls (376.89,112.31) and (377.72,113.75) .. (379.13,114.13) .. controls (380.53,114.5) and (381.96,113.67) .. (382.34,112.27) -- cycle ;
\draw   (529.52,121.81) .. controls (528.31,121.01) and (526.68,121.34) .. (525.88,122.55) .. controls (525.08,123.76) and (525.41,125.39) .. (526.62,126.19) .. controls (527.82,126.99) and (529.45,126.66) .. (530.25,125.45) .. controls (531.06,124.24) and (530.73,122.61) .. (529.52,121.81) -- cycle ;
\draw   (563.69,125.63) .. controls (563.69,124.18) and (562.52,123) .. (561.07,123) .. controls (559.62,123) and (558.44,124.18) .. (558.44,125.63) .. controls (558.44,127.07) and (559.62,128.25) .. (561.07,128.25) .. controls (562.52,128.25) and (563.69,127.07) .. (563.69,125.63) -- cycle ;
\draw   (144.82,122.06) -- (152.57,120.58) -- (151.04,127.71) ;

\draw (159.22,155.7) node [anchor=north west][inner sep=0.75pt]   [align=left] {L};
\draw (110.33,124.37) node [anchor=north west][inner sep=0.75pt]   [align=left] {R};
\draw (132.41,226) node [anchor=north west][inner sep=0.75pt]   [align=left] {$\displaystyle \zeta =0$};
\draw (93.41,181) node [anchor=north west][inner sep=0.75pt]   [align=left] {$\displaystyle \zeta =1$};
\draw (349.22,161.7) node [anchor=north west][inner sep=0.75pt]   [align=left] {L};
\draw (299.33,124.37) node [anchor=north west][inner sep=0.75pt]   [align=left] {R};
\draw (321.41,226) node [anchor=north west][inner sep=0.75pt]   [align=left] {$\displaystyle \zeta =0$};
\draw (282.41,181) node [anchor=north west][inner sep=0.75pt]   [align=left] {$\displaystyle \zeta =1$};
\draw (542.22,154.7) node [anchor=north west][inner sep=0.75pt]   [align=left] {L};
\draw (493.33,121.37) node [anchor=north west][inner sep=0.75pt]   [align=left] {R};
\draw (515.41,223) node [anchor=north west][inner sep=0.75pt]   [align=left] {$\displaystyle \zeta =0$};
\draw (476.41,178) node [anchor=north west][inner sep=0.75pt]   [align=left] {$\displaystyle \zeta =1$};
\draw (144,92) node [anchor=north west][inner sep=0.75pt]   [align=left] {$\displaystyle g_{R}$};
\draw (373,129) node [anchor=north west][inner sep=0.75pt]   [align=left] {$\displaystyle g_{L}$};
\draw (569,118) node [anchor=north west][inner sep=0.75pt]   [align=left] {$\displaystyle \mathbb{G}$};

\end{tikzpicture}
    \caption{Propagators in dS-SK geometry: the boundary to bulk propagators are denoted in red and the bulk to bulk propagator is denoted in brown.}
    \label{fig:Propagators}
\end{figure}

The coefficients $\JFb,\JPb$ appearing above can be linked to the left/right sources via the double Dirichlet condition, i.e., at the left/right copy of the worldlines, we  impose
\begin{equation}\label{eq:jdSPtdef}
	\begin{split}
		\multj_L(\w,\spL)\equiv \lim_{\zeta\to 0}r^{\nu+\frac{\nn-1}{2}}\phN &=-\JFb(\w,\spL)+e^{2\pi\w}\JPb(\w,\spL) \ ,\\
		\multj_R(\w,\spL)\equiv \lim_{\zeta\to 1}r^{\nu+\frac{\nn-1}{2}}\phN &=-\JFb(\w,\spL)+\JPb(\w,\spL) \ .
	\end{split}
\end{equation}
Using this, we can then rewrite Eq.\eqref{eq:phN_FPbasis} as $\phN(r,\w,\spL)=g_R\multj_R-g_L\multj_L$, where $g_{R,L}(\zeta,\w,\spL)$ denote the right/left boundary-to-bulk propagators on the dS-SK geometry (see \figurename\ \ref{fig:Propagators}). Our use of the symbol $\multj$  here is a deliberate allusion to the observer's multipole moments. Inverting 
the above relations, we obtain
\begin{align}\label{eq:jPjF}
	\begin{split}
		\JFb(\w,\spL)&\equiv-\Bigl\{(1+n_\w)\multj_R(\w,\spL)-n_\w \multj_L(\w,\spL)\Bigr\}=-\multj_A(\w,\spL)-\left(n_\w+\frac{1}{2}\right)\multj_D(\w,\spL)\\
		\JPb(\w,\spL)&\equiv-n_\w\Bigl\{\multj_R(\w,\spL)- \multj_L(\w,\spL)\Bigr\} =-n_\w\ \multj_D(\w,\spL)\ .
	\end{split}
\end{align}
Here we have introduced the average/difference sources $\multj_A\equiv \frac{1}{2}\multj_R+\frac{1}{2}\multj_L$ and $\multj_D\equiv\multj_R-\multj_L$. 
We note here the natural appearance of the Bose-Einstein factor 
	\begin{align}
	\begin{split}
		n_\w\equiv \frac{1}{e^{2\pi\w}-1} \ .
	\end{split}
\end{align}
Such a factor arises naturally by solving the detailed-balance constraint $1+n_\w=e^{2\pi\w}n_\w$
which equates the probability of spontaneous/stimulated emission by the source to the  absorption probability. The appearance of such a factor is an evidence that dS-SK contour naturally incorporates the thermality of Hawking radiation emitted from the dS horizon\cite{Gibbons:1977mu}. 

Given the solution determined in terms of the multipole moments, we can compute the on-shell action. A scalar system with a requisite exponent is given by an action
\begin{equation}\label{eq:ActPhiMain}
		S=-\frac{1}{2}\int d^{d+1}x \sqrt{-g}\ r^{\nn+1-d}\left\{(\partial\Phi_{\nn})^2+\frac{\Phi_{\nn}^2}{4r^2}\left[(d+\nn-3)(d-\nn-1)-r^2\left(4\mu^2-(\nn+1)^2\right)\right]\right\}\ .
\end{equation}
We will refer to this as a \emph{designer} scalar system with a radially varying dilaton, an appropriate centrifugal potential term, and a mass term. Our motivation to consider this class of actions is that, at specific values of $\nn$ and $\mu$, the above action captures the physics of different field theories. For example, a massive KG scalar of mass $m$ corresponds to setting $\nn=d-1$ and $4m^2=(\nn+1)^2-4\mu^2=d^2-4\mu^2$ in the above action. Another example is the KG scalar field with a conformal mass: this corresponds to setting $\nn=d-1$ and $\mu=\frac{1}{2}$. 

Further,  such actions with different values of $\nn$ and $\mu$ arise naturally   when considering scalar-vector-tensor spherical harmonic decompositions of Maxwell as well as linearised Einstein equations about dS background\cite{Mukohyama:2000ui,Kodama:2003kk,Ishibashi:2004wx}. More precisely, the radial ODEs  in all these sectors coincide with the radial ODE obtained from the above action for some value of $\nn$ and $\mu$ (See table \ref{table:Nmuvalues}).\footnote{We note that, for such massless fields, the exponent $\nn$ and the parameter $\mu$ in all sectors  are related by the condition $4\mu^2=(\nn+1)^2$, i.e., they do not have the last term given in Eq.\eqref{eq:ActPhiMain}.} For these reasons, we consider it worthwhile to study the influence phase obtained by integrating out such designer scalars.

	\begin{table}[H]\caption{$\nn,\mu$ values for different massless fields}
	\vspace{4pt}\centering 
	\begin{tabular}{| c | c | c | c | c | c | c |}
		\hline                        
		& KG Scalar & EM Vector & EM Scalar  & Gravity Tensor & Gravity Vector & Gravity Scalar\\ [0.5ex]
		\hline                  
		$\nn$ & $d-1$ & $d-3$ & $3-d$ & $d-1$  & $1-d$ & $3-d$\\ [1ex]      
		\hline                  
		$\mu$ & $\frac{d}{2}$ & $\frac{d}{2}-1$ & $\frac{d}{2}-2$  & $\frac{d}{2}$ & $\frac{d}{2}-1$ &$\frac{d}{2}-2$\\ [1ex]      
		\hline
	\end{tabular}
	\label{table:Nmuvalues}
\end{table}

The radial ODE for designer scalar systems can be solved exactly in terms of hypergeometric functions. The outgoing boundary to bulk Green function satisfying the boundary condition in Eq.\eqref{eq:PhiBCMain} is given by\cite{Bunch:1978yq,Lopez-Ortega:2006aal,Anninos:2011af}
\begin{equation}\label{eq:GoutImain}
	\begin{split}
		\gO(r,\w,\spL) &=r^{\nu-\frac{\nn}{2}}(1+r)^{-i\w}
		\\
		&\quad\times \frac{\Gamma\left(\frac{1+\nu-
				\mu-i\w}{2}\right)\Gamma\left(\frac{1+\nu+\mu-i\w}{2}\right)}{\Gamma(1-i\w)\Gamma\left(1+\nu \right)}\ {}_2F_1\left[\frac{1+\nu-
			\mu-i\w}{2},\frac{1+\nu+\mu-i\w}{2};1-i\w;1-r^2\right]\ ,
	\end{split}
\end{equation}
where we have  used $\nu\equiv \ell+\frac{d}{2}-1$. This solution is manifestly analytic near $r=1$ and thus this is an outgoing solution. The solution on the dS-SK geometry can then be written as 
\begin{equation}
\begin{split}
\Phi_{\nn}(\zeta,u,\Om)=\sum_\spL\int \frac{d\w}{2\pi}\phN(\zeta,\w,\spL)\mathscr{Y}_{\spL}(\Om)\ e^{-i\w u}
\end{split}
\end{equation}
with the radial part $\phN(\zeta,\w,\spL)$ being given by Eq.\eqref{eq:phN_FPbasis}. This expression can then be substituted back into the designer scalar action given in Eq.\eqref{eq:ActPhiMain}. The  resultant on-shell action itself is formally divergent but can be rendered finite with counterterms localised at the worldlines. We will refer the reader to 
appendices\ \ref{app:DesSc} and \ref{app:Inf} for the technical details of how this is done.
The end result of this evaluation can be cast into the form
\begin{equation}\label{eq:SCIPpt}
	\begin{split}
		\SCIP&=-\sum_\spL\int\frac{d\w}{2\pi} \kO(\w,\ell)\  [\multj_R-\multj_L]^*\ [(1+n_\w)\multj_R-n_\w\multj_L]\\
		&=\sum_\spL\int\frac{d\w}{2\pi} \frac{\kO(\w,\ell)}{1+n_\w}\  \JPb^*\JFb=-\sum_\spL\int\frac{d\w}{2\pi} \kO(\w,\ell)\  \multj_D^*\ \Bigl[\multj_A+\left(n_\w+\frac{1}{2}\right)\multj_D\Bigr]\ ,
	\end{split}
\end{equation}
where we have written the answer in right/left, past/future as well as the average/difference basis. Here, $\kO$ is the boundary 2-point function encoding the effects of radiation reaction (explicit expressions are provided below). For this reason, we will refer to it as the \emph{radiation reaction kernel}. Notable features of this action are as follows:
	\begin{enumerate}
		\item The absence of $\multj_A^{*}(\w,\spL)\multj_A(\w,\spL)$ term is an expected consequence of the collapse rule which demands that the influence phase go to zero when the sources on the two sides are the same.
		\item The $\multj_D^{*}(\w,\spL)\multj_A(\w,\spL)$ coefficient is imaginary, implying that this term is purely dissipative. We will show in the next section that this term captures the radiation reaction experienced by the observer.
		\item The noise term captured by the $\multj_D^{*}(\w,\spL)\multj_D(\w,\spL)$ coefficient is proportional to the dissipative term with a factor. This factor is correctly picked out by our geometry so as to satisfy the Kubo-Martin-Schwinger (KMS) condition for dS.
	\end{enumerate}
	Hence, we claim that this action correctly captures the effect of the environment on the observer. In fact, this is exactly the form expected out of two-point functions of thermal systems\cite{Chaudhuri:2018ymp} and matches with analogous expressions in holographic open systems\cite{Chakrabarty:2019aeu,Jana:2020vyx}.

 	\begin{table}[H]
	\centering
	\caption{$\tau_{dS}$ for $\mu=\frac{d}{2}$ (Massless KG scalar, Gravity tensor sector)}\label{tab:tau0}
	\setlength{\extrarowheight}{2pt}
	\begin{tabular}{|c|c|c|c|c|c|}
		\hline
		$\mu=\frac{d}{2}$&$\ell=1$&$\ell=2$&$\ell=3$&$\ell=4$&$\ell=5$\\[0.5ex]\hline
	$d=3$ & $4$ & $1$ & $\frac{64}{225}$ & $\frac{4}{49}$ & $\frac{256}{11025}$
	\\[0.5ex]
 $d=4$ & $\frac{9\pi^2}{16}$ & $1$ & $\frac{25\pi^2}{1024}$ & $\frac{1}{16}$ & $\frac{441\pi^2}{262144}$
	\\[0.5ex]
	$d=5$ &$\frac{64}{9}$ & $1$ &$ \frac{256}{1225}$ & $\frac{4}{81}$ &
	$\frac{16384}{1334025} $\\[0.5ex]
 $d=6$ & $\frac{225\pi^2}{256}$ & $1$ & $\frac{1225\pi^2}{65536}$ & $\frac{1}{25}$ & $\frac{3969\pi^2}{4194304}$
	\\[0.5ex]
	$d=7$ &$\frac{256}{25}$ & $1$ &$ \frac{16384}{99225} $& $\frac{4}{121}$ &
	$\frac{65536}{9018009}$ \\[0.5ex]
 $d=8$ & $\frac{1225\pi^2}{1024}$ & $1$ & $\frac{3969\pi^2}{262144}$ & $\frac{1}{36}$ & $\frac{9801\pi^2}{16777216}$
	\\[0.5ex]
	$d=9$ &$\frac{16384}{1225}$ & $1$ & $\frac{65536}{480249}$ &
	$\frac{4}{169}$ & $\frac{1048576}{225450225}$\\[0.5ex]
  $d=10$ & $\frac{99225\pi^2}{65536}$ & $1$ & $\frac{53361\pi^2}{4194304}$ & $\frac{1}{49}$ & $\frac{1656369\pi^2}{4294967296}$
	\\[0.5ex]
	$d=11$ &$\frac{65536}{3969} $& $1$ &$ \frac{1048576}{9018009}$ &
	$\frac{4}{225}$ & $\frac{4194304}{1329696225}$ \\[0.5ex]
\hline
	\end{tabular}
\end{table}

\begin{table}[H]
	\centering
	\caption{$\tau_{dS}$ for $\mu=\frac{d}{2}-1$ (EM/Gravity vector sector)}\label{tab:tau1}
	\setlength{\extrarowheight}{2pt}
	\begin{tabular}{|c|c|c|c|c|c|c|}
		\hline
		$\mu=\frac{d}{2}-1$&$\ell=0$&$\ell=1$&$\ell=2$&$\ell=3$&$\ell=4$&$\ell=5$\\[0.5ex]\hline
		$d=3$ & $1$ & $1$ & $\frac{4}{9}$ & $\frac{4}{25}$ & $\frac{64}{1225}$&$ \frac{64}{3969}$
		\\[0.5ex]
        $d=4$ & $\frac{\pi^2}{4}$ & $1$ & $\frac{9\pi^2}{256}$ & $\frac{1}{9}$ & $\frac{225\pi^2}{65536}$&$ \frac{1}{100}$
		\\[0.5ex]
		$d=5$ &$4$ & $1$ &$ \frac{64}{225}$ & $\frac{4}{49}$ &
		$\frac{256}{11025} $& $\frac{64}{9801}$\\[0.5ex]
        $d=6$ & $\frac{9\pi^2}{16}$ & $1$ & $\frac{25\pi^2}{1024}$ & $\frac{1}{16}$ & $\frac{441\pi^2}{262144}$&$ \frac{1}{225}$
		\\[0.5ex]
		$d=7$ &$\frac{64}{9}$ & $1$ &$ \frac{256}{1225} $& $\frac{4}{81}$ &
		$\frac{16384}{1334025}$& $\frac{64}{20449}$ \\[0.5ex]
        $d=8$ & $\frac{225\pi^2}{2566}$ & $1$ & $\frac{1225\pi^2}{65536}$ & $\frac{1}{25}$ & $\frac{3969\pi^2}{4194304}$&$ \frac{1}{441}$
		\\[0.5ex]
		$d=9$ &$ \frac{256}{25} $ & $1$ & $\frac{16384}{99225}$ &
		$\frac{4}{121}$ & $\frac{65536}{9018009}$ & $\frac{64}{38025}$ \\[0.5ex]
  
        $d=10$ & $\frac{1225\pi^2}{1024}$ & $1$ & $\frac{25\pi^2}{1024}$ & $\frac{1}{16}$ & $\frac{441\pi^2}{262144}$&$ \frac{1}{225}$
		\\[0.5ex]
		$d=11$ &$\frac{16384}{1225} $& $1$ &$\frac{65536}{480249} $ &
		$\frac{4}{169}$ & $\frac{1048576}{225450225}$ &$ \frac{64}{65025}$\\[0.5ex] \hline
	\end{tabular}
\end{table}

\begin{table}[H]
	\centering
	\caption{$\tau_{dS}$ for $\mu=\frac{d}{2}-2$ (EM/Gravity scalar sector)}\label{tab:tau2}
	\setlength{\extrarowheight}{2pt}
	\begin{tabular}{|c|c|c|c|c|c|c|}
		\hline
		$\mu=\frac{d}{2}-2$&$\ell=0$&$\ell=1$&$\ell=2$&$\ell=3$&$\ell=4$&$\ell=5$\\[0.5ex]\hline
		$d=3$ & $1$ & $1$ & $\frac{4}{9}$ & $\frac{4}{25}$ & $\frac{64}{1225}$&$ \frac{64}{3969}$
		\\[0.5ex]
        $d=4$ & $1$ &  $\frac{\pi^2}{16}$ & $\frac{1}{4}$ & $\frac{9\pi^2}{1024}$ & $\frac{1}{36}$ & $\frac{225\pi^2}{262144}$
		\\[0.5ex]
		$d=5$ &$1$ & $\frac{4}{9}$ &$  \frac{4}{25}$ & $\frac{64}{1225}$ &
		$\frac{64}{3969} $& $\frac{256}{53361}$\\[0.5ex]
        $d=6$ & $1$ &  $\frac{9\pi^2}{256}$ & $\frac{1}{9}$ & $\frac{225\pi^2}{65536}$ & $\frac{1}{100}$ & $\frac{1225\pi^2}{4194304}$
		\\[0.5ex]
		$d=7$ &$1$ & $\frac{64}{225}$ &$\frac{4}{49} $& $\frac{256}{11025}$ &
		$\frac{64}{9801}$& $\frac{16384}{9018009}$ \\[0.5ex]
        $d=8$ & $1$ &  $\frac{25\pi^2}{1024}$ & $\frac{1}{16}$ & $\frac{441\pi^2}{262144}$ & $\frac{1}{225}$ & $\frac{2025\pi^2}{16777216}$
		\\[0.5ex]
		$d=9$ &$1$ & $\frac{256}{1225}$ & $\frac{4}{81}$ &
		$\frac{16384}{1334025}$ & $\frac{64}{20449}$ & $\frac{65536}{81162081}$ \\[0.5ex]
        $d=10$ & $1$ &  $\frac{1225\pi^2}{65536}$ & $\frac{1}{25}$ & $\frac{3969\pi^2}{4194304}$ & $\frac{1}{441}$ & $\frac{245025\pi^2}{4294967296}$
		\\[0.5ex]
		$d=11$ &$1$& $\frac{16384}{99225} $ &$\frac{4}{121}$ &
		$\frac{65536}{9018009}$ & $\frac{64}{38025}$ &$ \frac{1048576}{2606204601}$\\[0.5ex] \hline
	\end{tabular}
\end{table}

The exact expressions for $\kO$ depend on whether the number of spatial dimensions $d$ is odd or even. For $d$ odd, we have \cite{Lopez-Ortega:2006aal,Anninos:2011af}
\begin{equation}\label{eq:KOutOddMain}
\begin{split}
		\kO|_{\text{Odd d}} &
   =-e^{i\nu\pi}\frac{2\pi i}{\Gamma(\nu)^2} \frac{\Gamma\left(\frac{1+\nu-\mu-i\w}{2}\right)\Gamma\left(\frac{1+\nu+\mu-i\w}{2}\right)}{\Gamma\left(\frac{1-\nu+\mu-i\w}{2}\right)\Gamma\left(\frac{1-\nu-\mu-i\w}{2}\right)} \ ,
	\end{split}
\end{equation}
and for $d$ even, we get
\begin{equation}\label{eq:KoutEvenMain}
	\begin{split}
		\kO|_\text{Even $d$} &=\Delta_\nn(\nu,\mu,\w)\left[\psi^{(0)}\left(\frac{1+\nu-\mu-i\w}{2}\right)+\psi^{(0)}\left(\frac{1+\nu+\mu-i\w}{2}\right)\right.\\ &\left.+\psi^{(0)}\left(\frac{1-\nu-\mu-i\w}{2}\right)+\psi^{(0)}\left(\frac{1-\nu+\mu-i\w}{2}\right)-4\psi^{(0)}(\nu)\right]\ ,
	\end{split}
\end{equation}
where $\psi^{(0)}(z)\equiv \frac{d}{dz}\ln\Gamma(z)$ is the di-gamma function and the function $\Delta_\nn$ is defined via
\begin{equation}\label{eq:ctEvenHMain}
\begin{split}
   \Delta_\nn(n,\mu,\w)&\equiv
  \frac{(-)^n}{\Gamma(n)^2} \frac{\Gamma\left(\frac{1+n-\mu-i\w}{2}\right)\Gamma\left(\frac{1+n+\mu-i\w}{2}\right)}{\Gamma\left(\frac{1-n+\mu-i\w}{2}\right)\Gamma\left(\frac{1-n-\mu-i\w}{2}\right)}=\frac{1}{\Gamma(n)^2}\prod\limits_{k=1}^{n}\left[\frac{\w^2}{4}+\frac{1}{4}(\mu-n+2k-1)^2\right]\\
  &=\Delta_\nn^\ast(n,\mu,\w)\ .
\end{split}
\end{equation} 
The important fact to note about these expressions is that, for all values of $\mu$ appearing in table \ref{table:Nmuvalues} except $\mu=\frac{d}{2}$, we get a nice small $\w$ expansion. For $\mu=\frac{d}{2}$, we still get a small $\w$ expansion for all $\ell>0$: only the $\ell=0$ term has a $1/\w$ behaviour at small $\w$. The physical interpretation of these statements is this: in all these cases except $\ell=0,\mu=\frac{d}{2}$, one obtains a Markovian open system at small $\w$, i.e., a cosmically old observer in dS does not retain any memory of its past.\footnote{\label{ftnt:KGScMemory}The  mild breakdown of small $\w$ expansion in $\mu=\frac{d}{2}$ gives a tail term in the radiation reaction. This has been previously noted in \cite{Burko:2002ge}. This tail term can be avoided either by turning off the monopole moment or by giving the scalar a small mass.} This is an interesting observation, especially in even $d$ where the corresponding flat spacetime problem has memory terms\cite{Birnholtz:2013nta}. This suggests that \emph{the radiation reaction problem in an expanding spacetime is perhaps better behaved than the one in flat spacetime.} In dual quantum mechanics, this predicts that a clean separation of slow/fast degrees of freedom should be possible, at least in the leading large $N$ approximation.

We will now argue that the fluctuations also admit a small $\w$ expansion. To this end, we use $1+n_\w+n_{-\w}=0$ to rewrite the cosmological influence phase as  
\begin{equation}
	\begin{split}
		\SCIP=-\sum_\spL\int\frac{d\w}{2\pi}  \Bigl[\kO(\w, \ell)\   \multj_D^*\multj_A+\frac{1}{2}\left(n_\w+\frac{1}{2}\right)[\kO(\w,\ell)-\kO(-\w,\ell)]\  \multj_D^*\multj_D\Bigr]\ .
	\end{split}
\end{equation}
Since $\w\ n_\w$ has a regular small $\w$ expansion, we conclude from the above expression that $\SCIP$ has a regular small frequency expansion provided  $\kO$ has such an expansion. 
Up to 1st order in $\w$, we have\begin{equation}\label{eq:KoutwexpLeading}
	\begin{split}
		\kO=\kO|_{\w=0}
		-i\ \w\ \tau_{dS}+\ldots
	\end{split}
\end{equation}
where $\tau_{dS}$ can be interpreted as the cosmological decay time-scale for slowly varying multipole moments in dS.\footnote{We tabulate $\tau_{dS}$ for various cases of interest in tables \ref{tab:tau0}, \ref{tab:tau1} and \ref{tab:tau2}.} Due to the dS version of  fluctuation-dissipation theorem, this is also proportional the variance of the Hubble Hawking noise. This fact can be gleaned from the leading $\multj_D^*\multj_D$ term in the cosmological influence phase:
\begin{equation}
	\begin{split}
		\SCIP\supset  i\sum_\spL\frac{\tau_{dS}}{2\pi}\int\frac{d\w}{2\pi}     \multj_D^*\multj_D\ .
	\end{split}
\end{equation}
Using the Hubbard-Stratonovich transformation, we can think of this term arising 
from integrating out a noise field with a time-domain action:
\begin{equation}
	\begin{split}
	\sum_\spL\int du    \left[\frac{i}{2}\frac{\pi}{\tau_{dS}}\nn^2(u)+\multj_D(u)\nn(u)\right]\ .
	\end{split}
\end{equation}
The first term here then shows that $\nn(u)$ behaves like a Gaussian noise field with variance $\frac{\tau_{dS}}{\pi}$.

We will conclude this section by describing how the above analysis can be readily generalised to extended sources in dS, modelled as a sequence of spherical shells. The main technical novelty is that we need the dS bulk-to-bulk propagator to compute the radial part of the field. The expression in Eq.\eqref{eq:phN_FPbasis} is then replaced by a radial contour integral
\begin{equation}\label{eq:dSSKBlkFMain}
	\begin{split}
		\phN(\zeta,\w,\spL)&=\oint  r_0^\nn dr_0\ \mathbb{G}(\zeta|\zeta_0,\w,\spL)\varrho_{_\nn}(\zeta_0,\w,\spL)\ ,
	\end{split}
\end{equation}
where $\varrho_{_\nn}(\zeta_0,\w,\spL)$ is a scalar source spread out over dS-SK geometry and $\mathbb{G}(\zeta|\zeta_0,\w,\spL)$ is the contour-ordered bulk-to-bulk propagator. It is regular everywhere except at $\zeta=\zeta_0$ where its radial derivative has a prescribed discontinuity. Further, we require regularity at the center of the right/left static patches, viz.,
\begin{equation}\label{eq:GbcMain}
	\begin{split}
		\lim_{\zeta\to 0}r^{\nu+\frac{\nn-1}{2}}\mathbb{G} =\lim_{\zeta\to 1}r^{\nu+\frac{\nn-1}{2}}\mathbb{G} =0.
	\end{split}
\end{equation}
These conditions uniquely determine the bulk-to-bulk propagator as specific combinations of the outgoing/incoming waves on either side of the source point $\zeta_0$. An explicit expression in terms of the right/left boundary-to-bulk propagators is (See  appendix \ref{app:Inf})
\begin{equation}\label{eq:BlkBlkdSMain}
	\begin{split}
	 \mathbb{G}(\zeta|\zeta_0,\w,\spL)&=\frac{1}{W_{LR}(\zeta_0,\w,\spL)} g_R(\zeta_\succ,\w,\spL)g_L(\zeta_\prec,\w,\spL)\\
  &\equiv\frac{1}{W_{LR}(\zeta_0,\w,\spL)}\begin{cases}
	 	 g_R(\zeta,\w,\spL)g_L(\zeta_0,\w,\spL) &\quad \text{if}\; \zeta\succ\zeta_0\\
         g_L(\zeta,\w,\spL)g_R(\zeta_0,\w,\spL) &\quad \text{if}\; \zeta\prec\zeta_0
	 \end{cases}	\ .
	\end{split}
\end{equation}
Here the symbols $\succ$ and $\prec$ denote  comparison using the radial contour ordering of dS-SK contour. The construction here is  analogous to the one in vacuum AdS\cite{Witten:1998qj}, as well as the contour-ordered bulk-to-bulk Green function in the SK contour corresponding to planar AdS black holes\cite{Loganayagam:2022zmq,OpenEFT}. 

Once we have the bulk-to-bulk Green function, the on-shell effective action in terms of the extended sources can be computed to be
\begin{equation}
\label{eq:SOnShellBlkSrcMain}
	\begin{split}
		S|_{\textbf{On-shell}}
		&= \frac{1}{2} \sum_\spL\int\frac{d\w}{2\pi}\oint  r^\nn dr\  \oint  r_0^\nn dr_0\ [\varrho_{_\nn}(\zeta,\w,\spL)]^*\mathbb{G}(\zeta|\zeta_0,\w,\spL)\varrho_{_\nn}(\zeta_0,\w,\spL) \ .
	\end{split}
\end{equation}
This is the familiar statement that, in free theories, on-shell action reduces to double integral over sources with an appropriate Green function serving as the kernel.
The above expression can then be evaluated for a sequence of shell sources by performing the radial contour integrals.

We find that the end result of this computation can be written as 
\begin{equation}
	\begin{split}
S|_{\textbf{On-shell}}=\SCIP^\text{Pt}+S_{\text{Int}}\ ,
	\end{split}
\end{equation}
where $\SCIP^\text{Pt}$ is the cosmological influence phase
of Eq.\eqref{eq:SCIPpt}, computed for the point-like source. To get this form, we should define 
 the multipole moments of the extended source via
\begin{equation}\label{eq:ShellMultDefMain}
	\begin{split}
	\multj_R(\w,\spL)\equiv \int_R dr\  r^\nn \Xi_n(r,\w,\spL)\ \left(\frac{1-r}{1+r}\right)^{-\frac{i\w}{2}}\varrho_{_\nn}(\zeta,\w,\spL)  \ ,\\
 \multj_L(\w,\spL) \equiv -\int_L dr\ r^\nn\Xi_n(r,\w,\spL)\ \left(\frac{1-r}{1+r}\right)^{-\frac{i\w}{2}}\varrho_{_\nn}(\zeta,\w,\spL) \  .
	\end{split}
\end{equation}
Here the integrals are over the right/left open static patches and  $\Xi_n(r,\w,\spL)$ is a smearing function given in Eq.\eqref{eq:XinExp}. The remaining terms in the on-shell action (denoted by $S_{\text{Int}}$) encode the conservative self-interactions of the extended source: 
\begin{equation}
\begin{split}
S_{\text{Int}}
  &=  \frac{1}{2}\sum_{\spL}\int\frac{d\w}{2\pi} [\multj_R^\ast \overline{\varphi}_{R,\text{Int}} -\multj_L^\ast\overline{\varphi}_{L,\text{Int}}]\ .
	\end{split}
\end{equation}
Here $\overline{\varphi}_{R/L,\text{Int}}$ denote appropriately radially-averaged mean fields in the right/left static patch which couple to the multipole moments defined in Eq.\eqref{eq:ShellMultDefMain}. Their explicit form is 
\begin{equation}
	\begin{split}
	\overline{\varphi}_{R,\text{Int}}(\w,\spL)\equiv \int_R dr\  r^\nn \Xi_{nn}(r,\w,\spL)\ \left(\frac{1-r}{1+r}\right)^{-\frac{i\w}{2}}\varrho_{_\nn}(\zeta,\w,\spL)  \ ,\\
 \overline{\varphi}_{L,\text{Int}}(\w,\spL) \equiv -\int_L dr\ r^\nn\Xi_{nn}(r,\w,\spL)\ \left(\frac{1-r}{1+r}\right)^{-\frac{i\w}{2}}\varrho_{_\nn}(\zeta,\w,\spL) \  .
	\end{split}
\end{equation}
Here $\Xi_{nn}(r,\w,\spL)$ is a time-reversal invariant Green function given in Eq.\eqref{eq:XinnExp}. 

In the next section, we will describe how these results for extended sources can be used to compute the radiation reaction force felt by a dS observer in arbitrary motion. To that end,
it is convenient to shift back to the standard time domain: we remind the reader that, till now, we have been working in the frequency domain dual to the outgoing EF time $u$. This is related to sources data on standard time slices via
\begin{equation}
	\begin{split}
 \varrho_{_\nn}(\zeta,\w,\spL)=
\int du\ e^{i\w u}\widetilde{\varrho}_{_\nn}(\zeta,t,\spL) =\int dt\ e^{i\w t}\left(\frac{1-r}{1+r}\right)^{\frac{i\w}{2}}\widetilde{\varrho}_{_\nn}(\zeta,t,\spL)\ ,
	\end{split}
\end{equation}
where we have used $u=t+\frac{1}{2}\ln\left(\frac{1-r}{1+r}\right)$. In other words, the combination $\left(\frac{1-r}{1+r}\right)^{-\frac{i\w}{2}}\varrho_{_\nn}(\zeta,\w,\spL)$ appearing in the above definitions is just the Fourier transform with respect to standard time.
Thus, the radiative multipole moments defined in Eq.\eqref{eq:ShellMultDefMain} can equivalently well be thought of as being computed via standard time slices, viz., 
\begin{equation}\label{eq:mulDefTime}
	\begin{split}
	\multj_R(\w,\spL)\equiv \int dt\ e^{i\w t}\int_R dr\  r^\nn \Xi_n(r,i\partial_t,\spL)\ \widetilde{\varrho}_{_\nn}(\zeta,t,\spL)  \ ,\\
 \multj_L(\w,\spL) \equiv -\int dt\ e^{i\w t}\int_L dr\ r^\nn\Xi_n(r,i\partial_t,\spL)\ \widetilde{\varrho}_{_\nn}(\zeta,t,\spL) \  .
	\end{split}
\end{equation}
Similar statements hold for the radially-averaged mean fields $\overline{\varphi}_{R/L,\text{Int}}$.

\section{Radiation reaction and flat space limit}\label{sec:flatRR}
We will turn to the physics of dS radiation reaction(RR), as encoded in the cosmological influence phase $\SCIP$. For simplicity, we will 
consider an arbitrarily moving point-like source of a KG scalar field (i.e., the  $\nn=d-1$ case). In particular, this means that we will no
longer consider the cases of scalars coming from the harmonic decomposition of EM/linearised gravity: the dS RR forces for such cases will be dealt with elsewhere\cite{dSSKvec}. The reason for this restriction is as follows: the analysis of RR force for EM/gravity requires extending the dS multipole expansion to vector/tensor symmetric-trace-free tensors, as well as keeping track of additional velocity dependences in the multipole moments, a task better done elsewhere. Further, we will confine ourselves to  $dS_{d+1}$ with odd values of $d$, where the flat spacetime RR force is known to be time-local\cite{Birnholtz:2013ffa}: in these cases, we can compute the RR force as a local expression in a low curvature (or small $H$) expansion.\footnote{We review the derivation of flat spacetime RR force in appendix\ \ref{app:FlatMult} of this work.}
To ensure clarity in the near-flat limit, we will restore Hubble constant $H$ explicitly in  what follows.

  \begin{figure}
     \centering

\tikzset{every picture/.style={line width=0.75pt}} 

\begin{tikzpicture}[x=0.75pt,y=0.75pt,yscale=-1,xscale=1]

\draw [line width=2.25]    (351,215) .. controls (391,185) and (324,89) .. (364,59) ;
\draw  [fill={rgb, 255:red, 255; green, 191; blue, 0 }  ,fill opacity=1 ] (360,197) .. controls (360,195.62) and (361.12,194.5) .. (362.5,194.5) .. controls (363.88,194.5) and (365,195.62) .. (365,197) .. controls (365,198.38) and (363.88,199.5) .. (362.5,199.5) .. controls (361.12,199.5) and (360,198.38) .. (360,197) -- cycle ;
\draw [color={rgb, 255:red, 255; green, 191; blue, 0 }  ,draw opacity=1 ][line width=1.5]    (362.5,194.5) .. controls (362.49,191.89) and (363.73,190.34) .. (366.24,189.85) .. controls (368.71,189.34) and (369.53,188.23) .. (368.72,186.51) .. controls (368.35,184.13) and (369.31,182.71) .. (371.62,182.25) .. controls (373.89,181.74) and (374.71,180.39) .. (374.06,178.2) .. controls (373.6,175.52) and (374.41,173.93) .. (376.5,173.43) .. controls (378.83,172.23) and (379.43,170.74) .. (378.31,168.96) .. controls (377.24,166.82) and (377.7,165.16) .. (379.7,163.98) .. controls (381.62,162.63) and (381.82,161.09) .. (380.3,159.38) .. controls (378.63,157.56) and (378.53,155.7) .. (379.99,153.8) .. controls (381.34,152.09) and (380.94,150.62) .. (378.81,149.38) .. controls (376.64,148.56) and (375.88,147.04) .. (376.54,144.81) .. controls (376.92,142.4) and (375.9,141.02) .. (373.48,140.67) .. controls (371.1,140.56) and (370.03,139.46) .. (370.26,137.36) .. controls (369.99,134.9) and (368.62,133.74) .. (366.13,133.87) .. controls (364.04,134.38) and (362.74,133.43) .. (362.23,131.02) .. controls (361.62,128.59) and (360.27,127.69) .. (358.19,128.31) .. controls (355.74,128.72) and (354.38,127.85) .. (354.11,125.71) .. controls (353.47,123.33) and (351.95,122.36) .. (349.55,122.79) .. controls (347.16,123.2) and (345.73,122.22) .. (345.26,119.87) .. controls (344.94,117.56) and (343.65,116.57) .. (341.39,116.89) .. controls (339.1,117.08) and (338,116.04) .. (338.11,113.78) .. controls (338.35,111.44) and (337.35,110.04) .. (335.11,109.59) .. controls (332.88,108.62) and (332.38,107.08) .. (333.6,104.97) .. controls (335.21,103.39) and (335.31,101.66) .. (333.92,99.77) .. controls (332.84,97.66) and (333.44,96.1) .. (335.71,95.1) .. controls (338.09,94.32) and (339.03,92.82) .. (338.53,90.6) .. controls (338.3,88.17) and (339.24,87) .. (341.33,87.11) .. controls (343.88,86.78) and (345.04,85.54) .. (344.82,83.38) .. controls (345.32,80.59) and (346.73,79.26) .. (349.06,79.39) .. controls (351.41,79.54) and (352.73,78.41) .. (353,76) -- (353,76) ;
\draw  [fill={rgb, 255:red, 255; green, 191; blue, 0 }  ,fill opacity=1 ] (350.5,76) .. controls (350.5,74.62) and (351.62,73.5) .. (353,73.5) .. controls (354.38,73.5) and (355.5,74.62) .. (355.5,76) .. controls (355.5,77.38) and (354.38,78.5) .. (353,78.5) .. controls (351.62,78.5) and (350.5,77.38) .. (350.5,76) -- cycle ;
\draw    (246,124.5) .. controls (285.8,94.65) and (260.26,45) .. (343.73,71.1) ;
\draw [shift={(345,71.5)}, rotate = 197.62] [color={rgb, 255:red, 0; green, 0; blue, 0 }  ][line width=0.75]    (10.93,-3.29) .. controls (6.95,-1.4) and (3.31,-0.3) .. (0,0) .. controls (3.31,0.3) and (6.95,1.4) .. (10.93,3.29)   ;
\draw    (245,150.5) .. controls (249.95,185.15) and (305.87,215.88) .. (352.59,201.94) ;
\draw [shift={(354,201.5)}, rotate = 162.3] [color={rgb, 255:red, 0; green, 0; blue, 0 }  ][line width=0.75]    (10.93,-3.29) .. controls (6.95,-1.4) and (3.31,-0.3) .. (0,0) .. controls (3.31,0.3) and (6.95,1.4) .. (10.93,3.29)   ;
\draw    (448,116) .. controls (414.51,78.08) and (400.42,99.82) .. (376.12,124.38) ;
\draw [shift={(375,125.5)}, rotate = 315] [color={rgb, 255:red, 0; green, 0; blue, 0 }  ][line width=0.75]    (10.93,-3.29) .. controls (6.95,-1.4) and (3.31,-0.3) .. (0,0) .. controls (3.31,0.3) and (6.95,1.4) .. (10.93,3.29)   ;

\draw (200,132) node [anchor=north west][inner sep=0.75pt]  [font=\scriptsize] [align=left] {Radiative Multipole moments};
\draw (413,118) node [anchor=north west][inner sep=0.75pt]  [font=\scriptsize] [align=left] {Two-point function};

\end{tikzpicture}
     \caption{RR computation has two main ingredients: radiative multipole moments and a two-point function describing how one multipole moment affects another. The black solid line denotes the trajectory of the source.}
     \label{fig:RRcomponents}
 \end{figure}
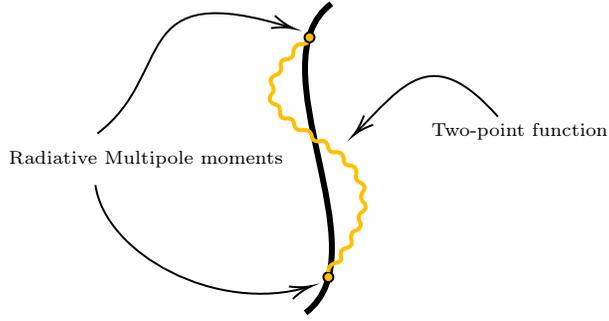

Scalar self-force calculations in curved spacetime are a simple setting to understand the intricacies of 
how RR is affected by curvature back-scattering\cite{Galtsov:1982hwm,Ori:1995ed,Ori:1997be,Quinn:2000wa,EricPoisson_2004}. It is a simple version of the more observationally relevant problem of  gravitational RR force in BH backgrounds, especially in the context of  extreme mass ratio inspirals\cite{Barack:2018yvs}. 
	
Before discussing the results, let us first understand the different components of an action that describe RR. RR is a dissipative term that is found in the $\multj^*_D(\w,\spL)\multj_A(\w,\spL)$ term of the action. This term is a product of two radiative multipole moments and a two-point function connecting them, as shown in Figure \ref{fig:RRcomponents}. In Equation \eqref{eq:SCIPpt}, this retarded two-point function is defined as $\kO$. The radiative multipole moment $\multj$ is obtained by smearing the source distribution with an appropriate function $\Xi_n$ (See Eq.\eqref{eq:mulDefTime}).

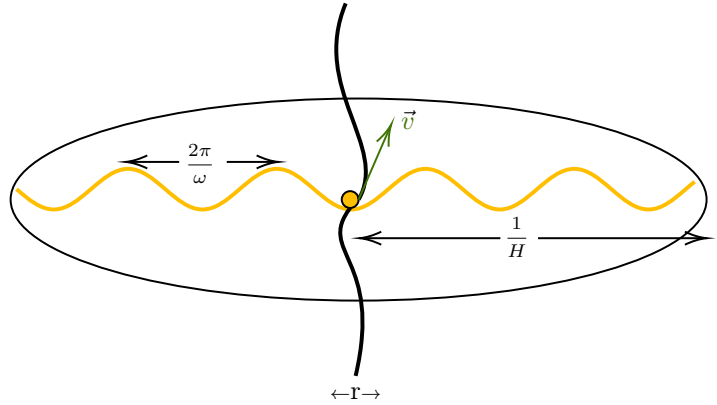
\begin{figure}
     \centering

\tikzset{every picture/.style={line width=0.75pt}} 

\begin{tikzpicture}[x=0.75pt,y=0.75pt,yscale=-1,xscale=1]

\draw   (128.5,179.5) .. controls (128.5,151.33) and (207.07,128.5) .. (304,128.5) .. controls (400.93,128.5) and (479.5,151.33) .. (479.5,179.5) .. controls (479.5,207.67) and (400.93,230.5) .. (304,230.5) .. controls (207.07,230.5) and (128.5,207.67) .. (128.5,179.5) -- cycle ;
\draw  [color={rgb, 255:red, 255; green, 191; blue, 0 }  ,draw opacity=1 ][line width=1.5]  (131.5,174.25) .. controls (137.61,179.5) and (143.46,184.5) .. (150.25,184.5) .. controls (157.04,184.5) and (162.89,179.5) .. (169,174.25) .. controls (175.11,169) and (180.96,164) .. (187.75,164) .. controls (194.54,164) and (200.39,169) .. (206.5,174.25) .. controls (212.61,179.5) and (218.46,184.5) .. (225.25,184.5) .. controls (232.04,184.5) and (237.89,179.5) .. (244,174.25) .. controls (250.11,169) and (255.96,164) .. (262.75,164) .. controls (269.54,164) and (275.39,169) .. (281.5,174.25) .. controls (287.61,179.5) and (293.46,184.5) .. (300.25,184.5) .. controls (307.04,184.5) and (312.89,179.5) .. (319,174.25) .. controls (325.11,169) and (330.96,164) .. (337.75,164) .. controls (344.54,164) and (350.39,169) .. (356.5,174.25) .. controls (362.61,179.5) and (368.46,184.5) .. (375.25,184.5) .. controls (382.04,184.5) and (387.89,179.5) .. (394,174.25) .. controls (400.11,169) and (405.96,164) .. (412.75,164) .. controls (419.54,164) and (425.39,169) .. (431.5,174.25) .. controls (437.61,179.5) and (443.46,184.5) .. (450.25,184.5) .. controls (457.04,184.5) and (462.89,179.5) .. (469,174.25) .. controls (470.51,172.95) and (472,171.67) .. (473.5,170.48) ;
\draw  [fill={rgb, 255:red, 255; green, 191; blue, 0 }  ,fill opacity=1 ] (295.5,179.5) .. controls (295.5,177.15) and (297.4,175.25) .. (299.75,175.25) .. controls (302.1,175.25) and (304,177.15) .. (304,179.5) .. controls (304,181.85) and (302.1,183.75) .. (299.75,183.75) .. controls (297.4,183.75) and (295.5,181.85) .. (295.5,179.5) -- cycle ;
\draw [line width=1.5]    (302.5,268.5) .. controls (316.5,208.5) and (282.5,206.5) .. (299.75,183.75) ;
\draw [line width=1.5]    (297.5,80.5) .. controls (279.5,123.5) and (319.5,152.5) .. (304,179.5) ;
\draw [color={rgb, 255:red, 65; green, 117; blue, 5 }  ,draw opacity=1 ]   (304,179.5) -- (319.7,143.33) ;
\draw [shift={(320.5,141.5)}, rotate = 113.47] [color={rgb, 255:red, 65; green, 117; blue, 5 }  ,draw opacity=1 ][line width=0.75]    (10.93,-3.29) .. controls (6.95,-1.4) and (3.31,-0.3) .. (0,0) .. controls (3.31,0.3) and (6.95,1.4) .. (10.93,3.29)   ;
\draw    (237.5,160.5) -- (261.5,160.5) ;
\draw [shift={(263.5,160.5)}, rotate = 180] [color={rgb, 255:red, 0; green, 0; blue, 0 }  ][line width=0.75]    (10.93,-3.29) .. controls (6.95,-1.4) and (3.31,-0.3) .. (0,0) .. controls (3.31,0.3) and (6.95,1.4) .. (10.93,3.29)   ;
\draw    (213.5,160.5) -- (188.5,160.5) ;
\draw [shift={(186.5,160.5)}, rotate = 360] [color={rgb, 255:red, 0; green, 0; blue, 0 }  ][line width=0.75]    (10.93,-3.29) .. controls (6.95,-1.4) and (3.31,-0.3) .. (0,0) .. controls (3.31,0.3) and (6.95,1.4) .. (10.93,3.29)   ;
\draw    (394.5,199) -- (477.5,199) ;
\draw [shift={(479.5,199)}, rotate = 180] [color={rgb, 255:red, 0; green, 0; blue, 0 }  ][line width=0.75]    (10.93,-3.29) .. controls (6.95,-1.4) and (3.31,-0.3) .. (0,0) .. controls (3.31,0.3) and (6.95,1.4) .. (10.93,3.29)   ;
\draw    (375.5,199) -- (306.5,199) ;
\draw [shift={(304.5,199)}, rotate = 360] [color={rgb, 255:red, 0; green, 0; blue, 0 }  ][line width=0.75]    (10.93,-3.29) .. controls (6.95,-1.4) and (3.31,-0.3) .. (0,0) .. controls (3.31,0.3) and (6.95,1.4) .. (10.93,3.29)   ;

\draw (288,273) node [anchor=north west][inner sep=0.75pt]  [font=\normalsize] [align=left] {{\scriptsize ←}r{\scriptsize →}};
\draw (215,150) node [anchor=north west][inner sep=0.75pt]  [font=\scriptsize] [align=left] {$\displaystyle \frac{2\pi }{\omega }$};
\draw (376,187) node [anchor=north west][inner sep=0.75pt]  [font=\scriptsize] [align=left] {$\displaystyle \frac{1}{H}$};
\draw (324,133) node [anchor=north west][inner sep=0.75pt]   [align=left] {$\displaystyle \vec{\textcolor[rgb]{0.25,0.46,0.02}{v}}$};

\end{tikzpicture}
     \caption{The RR is given in a post-Newtonian expansion: the velocity is taken to be small ($v\ll 1$) and the trajectory is centred about the south pole ($\w r \ll 1$) while the near-flat expansion requires that the curvature effects are small(i.e. $H\ll \w$ and $rH\ll 1$).}
     \label{fig:nearflatexp}
 \end{figure}

Now, let us consider the RR on a single particle. For this, we assume that the particle's trajectory is close to the south pole ($rH\ll 1$), the velocity is small ($v\ll c$), and the wavelength of the radiation is much smaller than the cosmological scale ($\w\gg H$), but much larger than the observer's length scales ($r\w\ll1$). Together with the multipole expansion, these approximations lead to the post-Newtonian expansion of the RR, as shown in Figure \ref{fig:nearflatexp}. In the doubled geometry, this corresponds to two charges moving along their trajectories near their respective south poles. The influence phase on a point charge is then written in terms of the average and difference of positions of the two charges, as well as their time derivatives.

	To obtain the correct flat space answer, we need to check if both the two-point function and the smearing function reduce to their flat space analogues as $H\to 0$. This can be checked from the $H$ expansions of those quantities. We can also obtain further curvature corrections and establish a near-flat expansion. To illustrate how to obtain this expansion easily for any desired order in the Hubble constant, we provide detailed calculations of these expansions in Appendix \ref{app:NearFlatExp}. For the reader's convenience, we also give a review of flat space RR in Appendix \ref{app:FlatGreenFns}.
	
	Given the near expansion of the action, we can, in a controlled fashion, calculate the curvature corrections to the RR force. This force is given by the variation of the Lagrangian with respect to $x_D$. The leading term in the PN expansion of the flat space RR is a scalar version of the Abraham-Lorentz-Dirac force and stems from the dipole moment of the particle. This term, in arbitrary $d$, with the first Hubble correction, we find to be:
\begin{align}
	\begin{split}
		F^i_\text{ALD}&=\frac{(-1)^{\frac{d+1}{2}}}{|\mathbb{S}^{d-1}|d!!(d-2)!!}\Bigg\{ \partial^{d}_t x^i -H^2\frac{d}{6}\left(d^2-1\right)\partial^{d-2}_t x^i\Bigg\}\ .
	\end{split}
\end{align}

    This expression gives an equation of motion that is third-order in derivative for $dS_4$. Even higher time derivatives show up if we include higher-order post-Newtonian corrections. This is a known effect in flat space calculations. In the appendix \ref{app:RadReact}, we give a detailed calculation of 2nd order post-Newtonian corrections along with Hubble corrections. 
    
    The overall sign of the leading term in the force agrees with the fact that this force is dissipative rather than anti-dissipative. To understand this, consider a 1d oscillator with an RR force of the form
\begin{equation}
	\frac{d^2x}{dt^2}+\w_0^2x=\lambda (-1)^\frac{d+1}{2}\frac{d^dx}{dt^d}\ ,
\end{equation}
where we will assume that $d$ is odd and $\lambda>0$. We would now like to argue that the RR force is dissipative. To see this, we note that the above equation is equivalent to a dispersion relation $\w^2=\w_0^2-i\lambda \w^d$, which can be solved approximately to give $\w\approx \w_0-i\lambda\w_0^d$. Since the imaginary part of $\w$ is negative, we conclude that  the above force is indeed dissipative.
 
    As noted in \cite{Birnholtz:2013nta}, the terms in the flat space PN expansion of the RR force add up to give a Poincare covariant expression. This is a non-trivial check for the accuracy of the result as both the structure of the multipole PN expansion and the requirement that it sums up to a covariant result leave little room for error. We similarly find that the curvature corrections obtained along with the flat space results are also tightly constrained: the contributions from our influence phase  non-trivially sum up to expressions covariant under dS metric.

    We refer the reader to appendix\ \ref{app:RadReact} for a detailed enumeration of the terms in the PN expansion. The final RR force then takes the form 
    \[ F^\mu_\text{RR}\equiv \frac{(-)^{\frac{d-1}{2}}}{|\mathbb{S}^{d-1}|(d-2)!!} f^\mu\]
    where $f^\mu$ has an expansion of the form
    \begin{equation}\label{eq:RRrecursion}
\begin{split}
f^\mu_d &= {}^0f^\mu_d -\frac{H^2}{4\times 3!}c_h\ {}^0f^\mu_{d-2} +\frac{H^4}{8\times 6!}[5c_h^2-40(d+2)c_h+32(d+2)(d^2-1)]\ {}^0f^\mu_{d-4}+O(H^6)\ .
\end{split}
\end{equation}	
   Here  $c_h\equiv 12\mu^2+d^2-4$ contains the information about the mass of the scalar and the combinations ${}^0f_d^\mu$'s for odd values of $d$ are (we have listed the expressions up to $d=11$ in appendix \ref{app:RadReact}): 
\begin{equation}
\begin{split}
{}^0f^\mu_1 &\equiv-v^\mu\ ,\\	
{}^0f^\mu_3 &\equiv \frac{P^{\mu\nu}}{3!!}\left\{-a_\nu^{(1)}\right\}-\frac{H^2}{3!!} \left\{v^\mu\right\}\ ,\\
{}^0f^\mu_5&\equiv \frac{P^{\mu\nu}}{5!!}\left\{-a_\nu^{(3)}+5\ (a\cdot a)\  a_\nu^{(1)}+10\ (a\cdot a^{(1)})\ a_\nu\right\}-H^2\frac{P^{\mu\nu}}{5!!}\left\{a_\nu^{(1)}\right\}+\frac{H^4}{5!!} \left\{-v^\mu\right\}\ ,\\
{}^0f^\mu_7&\equiv \frac{P^{\mu\nu}}{7!!}\left\{-a_\nu^{(5)}+14\ (a\cdot a)\  a_\nu^{(3)}+70\ (a\cdot a^{(1)})\ a_\nu^{(2)}+84\ (a\cdot a^{(2)})\ a_\nu^{(1)}+42\ (a\cdot a^{(3)})\ a_\nu\right.\\
&\left.\qquad\qquad+\frac{224}{3}\ (a^{(1)}\cdot a^{(1)})\  a_\nu^{(1)}+105\ (a^{(1)}\cdot a^{(2)})\ a_\nu+O(a^5)\right\}\\
&\quad-H^2\frac{P^{\mu\nu}}{7!!}\left\{a_\nu^{(3)}+15\ (a\cdot a)\  a_\nu^{(1)}+37\ (a\cdot a^{(1)})\ a_\nu\right\}+H^4\frac{P^{\mu\nu}}{7!!}\left\{-a_\nu^{(1)}\right\}-\frac{H^6}{7!!} \left\{v^\mu\right\}\ .\\
\end{split}
\end{equation}
Here $v^\mu=\frac{dx^\mu}{d\tau}$ is the proper velocity of the particle computed using dS metric, $a^\mu\equiv \frac{D^2x^\mu}{D\tau^2}$ is its proper acceleration and $P^{\mu\nu}\equiv g^{\mu\nu}+v^\mu v^\nu$ is the transverse projector to the worldline. We use  $a_\mu^{(k)}\equiv \frac{D^ka^\mu}{D\tau^k}$ to denote the proper-time derivatives of the velocity. All the spacetime dot products are computed using dS metric.

One remarkable feature of the above formula for radiation reaction is the recursive nature of the Hubble corrections. One can see that the $O(H^{2k})$ correction to the force in $d$ dimensions is related to the RR force in $d-2k$ dimensions. It would be interesting to see whether there are specific quantum mechanical models which can reproduce such a recursive structure.
    
One of the consequences of this recurrence is that the $H^{d-1}$ terms in $dS_{d+1}$ resembles the RR effects in $d=1$ flat space. The flat space $d=1$ massless scalar RR was explored in \cite{Burko:2002gf}. However, as noted there, it is inconsistent to assume a constant coupling for a particle coupled to a massless scalar in 2D flat space. Similar issues emerge at $O(H^2)$ in $d=3$ dS\cite{Burko:2002ge} and in general in any $d$ at $O(H^{d-1})$ due to the aforementioned recurrence relation. This is, in turn, related to the breakdown of the small $\w$ expansion of the $\kO$ noted in 
    footnote \ref{ftnt:KGScMemory}: an issue that can be cured by turning on a small mass for the scalar.

    We have checked that the flat limit of the RR force coincides with  the covariant expressions derived in \cite{Birnholtz:2013nta}. However, there are sign mismatches with expressions of \cite{Galtsov:2007zz}\footnote{This sign mismatch was noted by \cite{Birnholtz:2013nta} as well.}. The expressions at order $H^2$ do not match the general curved space force in \cite{Galtsov:2007zz} restricted to dS. Since our methods differ significantly from \cite{Galtsov:2007zz}, we are unable to comment further on the specific source of disagreement.

\section{Interactions}\label{sec:interactions}
    In this section, we will describe how the computation of on-shell action can be extended beyond the free-field examples. In particular, we would like to check that our prescription in Eq.\eqref{eq:dSSKPresc} equating the cosmological influence phase to on-shell action, still works after we  include interactions. We will check this in a simple example: $\phN^3$  theory in dS$_4$. However, as will be clear below, our arguments can be easily adapted to set up perturbative diagrammatics for arbitrary  interactions.

    For $\phN^3$  theory in dS$_4$, we should simply evaluate the bulk on-shell action with the cubic interaction term. At leading order in perturbation theory, the cubic contribution to $\SCIP$ is obtained by substituting the free field solutions into interaction terms of the action.\footnote{The argument here is similar to the one given in appendix $C$ of \cite{Jana:2020vyx} for gr-SK geometry.} The interaction term should then be integrated over the full dS-SK geometry: this is the dS version of a Witten diagram vertex. 
    
    We saw in Sec\S\ref{sec:CIP} that the $\SCIP$ we derived satisfies constraints due to SK collapse and KMS conditions. That this should be true for interacting theories as well is not clear a priori, but we will now show that  these constraints are still satisfied at least at the level of contact diagrams. This is most easily seen in terms of the $P-F$ basis multipole moments defined in Eq.\eqref{eq:jPjF}. In terms of these multipole moments, SK collapse and KMS conditions are equivalent to showing that there are no terms in the action with only  $\JFb$ or only $\JPb$\cite{Chaudhuri:2018ymp}. 
    
    To check these conditions,  we use Eq.\eqref{eq:phN_FPbasis} and write the vertex contribution to the on-shell action as
\begin{align}
\begin{split}
	-\frac{\lambda_3}{3!}\int d^{3+1}x&\;\phN^3=\sum_{\ell_i,m_i}\text{Gaunt}(\ell_i,m_i)\int_{\w_1,\w_2,\w_3}\delta(\w_1+\w_2+\w_3)\\ &\times\left[\mathcal{I}_{FFF}(\w_i,\ell_i,m_i)+\mathcal{I}_{FFP}(\w_i,\ell_i,m_i)+\mathcal{I}_{FPP}(\w_i,\ell_i,m_i)+\mathcal{I}_{PPP}(\w_i,\ell_i,m_i) \right]\ .
\end{split}
\end{align}
	Here the index $i$ runs over $\{1,2,3\}$ and $\text{Gaunt}(\ell_i,m_i)$ are the Gaunt coefficients coming from the integral of 3 spherical harmonics over the sphere (see equation 34.3.22 of \cite{NIST:DLMF}). Time-translation invariance implies that the three frequencies $\w_1$, $\w_2$ and $\w_3$ are constrained by an energy-conserving $\delta$ function. The contributions to the cubic influence phase are given by radial contour integrals, viz., 
\begin{align}
\begin{split}
    \mathcal{I}_{FFF}(\w_i,\ell_i,m_i) &\equiv \frac{\lambda_3}{3!}\oint_{\zeta}\gO (\zeta,\w_1,\ell_1)\ \gO (\zeta,\w_2,\ell_2)\ \gO(\zeta,\w_3, \ell_3) \ \\ 
    & \hspace{3cm} \times \JFb(\w_1,\spL_1) \JFb(\w_2,\spL_2) \JFb(\w_3,\spL_3) \ ,\\
   \mathcal{I}_{FFP}(\w_i,\ell_i,m_i) &\equiv -\frac{\lambda_3}{2!}\oint_{\zeta}e^{2\pi\w_3(1-\zeta)}\ \gO (\zeta,\w_1,\ell_1)\ \gs (\zeta,\w_2,\ell_2)\ \gs(\zeta,\w_3, \ell_3)\ \\
    &\hspace{3cm}\times\JFb(\w_1,\spL_1)\JPb(\w_2,\spL_2)\JPb(\w_3,\spL_3) \ ,  \\
    \mathcal{I}_{FPP}(\w_i,\ell_i,m_i) &\equiv\frac{\lambda_3}{2!}\oint_{\zeta}e^{2\pi(\w_2+\w_3)(1-\zeta)}\ \gO (\zeta,\w_1,\ell_1)\ \gs (\zeta,\w_2,\ell_2)\ \gs(\zeta,\w_3, \ell_3)\ \\
	&\hspace{3cm}\times\JFb(\w_1,\spL_1)\JPb(\w_2,\spL_2)\JPb(\w_3,\spL_3) \ , \\
    \mathcal{I}_{PPP}(\w_i,\ell_i,m_i) &\equiv-\frac{\lambda_3}{3!} \oint_{\zeta}e^{2\pi(\w_1+\w_2+\w_3)(1-\zeta)}\ \gs (\zeta,\w_1,\ell_1)\ \gs (\zeta,\w_2,\ell_2)\ \gs(\zeta,\w_3, \ell_3)\ \\
    &\hspace{3cm}\times\JPb(\w_1,\spL_1)\JPb(\w_2,\spL_2)\JPb(\w_3,\spL_3) \ .\\[2ex]
\end{split}
\end{align}
We now note that, since  $\gO$ is analytic, the integrands in $\mathcal{I}_{FFF}$ and $\mathcal{I}_{PPP}$ are analytic (to see the latter, we use energy conservation). This, in turn, implies that these integrals evaluate to zero by Cauchy's theorem. It is now evident that
this argument generalises  to all contact diagrams of $\phN^n$ type, thus demonstrating our claim about  SK collapse and KMS conditions. 
A similar argument in the AdS blackhole case has been checked also for 
exchange diagrams\cite{Loganayagam:2022zmq,OpenEFT} and it would be interesting to check whether a similar claim holds here. Further, it would also be interesting to study the correction to the radiation reaction due to such non-linear interactions\cite{dSSKint}.

	\section{Summary and Discussion}\label{sec:conc}
	In this work, we have proposed  a de Sitter-Schwinger Keldysh(dS-SK) geometry formed by two copies of the static patch stitched together at their future horizons. We then showed how  the influence phase of a dS observer could be obtained by evaluating the on-shell action on this geometry. Our proposal yields results that pass a variety of checks: first, from a broad structural point of view, it satisfies  the constraints imposed on it from bulk unitarity (SK collapse) and the dS version of Kubo-Martin-Schwinger (KMS) conditions. 
 
    Another check is the flat space limit,  where we showed that, for point-like sources, the dissipative part of the action correctly produces the flat space radiation reaction. This also allows us to calculate Hubble corrections to the radiation reaction in odd spatial dimensions, and show that they combine into generally covariant expressions on the dS background, which serves another non-trivial check on our computation. As a technical aside, we have also shown how we can counter-term the influence phase for localised sources with multipole moments by using a Dirac-Deitweiler-Whiting type  decomposition of the dS Green functions.

    Our analysis can readily be extended to interactions, following techniques invented in the AdS context\cite{Jana:2020vyx,Jana:2021xyn,Loganayagam:2022zmq,OpenEFT}, as we sketched in the main text. This aspect will be explored in detail elsewhere\cite{dSSKint}. It would also be interesting to explore whether the familiar tools of conformal invariance, e.g., conformal block decomposition, can shed more light on the structure of radiation reaction at a non-linear level. On the face of it, the presence of the observer breaks the dS isometries to just rotations/time-translations around the observer's worldline. But the re-emergence of the full dS isometry in the effective action that we described above, suggests that conformal techniques could be fruitfully exploited to understand the structure of Hubble corrections to the radiation reaction. There is also the question of extending our analysis to gauge theories and linearised gravity, which we shall pursue in a subsequent work\cite{dSSKvec}. 

    In this work, we have advocated a point of view that the real-world cosmology is fruitfully framed in terms of a cosmological influence phase $\SCIP$ for an observer's worldline. It is interesting to ask whether realistic FLRW cosmology from $\Lambda$-CDM and the CMB phenomenology can indeed be rewritten in these terms. To this end, it would be interesting to extend our analysis to time-varying cosmological spacetimes: perhaps, one should begin by extending our framework to simpler time-dependent extensions involving sudden/adiabatic approximations. More broadly, we can enquire of the role played by radiation reaction in cosmology. Understanding the gravitational radiation reaction at the galactic/extra-galactic scales might be crucial to predict the stochastic gravitational wave background\cite{Yan:2023slm,Yang:2023eqi,Saeedzadeh:2023biq}.  

    Much of what we say about dS radiation reaction can readily be adapted to the AdS case, with a change of signs. This statement is expected to be true at short times, where the cosmological constant can be treated perturbatively, and its sign does not result in any qualitatively new features. Thus, at short times, we expect generally covariant expressions for the radiation reaction felt by an AdS observer, very similar to the ones we derive in this work. However, we expect qualitative differences at long time-scales due to reflection at AdS asymptotia, resulting in long-time tails in radiation reaction. Further, we do not expect an analogue of dS Hawking radiation in AdS. It might be worthwhile to make these intuitions more precise and understand the dual CFT interpretation of these statements. This would be a good test of  the existing proposals describing bulk observers within AdS/CFT\cite{Papadodimas:2012aq,Maxfield:2017rkn,Jafferis:2020ora}.

    We began this note by motivating our work in the context of solipsistic holography. We see the results here as a first step towards constructing an open system whose details can be compared against proposed dual quantum mechanical models. Following the examples like BFSS
    matrix model\cite{Banks:1996vh}\footnote{See \cite{Taylor:2001vb,Ydri:2017ncg} for a review and \cite{Susskind:2021dfc,Brahma:2022ikl} for matrix model proposals in dS. }, it is natural to expect some sort of a large $N$ matrix quantum mechanics to give rise to the same influence phase as what we derive here.\footnote{An especially interesting avenue is to cleverly use known AdS/CFT to derive dS duals: this can be done either by embedding a dS bubble within AdS\cite{Anninos:2017hhn,Anninos:2018svg,Anninos:2022ujl,Anninos:2022hqo,Sahu:2023fbx} or by $TT$-like deformation of the dual CFT\cite{Gorbenko:2018oov,Lewkowycz:2019xse,Shyam:2021ciy,Coleman:2021nor,Torroba:2022jrk}. It would be interesting to see how the radiation reaction viewpoint we advocate here fits within such proposals. We thank the referee for bringing some of these works to our attention.} To check this, it would be good to construct a formalism for computing the influence phase of slow macroscopic observables in a large $N$ matrix model: our computations suggest that a clean separation of slow/fast modes is possible at least when there is a dual gravity description. These slow observables describing the dS observer should not be entirely gauge-invariant but rather have the structure of  partially gauge-fixed probes\cite{Ferrari:2013aba,Ferrari:2013waa,Ferrari:2016vcl}.
   Whether this is so is yet to be seen.

\section*{Acknowledgements}
We would like to thank Ofek Birnholtz, Tuneer Chakraborty, Chandramouli Chowdhury, Victor Godet, Chandan Jana, Godwin Martin, Shiraz Minwalla, Priyadarshi Paul, Suvrat Raju, Mukund Rangamani, Joseph Samuel, Ashoke Sen, Shivam Sharma, Akhil Sivakumar, Sandip Trivedi and Spenta Wadia for valuable discussions. RL would like to thank the organisers of All Lambdas Holography @ Prague 2021 online workshop for discussions related to this work. We acknowledge support of the Department of Atomic Energy, Government of India, under project no. RTI4001, and would also like to acknowledge our debt to the people of India for
their steady and generous support to research in the basic sciences.

\appendix
\section{STF tensors and multipole expansion}\label{app:FlatMult}

We will begin by reviewing the notion of symmetric trace-free tensors, which are the appropriate tools to discuss multipole expansion. The $d=3$ version of this story is discussed in a variety of places.\footnote{See \cite{poisson2014gravity} for a textbook discussion. We will refer the reader to \cite{higuchi1986A,Rubin:1984tc,rubin1984eigenvalues,Frye:2012jj,Birnholtz:2013ffa,Birnholtz:2013nta,Bhattacharyya:2016nhn,PhysRevD.107.104031} for a discussion of STF tensors in general dimensions.} The generalisation to arbitrary dimensions is straightforward, if somewhat involved. In the course of this work, we had to use a variety of identities involving STF tensors in arbitrary dimensions scattered across these references. The goal of this section is to  review this theory for the reader's benefit.

We will begin with a more traditional account of electrostatic multipole expansion in $\mathbb{R}^d$ via orthonormal spherical harmonics on $\mathbb{S}^{d-1}$. This is the generalisation of familiar multipole expansion in $d=3$, and we will use it to  set the stage for a more modern account of multipole expansion using symmetric, trace-free (STF) tensors in the later subsections. We conclude this appendix with a discussion of radiation reaction in flat spacetime using these tools.

\subsection{Orthonormal Spherical harmonics on \mathinhead{\mathbb{S}^{d-1}}{S(d-1)}}
Let us begin by considering the problem of electrostatics in $\mathbb{R}^d$. Our goal in this subsection would be to describe the multipole expansion in this case, given an orthonormal basis of spherical harmonics on $\mathbb{S}^{d-1}$. Later in this subsection, we will give an explicit construction of such an orthonormal basis, which can, in principle, be used in explicit computations.

Given a charge distribution $\rho(\vec{r})$, the electric potential produced by such a distribution is given in terms of the Newton-Coulomb integral
\begin{equation}
\phi(\vec{r})=\int d^dr_0 \frac{\rho(\vec{r}_0)}{(d-2)|\mathbb{S}^{d-1}||\vec{r}-\vec{r}_0|^{d-2}}\ .
\end{equation}
Here, we have denoted the volume of a unit sphere $\mathbb{S}^{d-1}$ via
\begin{equation}|\mathbb{S}^{d-1}|\equiv \frac{2\pi^{\frac{d}{2}}}{\Gamma(\frac{d}{2})}\ ,
\end{equation}
 and have fixed our normalisations such that the Poisson equation takes the form $\nabla^2\phi=-\rho$. While the above integral is indeed the right, a more useful answer is obtained by performing
a multipole expansion of the Newton-Coulomb potential in terms of Legendre polynomials. In $d=3$, this is a well-known statement from undergraduate physics courses, and we will now describe a quick way to generalise this statement to arbitrary dimensions.

To this end, consider a simple problem where the answer due to multipole expansion is straightforward: we imagine a spherical shell of radius $R$ in $\mathbb{R}^d$ carrying a surface charge density $\sigma_{\ell\vec{m}}(\hat{r})$ proportional to a spherical harmonic $\mathscr{Y}_{\ell\vec{m}}(\hat{r})$, i.e., a spherical harmonic which under the sphere laplacian has an eigenvalue $-\ell(\ell+d-2)$ and we use $\vec{m}$ to denote the additional labels required to furnish an orthonormal basis within this eigenspace. The above eigenvalue follows from demanding that $r^\ell\mathscr{Y}_{\ell\vec{m}}(\hat{r})$
be a harmonic function annihilated by the Laplacian operator
\begin{equation}\begin{split}
 \nabla^2_{\mathbb{R}^d}\equiv \frac{1}{r^{d-1}}\frac{\partial}{\partial r} r^{d-1}\frac{\partial}{\partial r}+\frac{1}{r^2}\nabla^2_{\mathbb{S}^{d-1}}\ .
\end{split}\end{equation}

By symmetry, the potential due to such a problem should also be proportional to the same spherical harmonic as the charge distribution. The potential  should be a harmonic function for $r\neq R$, regular at the origin, vanishing at infinity, be continuous at $r=R$ but have a derivative discontinuity at the shell equal to the charge density. These requirements uniquely determine the solution to be 
\begin{equation}\begin{split}
 \frac{R\sigma_{\ell\vec{m}}(\hat{r})}{(2\ell+d-2)}\Bigl\{\Theta(r<R)\frac{r^\ell}{R^\ell}+\Theta(r>R)\frac{R^{\ell+d-2}}{r^{\ell+d-2}}\Bigr\} \ .
\end{split}\end{equation}
This answer can be generalised to an arbitrary charge distribution, once it is realised that any distribution can be built shell by shell and $\ell$ by $\ell$. Using an orthonormal basis of spherical harmonics to do the projection to every $\ell$, we can then write the potential  for an arbitrary charge distribution as
\begin{equation}\begin{split}
\int d^dr_0\ \rho(\vec{r}_0)\ \sum_{\ell\vec{m}} \frac{\mathscr{Y}_{\ell\vec{m}}(\hat{r})\mathscr{Y}^\ast_{\ell\vec{m}}(\hat{r}_0)}{(2\ell+d-2)r_0^{d-2}}\Bigl\{\Theta(r<r_0)\frac{r^\ell}{r_0^\ell}+\Theta(r>R)\frac{r_0^{\ell+d-2}}{r^{\ell+d-2}}\Bigr\} \ .
\end{split}\end{equation}
Comparing this against the Newton-Coulomb integral, we obtain the multipole expansion formula in $\mathbb{R}^d$ :
\begin{equation}\begin{split}
 &\frac{1}{(d-2)|\mathbb{S}^{d-1}||\vec{r}-\vec{r}_0|^{d-2}}\ \\
&\quad= \sum_{\ell\vec{m}} \frac{\mathscr{Y}_{\ell\vec{m}}(\hat{r})\mathscr{Y}^\ast_{\ell\vec{m}}(\hat{r}_0)}{(2\ell+d-2)r_0^{d-2}}\Bigl\{\Theta(r<r_0)\frac{r^\ell}{r_0^\ell}+\Theta(r>R)\frac{r_0^{\ell+d-2}}{r^{\ell+d-2}}\Bigr\}\ .
\end{split}\end{equation}
If we define the spherical multipole moments of the charge distribution $\rho(\vec{r})$ by 
\begin{equation}\label{eq:EstaticMp}\begin{split}
q_{_{\ell\vec{m}}}\equiv \frac{1}{2\ell+d-2}\int d^dr_0\ \rho(\vec{r}_0)\ r_0^\ell \mathscr{Y}^\ast_{\ell\vec{m}}(\hat{r}_0) \ ,
\end{split}\end{equation}
we can write the potential far-outside the charge distribution as 
\begin{equation}\begin{split}
 \sum_{\ell\vec{m}} \frac{1}{r^{\ell+d-2}}\ q_{_{\ell\vec{m}}}\mathscr{Y}_{\ell\vec{m}}(\hat{r})\ .
\end{split}\end{equation}
This is the basic content of multipole expansion in electrostatics. However, to actually compute these multipole moments for a give charge distribution $\rho(\vec{r})$, we will need the explicit form of the spherical harmonics  $\mathscr{Y}_{\ell\vec{m}}(\hat{r})$ on $\mathbb{S}^{d-1}$:  we will now proceed to address this in the rest the subsection. 

The first step in constructing the spherical harmonics is to derive the most symmetric among them: the Legendre polynomials. We will do this by recasting the above expansion in terms of the Legendre polynomial. In the formula above, the sum over orthonormal spherical harmonics of a given $\ell$ can be performed through a higher dimensional generalisation of the addition theorem for spherical harmonics, viz., 
\begin{equation}\begin{split}
\sum_{\vec{m}} \mathscr{Y}_{\ell\vec{m}}(\hat{r})\mathscr{Y}^\ast_{\ell\vec{m}}(\hat{r}_0)=\frac{N_{HH}(d,\ell)}{|\mathbb{S}^{d-1}|}P_\ell(d,\hat{r}\cdot\hat{r}_0)\ .
\end{split}\end{equation}
Here $N_{HH}(d,\ell)$ is the number of orthonormal spherical harmonics of degree $\ell$, with the notation here inspired by the fact that it is also the number of linearly independent, homogeneous, harmonic polynomials (HHPs) of degree $\ell$ in $\mathbb{R}^d$. We will elaborate on this and get an explicit expression for $N_{HH}(d,\ell)$  below. For now, we move on to note that $P_\ell(d,x)$ is the generalisation of the Legendre polynomial to $\mathbb{R}^d$: it is the unique spherical harmonic  invariant under $SO(d-1)$ rotations which keep two poles of $\mathbb{S}^{d-1}$ fixed and is normalised to unity at the north pole, i.e., $P_\ell(d,x=1)\equiv 1$. 

With the above definitions, we can argue for the above addition theorem as follows: first of all, the  sum over orthonormal spherical harmonics of a given $\ell$ should be a spherical harmonic which only depends on the relative orientation of $\hat{r}$ and $\hat{r}_0$ and hence, the above sum should be proportional to $P_\ell(d,\hat{r}\cdot\hat{r}_0)$. The constant of proportionality can then be fixed by setting $\hat{r}=\hat{r}_0$ and integrating over the sphere  $\mathbb{S}^{d-1}$ using orthonormality. 

As a corollary of the above addition theorem, we note the following formula for the inner product between Legendre harmonics of two different orientations: 
\begin{equation}\begin{split}
\int_{\mathbb{S}^{d-1}} P_\ell(d,\hat{r}\cdot\hat{r}_0)P_{\ell'}(d,\hat{r}\cdot\hat{r}'_0)
=\delta_{\ell\ell'}\frac{|\mathbb{S}^{d-1}|}{N_{HH}(d,\ell)}P_\ell(d,\hat{r}_0\cdot\hat{r}'_0)\ .
\end{split}\end{equation}
This statement follows directly by the use of addition theorem followed by the fact that $\mathscr{Y}_{\ell\vec{m}}(\hat{r})$ are assumed to be orthonormal.For $\hat{r}_0=\hat{r}_0'$, we get the Legendre orthogonality relation
\begin{equation}\label{Eq:LegOrtho}\begin{split}
\int_0^\pi d\vartheta\ \sin^{d-2}\vartheta\  P_\ell(d,\cos\vartheta)P_{\ell'}(d,\cos\vartheta)
=\delta_{\ell\ell'}\frac{|\mathbb{S}^{d-1}|}{|\mathbb{S}^{d-2}|N_{HH}(d,\ell)}\ .
\end{split}\end{equation}

With the addition theorem, we can recast the multipole expansion in terms of  the Legendre polynomial as\footnote{This expansion is often used to define Gegenbauer polynomials $C^\mu_\ell(z)$, which differ from the generalised Legendre polynomials introduced here merely by an overall normalisation. These polynomials are also proportional to the associated Legendre functions. The explicit relations are given by
\begin{equation}\begin{split}
C^{\frac{d}{2}-1}_\ell(z)\equiv (d-2)\frac{N_{HH}(d,\ell)}{2\ell+d-2}P_\ell(d,z)\ ,\quad P_\lambda^{-\mu}(z)\equiv \frac{\left(\sqrt{1-z^2}\right)^{\mu}}{2^\mu \mu!}P_{\lambda-\mu}(2\mu+3;z)\ .
\end{split}\end{equation}} 
\begin{equation}\begin{split}
 &\frac{1}{(d-2)|\vec{r}-\vec{r}_0|^{d-2}}\ \\
&\quad= \sum_{\ell} \frac{N_{HH}(d,\ell)P_\ell(d,\hat{r}\cdot\hat{r}_0)}{(2\ell+d-2)r_0^{d-2}}\Bigl\{\Theta(r<r_0)\frac{r^\ell}{r_0^\ell}+\Theta(r>R)\frac{r_0^{\ell+d-2}}{r^{\ell+d-2}}\Bigr\}\ .
\end{split}\end{equation}
As is well-known in $d=3$ case, this series expansion can be used to derive an explicit expression for $P_\ell(d,x)$.

The steps involved are as follows: we take the case $r_0<r$, set  $t=\frac{r_0}{r}<1$ and $x=\hat{r}\cdot\hat{r}_0$  to write
\begin{equation}\begin{split}
N_{HH}(d,\ell)P_\ell(d,x)&=(2\ell+d-2)\times \text{Coefficient of $t^\ell$ in } \frac{1}{(d-2)(1-2xt+t^2)^{\frac{d}{2}-1}} \ .
\end{split}\end{equation}
To extract the $t^\ell$ coefficient, we use 
\begin{equation}\begin{split}
\frac{(2\ell+d-2)}{(d-2)(1-2xt+t^2)^{\frac{d}{2}-1}}
&=\frac{\ell+\frac{d}{2}-1}{\Gamma(\frac{d}{2})}\int_0^\infty ds\  s^{\frac{d}{2}-2}\ e^{-s+2xst-st^2}\ ,  \end{split}\end{equation}
expand the exponentials involving $t$ and integrate to obtain
\begin{equation}\begin{split}
N_{HH}(d,\ell)P_\ell(d,x)
&=\frac{\ell+\frac{d}{2}-1}{\Gamma(\frac{d}{2})}\sum_k \int_0^\infty ds\  s^{\frac{d}{2}-2}\ e^{-s}
\frac{(2xs)^{\ell-2k}}{(\ell-2k)!}\frac{(-s)^k}{k!}\\
&=\frac{2^\ell\Gamma\left(\ell+\frac{d}{2}\right)}{\Gamma(\frac{d}{2})}\sum_k \frac{\Gamma\left(\ell+\frac{d}{2}-1-k\right)}{\Gamma(\ell+\frac{d}{2}-1)}
\frac{(-)^k}{2^{2k}k!}\frac{x^{\ell-2k}}{(\ell-2k)!} \ .
\end{split}\end{equation}
Here, the sum over $k$ runs from $k=0$ and until the combination $\ell-2k$ is non-negative. Defining the normalisation factor\footnote{The interpretation of this ubiquitous normalisation factor will become clearer when we describe STF tensors in the next subsection. For now, we will note that $\nn_{d,\ell}$ is an inverse integer which has the following alternate forms \begin{equation}\begin{split}
\nn_{d,\ell}\equiv \frac{|\mathbb{S}^{d+2\ell-1}|}{|\mathbb{S}^1|^\ell|\mathbb{S}^{d-1}|}=\frac{(d-2)!!}{(d+2\ell-2)!!}\ .
\end{split}\end{equation}}
\begin{equation}\begin{split}
\nn_{d,\ell}\equiv \frac{\Gamma(\frac{d}{2})}{2^\ell\Gamma\left(\ell+\frac{d}{2}\right)}  \ ,\quad 
\nu\equiv \frac{d}{2}+\ell-1\ ,
\end{split}\end{equation}
we finally obtain an explicit expression for the generalised Legendre polynomial as
\begin{equation}\label{Eq:LegGegen}\begin{split}
\nn_{d,\ell}N_{HH}(d,\ell)P_\ell(d,x)&=\sum_k \frac{\Gamma\left(\nu-k\right)}{2^{2k}k!\Gamma(\nu)}
\frac{(-)^k x^{\ell-2k}}{(\ell-2k)!} \ .
\end{split}\end{equation}
Incidentally, the same expansion at $x=1$ also gives the number of orthonormal spherical harmonics of degree $\ell$ as 
\begin{equation}\label{Eq:NHHdef}\begin{split}
N_{HH}(d,\ell)&=\frac{2\ell+d-2}{d-2}\times \text{Coefficient of $t^\ell$ in } \frac{1}{(1-t)^{d-2}}=\frac{2\ell+d-2}{d-2}\binom{\ell+d-3}{\ell} \ .
\end{split}\end{equation}
This finishes our construction of the Legendre harmonic on $\mathbb{S}^{d-1}$.

Next, we will give a recursive construction of a complete orthonormal basis of spherical harmonics on $\mathbb{S}^{d-1}$, just using the Legendre polynomials constructed above. We begin with an explicit spherical coordinate system in $\mathbb{R}^d$ given by
\begin{equation}\begin{split}
x_1&=r\ \sin\vartheta_{d-2}\ \sin\vartheta_{d-3}\ \ldots\ \sin\vartheta_2\ \sin\vartheta_1\ \cos\varphi\ ,\\
x_2&=r\ \sin\vartheta_{d-2}\ \sin\vartheta_{d-3}\ \ldots\ \sin\vartheta_2\ \sin\vartheta_1\ \sin\varphi\ ,\\
x_3&=r\ \sin\vartheta_{d-2}\ \sin\vartheta_{d-3}\ \ldots\ \sin\vartheta_2\ \cos\vartheta_1\ ,\\
x_4&=r\ \sin\vartheta_{d-2}\ \sin\vartheta_{d-3}\ \ldots\ \cos\vartheta_2\  ,\\
& \ldots\ ,\\
x_{d-2}&=r\ \sin\vartheta_{d-2}\ \sin\vartheta_{d-3}\ \cos\vartheta_{d-4}\ ,\\
x_{d-1}&=r\ \sin\vartheta_{d-2}\  \cos\vartheta_{d-3}\ ,\\
x_{d}&=r\ \cos\vartheta_{d-2}\ .
\end{split}\end{equation}
Here the radius $r$ varies from $0$ to $\infty$ whereas the allowed values of  angles is $\vartheta_i\in[0,\pi]$ and $\varphi\in[0,2\pi)$.
In these coordinates, we can write the metric of 
$\mathbb{S}^{d-1}$ as 
\begin{equation}\begin{split} d\Om_{d-1}^2&= d\vartheta_{d-2}^2 
+\sin^2\vartheta_{d-2}d\Om_{d-2}^2\\
&= d\vartheta_{d-2}^2 
+\sin^2\vartheta_{d-2} d\vartheta_{d-3}^2 
+ \sin^2\vartheta_{d-2} \sin^2\vartheta_{d-3} d\vartheta_{d-4}^2 +\ldots\\
&\quad+\prod_{k=j+1}^{d-2}\sin^2\vartheta_k\ d\vartheta_j^2+\ldots+\prod_{k=1}^{d-2}\sin^2\vartheta_k\ d\varphi^2
.
\end{split}\end{equation}
The volume form 
\begin{equation}\begin{split} 
\int_{\mathbb{S}^{d-1}}(\ldots)\equiv \int d\vartheta_1\wedge d\vartheta_2\ldots d\vartheta_{d-2}\wedge d\varphi\ \prod_{k=1}^{d-2}  \sin^k\vartheta_k\ (\ldots)\ .
\end{split}\end{equation}

We are interested in constructing an orthonormal basis of spherical harmonics in these coordinates. As we described above, the simplest  spherical harmonic is the Legendre harmonic $P_\ell(d,\cos\vartheta_{d-2})$ which depends only on $\vartheta_{d-2}$. It obeys the second-order ODE
\begin{equation}
\begin{split}
\left[\frac{1}{\sin^{d-2}\vartheta}\frac{d}{d\vartheta}\sin^{d-2}\vartheta\frac{d}{d\vartheta}+\ell(\ell+d-2)\right]
P_\ell(d,\cos\vartheta)=0\ .
\end{split}\end{equation}
The function $P_\ell(d,\cos\vartheta)$
is the unique $\ell^{th}$ degree polynomial in $\cos\vartheta$ that solves the above ODE and is normalised to $P_\ell(d,\cos\vartheta=1)=1$. In general, spherical harmonics of degree $\ell$ obey the eigenvalue equation $\left[\nabla^2_{\mathbb{S}^{d-1}}+\ell(\ell+d-2)\right]
\mathcal{S}_\ell(\Om_{d-1})=0$, or in more detail
\begin{equation}
\begin{split}
\left[\frac{1}{\sin^{d-2}\vartheta_{d-2}}\frac{\partial}{\partial\vartheta_{d-2}}\sin^{d-2}\vartheta_{d-2}
\frac{\partial}{\partial\vartheta_{d-2}}+\frac{1}{\sin^2\vartheta_{d-2}}\nabla^2_{\mathbb{S}^{d-2}}+\ell(\ell+d-2)\right]
\mathcal{S}_\ell(\Om_{d-1})=0\ .
\end{split}\end{equation}
This equation can be solved via a separation of variables ansatz
\begin{equation}
\begin{split}
\mathcal{S}_\ell = (\sin\vartheta_{d-2})^m P_{\ell-m}(d+2m,\cos\vartheta_{d-2})\widehat{\mathcal{S}}_m(\Om_{d-2})\ ,
\end{split}\end{equation}
for a non-negative integer $0\leq m\leq \ell$. Substituting this ansatz into the equation above yields the eigenvalue equation
$\left[\nabla^2_{\mathbb{S}^{d-2}}+m(m+d-3)\right]
\widehat{\mathcal{S}}_m(\Om_{d-2})=0$ in the lower dimensional sphere, i.e., the function $\widehat{\mathcal{S}}_m(\Om_{d-2})$
is actually a spherical harmonic of degree $m$ on 
$\mathbb{S}^{d-2}$. This gives rise to 
\begin{equation}\label{Eq:NHHRec}
\begin{split}
\sum_{m=0}^\ell N_{HH}(d-1,m)=N_{HH}(d,\ell)
\end{split}\end{equation}
number of spherical harmonics of degree $\ell$ on $\mathbb{S}^{d-1}$ (to get the above equality, we have used Eq.\eqref{Eq:NHHdef}). Recursing this construction, we get a set of spherical harmonics of the form
\begin{equation}\label{Eq:REcSphHarm}
\begin{split}
\mathcal{C}_{\ell\vec{m}}\ e^{\pm im_{_1}\varphi }\left[\prod_{k=1}^{d-2}(\sin\vartheta_k)^{m_k} P_{m_{k+1}-m_k}(k+2+2m_k,\cos\vartheta_k)\right]_{m_{d-1}=\ell}\ ,
\end{split}\end{equation}
one for every non-decreasing sequence of non-negative integers\begin{equation}
\begin{split}
0\leq m_{_1}\leq m_2\ldots\leq m_{d-2}\leq m_{d-1}=\ell\ .
\end{split}\end{equation}
Here $\mathcal{C}_{\ell\vec{m}}$ is a normalisation constant which we shall determine below.

We will now argue that these spherical harmonics form an orthonormal set: any two harmonics with distinct $e^{i\varphi }$ factors are evidently orthogonal. Thus, we need to address only the case where $e^{i\varphi }$  factors are the same. Without loss of generality, let us assume that the dependences on $\vartheta_k$ for all $k<i$ are also the same between the two spherical harmonics for some $i<d-1$, and they differ first on their $\vartheta_i$ dependence, i.e., we consider two spherical harmonics in the above set with $m_k=m'_k$ for all $k\leq i$, but have $m_{i+1}\neq m_{i+1}'$. The inner product between these two spherical harmonics then has a factor
\begin{equation}
\begin{split}\int_0^\pi d\vartheta_i
(\sin\vartheta_i)^{i+2m_i} P_{m_{i+1}-m_i}(i+2+2m_i,\cos\vartheta_i)P_{m'_{i+1}-m_i}(i+2+2m_i,\cos\vartheta_i)\ ,
\end{split}\end{equation}
which then vanishes using Legendre orthogonality (see Eq.\eqref{Eq:LegOrtho}) on $\mathbb{S}^{i+2m_i+1}$. The mutual orthogonality along with the counting in Eq.\eqref{Eq:NHHRec} proves then that we have indeed constructed a complete set of spherical harmonics of degree $\ell$ on $\mathbb{S}^{d-1}$.

We will conclude this discussion by normalising the spherical harmonics constructed above. The norm computation reduces to a product integral like the one above, which can then be evaluated using Eq.\eqref{Eq:LegOrtho}. Thus, the normalisation of the spherical harmonic given in Eq.\eqref{Eq:REcSphHarm} is given by
\begin{equation}
\begin{split}
|\mathcal{C}_{\ell\vec{m}}|^{-2}&\equiv 2\pi\prod_{i=1}^{d-2}\int_0^\pi d\vartheta_i
(\sin\vartheta_i)^{i+2m_i} P^2_{m_{i+1}-m_i}(i+2+2m_i,\cos\vartheta_i)\\
&=2\pi\prod_{i=1}^{d-2}\frac{|\mathbb{S}^{i+2m_i+1}|}{|\mathbb{S}^{i+2m_i}|N_{HH}(i+2m_i+2,m_{i+1}-m_i)}\ ,
\end{split}\end{equation}
With this, we have a concrete realisation of the orthonormal spherical harmonics $\mathscr{Y}_{\ell\vec{m}}(\hat{r})$, using which multipole moments could be computed for a given charge distribution.

\subsection{STF tensors in \mathinhead{\mathbb{R}^{d}}{Rd} and cartesian multipole moments}
Till now, we have described the multipole expansion in terms of an orthonormal basis of spherical harmonics $\mathscr{Y}_{\ell\vec{m}}(\hat{r})$ and the corresponding spherical multipole moments $q_{_{\ell\vec{m}}}$. We will now describe an alternate formalism based on a more symmetric, but over-complete basis of spherical harmonics made of Legendre polynomials about arbitrary directions (we will call this basis an STF basis). A general spherical harmonic in  STF basis is naturally described by symmetric trace-free (STF) tensors with constant cartesian components. 

For definiteness, we consider spherical harmonics of the form 
\begin{equation}\label{eq:LegSTFReln}\begin{split}
\nn_{d,\ell}N_{HH}(d,\ell)P_\ell(d,\hat{\kappa}\cdot\hat{r})&=\sum_k \frac{\Gamma\left(\nu-k\right)}{2^{2k}k!\Gamma(\nu)}
\frac{(-)^k (\hat{\kappa}\cdot\hat{r})^{\ell-2k}}{(\ell-2k)!} \\
&=\frac{1}{\ell !}\hat{\kappa}_{i_1}\hat{\kappa}_{i_2}\ldots\hat{\kappa}_{i_\ell}\hat{r}^{<i_1}\hat{r}^{i_2}\ldots \hat{r}^{i_\ell>}\\
&=\frac{1}{\ell !}\hat{\kappa}_{i_1}\hat{\kappa}_{i_2}\ldots\hat{\kappa}_{i_\ell}\hat{r}^{j_1}\hat{r}^{j_2}\ldots \hat{r}^{j_\ell} \Pi^{<i_1i_2\ldots i_\ell>}_{ <j_1j_2\ldots j_\ell>}\ ,
\end{split}\end{equation}
where $\hat{\kappa}$ is an arbitrary unit vector, and in the last line we have written the spherical harmonic as a projected contraction of two tensors. The angular bracket here denotes the  symmetric trace-free (STF) projection and $\Pi$ is the STF-projector. An explicit expression that follows from the above definition is
\begin{equation}\begin{split}
\hat{r}^{<i_1}\hat{r}^{i_2}\ldots \hat{r}^{i_\ell>}&=\sum_k \frac{(-)^k \Gamma\left(\nu-k\right)}{2^{k}\Gamma(\nu)}\\
&\times \Bigl\{\hat{r}^{i_1}\hat{r}^{i_2}\ldots \hat{r}^{i_{\ell-2k}}\delta^{i_{\ell+1-2k}i_{\ell+2-2k}}\ldots \delta^{i_{\ell-1}i_{\ell}}+\text{distinct index permutations}\Bigr\}\ .
\end{split}\end{equation}
Here the sum within the curly  braces sums over all index permutations of the set $\{i_1,\ldots,i_\ell\}$ which give distinct answers. The number of such distinct permutations can be counted as follows: there are $\binom{\ell}{2k}$ ways of choosing the subset of indices that go into Kronecker deltas, and $\frac{(2k)!}{2^k k!}=(2k-1)!!$ distinct ways of  pairing a given subset.\footnote{The number of pairings can be counted as follows: the $(2k)!$ ways to permute the subset of indices on Kronecker deltas. Exchanging an index within a pair, as well as permuting the pair as a whole does not change the final resultant pairings, i.e., there is a $(\mathbb{Z}_2)^k\times \mathbb{S}_k$ automorphism group which acts freely and transitively on the equivalence class of permutations which result in a given pairing. We hence obtain the number of distinct pairings  by dividing out the cardinality of the automorphism group. }Thus, the total number of distinct permutations is $(2k-1)!!\binom{\ell}{2k}=\frac{\ell !}{2^k k!(\ell-2k)!}$. With this counting of distinct permutations, it is then easy to check that contracting $\hat{r}^{<i_1}\hat{r}^{i_2}\ldots \hat{r}^{i_\ell>}$ with $\frac{1}{\ell !}\hat{\kappa}_{i_1}\hat{\kappa}_{i_2}\ldots\hat{\kappa}_{i_\ell}$ does give $\nn_{d,\ell}\nn_{HH}(d,\ell)P_\ell(d,\hat{\kappa}\cdot\hat{r})$. The STF projector
can also be given a closed-form expression as:
\begin{equation}\begin{split}
\Pi^{<i_1i_2\ldots i_\ell>}_{ <j_1j_2\ldots j_\ell>}&=\sum_k \frac{(-)^k \Gamma\left(\nu-k\right)}{2^{k}k!(\ell-2k)!\Gamma(\nu)}\\
&\times \delta^{(i_1}_{(j_1}\delta^{i_2}_{j_2}\ldots \delta^{i_{\ell-2k}}_{j_{\ell-2k}}\delta^{i_{\ell+1-2k}i_{\ell+2-2k}}\ldots \delta^{i_{\ell-1}i_{\ell})}\delta_{j_{\ell+1-2k}j_{\ell+2-2k}}\ldots \delta_{j_{\ell-1}j_{\ell})}\ ,
\end{split}\end{equation}
where the $(i_1\ldots i_\ell)$ denotes a symmetric projection. To elucidate the arguments above, we will now write down the explicit expressions of $\hat{r}^{<i_1}\hat{r}^{i_2}\ldots \hat{r}^{i_\ell>}$ for $\ell\leq 5$. We have 
\begin{equation}\begin{split}
&\hat{r}^{<i_1>}\equiv \hat{r}^{<i_1>}\ ,\ 
\hat{r}^{<i_1}\hat{r}^{i_2>}\equiv \hat{r}^{i_1}\hat{r}^{i_2}-\frac{1}{d}\delta^{i_1i_2}\\
&\hat{r}^{<i_1}\hat{r}^{i_2}\hat{r}^{i_3>}\equiv \hat{r}^{i_1}\hat{r}^{i_2}\hat{r}^{i_3}-\frac{1}{d+2} \left(\hat{r}^{i_1}\delta^{i_2i_3}+\hat{r}^{i_2}\delta^{i_1i_3}+\hat{r}^{i_3}\delta^{i_1i_2}\right)\ ,\\
\end{split}\end{equation}
for $\ell\leq 3$. For $\ell=4$, we have 
\begin{equation}\begin{split}
&\hat{r}^{<i_1}\hat{r}^{i_2}\hat{r}^{i_3}\hat{r}^{i_4>}\equiv \hat{r}^{i_1}\hat{r}^{i_2}\hat{r}^{i_3}\hat{r}^{i_4}\\
&\quad -\frac{1}{d+4}\left(\hat{r}^{i_1}\hat{r}^{i_2}\delta^{i_3i_4}+\hat{r}^{i_1}\hat{r}^{i_3}\delta^{i_2i_4}+\hat{r}^{i_1}\hat{r}^{i_4}\delta^{i_2i_3}+\hat{r}^{i_2}\hat{r}^{i_3}\delta^{i_1i_4}+\hat{r}^{i_2}\hat{r}^{i_4}\delta^{i_1i_3}+\hat{r}^{i_3}\hat{r}^{i_4}\delta^{i_1i_2}\right)\\
&\quad +\frac{1}{(d+4)(d+2)}\left(\delta^{i_1i_2}\delta^{i_3i_4}+\delta^{i_1i_3}\delta^{i_2i_4}+\delta^{i_1i_4}\delta^{i_2i_3}\right)\ ,\\
\end{split}\end{equation}
and for $\ell=5$, we get 
\begin{equation}\begin{split}
&\hat{r}^{<i_1}\hat{r}^{i_2}\hat{r}^{i_3}\hat{r}^{i_4}\hat{r}^{i_5>}\equiv \hat{r}^{i_1}\hat{r}^{i_2}\hat{r}^{i_3}\hat{r}^{i_4}\hat{r}^{i_5}\\
&\quad -\frac{1}{d+6}\Bigl(\hat{r}^{i_1}\hat{r}^{i_2}\hat{r}^{i_3}\delta^{i_4i_5}+\hat{r}^{i_1}\hat{r}^{i_2}\hat{r}^{i_4}\delta^{i_3i_5}+\hat{r}^{i_1}\hat{r}^{i_3}\hat{r}^{i_4}\delta^{i_2i_5}+\hat{r}^{i_2}\hat{r}^{i_3}\hat{r}^{i_4}\delta^{i_1i_5}\Bigr.\\
&\quad \Bigl.+\hat{r}^{i_5}\hat{r}^{i_1}\hat{r}^{i_2}\delta^{i_3i_4}+\hat{r}^{i_5}\hat{r}^{i_1}\hat{r}^{i_3}\delta^{i_2i_4}+\hat{r}^{i_5}\hat{r}^{i_1}\hat{r}^{i_4}\delta^{i_2i_3}+\hat{r}^{i_5}\hat{r}^{i_2}\hat{r}^{i_3}\delta^{i_1i_4}+\hat{r}^{i_5}\hat{r}^{i_3}\hat{r}^{i_4}\delta^{i_1i_2}\Bigr)\\
&\ +\frac{1}{(d+6)(d+4)}\Bigl(\hat{r}^{i_1}\delta^{i_2i_3}\delta^{i_4i_5}+\hat{r}^{i_1}\delta^{i_2i_4}\delta^{i_3i_5}+\hat{r}^{i_1}\delta^{i_2i_5}\delta^{i_3i_4}\Bigr.\\
&\ +\Bigl.\hat{r}^{i_2}\delta^{i_1i_3}\delta^{i_4i_5}+\hat{r}^{i_2}\delta^{i_1i_4}\delta^{i_3i_5}+\hat{r}^{i_2}\delta^{i_1i_5}\delta^{i_3i_4}+\hat{r}^{i_3}\delta^{i_1i_2}\delta^{i_4i_5}+\hat{r}^{i_3}\delta^{i_1i_4}\delta^{i_2i_5}+\hat{r}^{i_3}\delta^{i_1i_5}\delta^{i_2i_4}\Bigr.\\
&\
\Bigl.+\hat{r}^{i_4}\delta^{i_1i_2}\delta^{i_3i_5}+\hat{r}^{i_4}\delta^{i_1i_3}\delta^{i_2i_5}+\hat{r}^{i_4}\delta^{i_1i_5}\delta^{i_2i_3}+\hat{r}^{i_5}\delta^{i_1i_2}\delta^{i_3i_4}+\hat{r}^{i_5}\delta^{i_1i_3}\delta^{i_2i_4}+\hat{r}^{i_5}\delta^{i_1i_4}\delta^{i_2i_3}\Bigr)\ .
\end{split}\end{equation}
The reader can check that the expressions in RHS are completely symmetric under permutations of indices, and vanish if we take a trace over any two indices. Further, our counting of distinct permutations can also be checked for every term written above.

A more succinct way to summarise the permutations/symmetrisations described above is to work instead with the homogeneous harmonic polynomials (HHPs) in cartesian coordinates
\begin{equation}
\begin{split}
x^{<i_1}x^{i_2}\ldots x^{i_\ell>}&\equiv r^\ell\ \hat{r}^{<i_1}\hat{r}^{i_2}\ldots \hat{r}^{i_\ell>}= \left[\sum_{k=0}^{\lfloor\frac{\ell}{2}\rfloor}\ \frac{\Gamma\left(\nu-k\right)}{k!\ \Gamma\left(\nu\right)}  \left(\frac{r}{2}\right)^{2k}(-\nabla^2)^k\right]_{\nu=\frac{d}{2}+\ell-1}x^{i_1}x^{i_2}\ldots x^{i_\ell}\ .
 \end{split}   
\end{equation}
The relation to generalised  Legendre polynomials then follows from
 \begin{equation}
\begin{split}
\frac{1}{\ell!}\kappa_{i_1}\kappa_{i_2}\ldots\kappa_{i_\ell}x^{<i_1}x^{i_2}\ldots x^{i_\ell>}&\equiv \left[\sum_{k=0}^{\lfloor\frac{\ell}{2}\rfloor}\ \frac{\Gamma\left(\nu-k\right)}{k!\ \Gamma\left(\nu\right)}  \left(\frac{r}{2}\right)^{2k}(-\nabla^2)^k\right]_{\nu=\frac{d}{2}+\ell-1}\frac{(\vec{\kappa}\cdot\vec{r})^\ell}{\ell!}\\
&= \left[\sum_{k=0}^{\lfloor\frac{\ell}{2}\rfloor}\ \frac{\Gamma\left(\nu-k\right)}{k!\ \Gamma\left(\nu\right)}  \left(-\frac{\kappa^2r^2}{4}\right)^{k}\frac{(\vec{\kappa}\cdot\vec{r})^{\ell-2k}}{(\ell-2k)!}\ \right]_{\nu=\frac{d}{2}+\ell-1}\\
&=\nn_{d,\ell}N_{HH}(d,\ell)(\kappa r)^\ell P_\ell\left(d,\hat{\kappa}\cdot\hat{r}\right)\ ,
 \end{split}   
\end{equation}
where, in the last step, we have used Eq.\eqref{Eq:LegGegen}. The STF basis for multipole expansion in flat spacetime is often introduced in terms of these cartesian HHPs (See e.g.\cite{poisson2014gravity}). In dS spacetime (and more generally in cosmology), the absence of global cartesian coordinates limits their scope. The STF basis for spherical harmonics is, however, a useful tool for multipole expansion in such spacetimes, since isotropy is still a true symmetry.
 
We will now describe how the STF basis relates to the description of spherical harmonics given before. 
For any given $\ell$, we can form 
\begin{equation}\label{Eq:NHdef}
\begin{split}
N_H(d,\ell)\equiv \binom{\ell+d-1}{\ell}
\end{split}   
\end{equation}
number of STF harmonics of the form  $\hat{r}^{<i_1}\hat{r}^{i_2}\ldots \hat{r}^{i_\ell>}$. The above binomial coefficient counts the number of ways $d$ directions can be filled into $\ell$ indices. The combinatorics here is identical to the bose-counting problem familiar from elementary statistical mechanics, where one counts the ways in which $d$ bosons could be filled into $\ell$ degenerate energy levels. All the STF harmonics are not however linearly independent, they obey $N_H(d,\ell-2)$ number of  conditions of the form 
\begin{equation}
\begin{split}
\delta_{i_1i_2}\hat{r}^{<i_1}\hat{r}^{i_2}\ldots \hat{r}^{i_\ell>}=0\ .
\end{split}   
\end{equation}
They hence span a vector space of spherical harmonics of dimension
\begin{equation}
\begin{split}
N_H(d,\ell)-N_H(d,\ell-2)=N_{HH}(d,\ell)\ ,
\end{split}   
\end{equation}    
where the equality folows by using the explicit forms in Eqs.\eqref{Eq:NHHdef} and \eqref{Eq:NHdef}. This shows that the harmonics $\hat{r}^{<i_1}\hat{r}^{i_2}\ldots \hat{r}^{i_\ell>}$ indeed form an overcomplete basis of spherical harmonics of degree $\ell$.

The completeness means the following: say we are given  a spherical harmonic $\mathscr{Y}_\ell(\hat{r})$ of degree $\ell$ on $\mathbb{S}^{d-1}$. We can then define a symmetric trace-free (STF) tensor $\mathscr{Y}_{i_1i_2\ldots i_\ell}$ of rank $\ell$ in $\mathbb{R}^d$ such that 
\begin{equation}
\begin{split}
\mathscr{Y}_\ell(\hat{r})=\frac{1}{\ell !}\mathscr{Y}_{i_1i_2\ldots i_\ell}\hat{r}^{<i_1}\hat{r}^{i_2}\ldots \hat{r}^{i_\ell>}\ .
\end{split}
\end{equation}
The orthonormal basis of spherical harmonics constructed in the previous subsection then defines
an orthonormal set of STF tensors 
\begin{equation}\label{eq:OrthoToSTF}
\begin{split}
\mathscr{Y}_{\ell\vec{m}}(\hat{r})=\frac{1}{\ell !}\mathscr{Y}^{(\ell\vec{m})}_{i_1i_2\ldots i_\ell}\hat{r}^{<i_1}\hat{r}^{i_2}\ldots \hat{r}^{i_\ell>}\ .
\end{split}
\end{equation}
Further, the inner product on the space of STF tensors is induced from the standard inner product on the space of functions on $\mathbb{S}^{d-1}$. To get an explicit expression, consider the following integral
\begin{equation}\begin{split}
\int_{\hat{r}\in\mathbb{S}^{d-1}}\frac{1}{\ell !}&\kappa_{i_1}\kappa_{i_2}\ldots\kappa_{i_\ell}\hat{r}^{<i_1}\hat{r}^{i_2}\ldots \hat{r}^{i_\ell>}\times \frac{1}{\ell !}\bar{\kappa}_{j_1}\bar{\kappa}_{j_2}\ldots\bar{\kappa}_{j_\ell}\hat{r}^{<j_1}\hat{r}^{j_2}\ldots \hat{r}^{j_\ell>}\\
&=[\nn_{d,\ell}N_{HH}(d,\ell)]^2 (\kappa\bar{\kappa})^\ell \int_{\hat{r}\in\mathbb{S}^{d-1}} P_\ell(d,\hat{\kappa}\cdot\hat{r}) P_\ell(d,\hat{\bar{\kappa}}\cdot\hat{r})\\
&=\nn^2_{d,\ell}N_{HH}(d,\ell)|\mathbb{S}^{d-1}| (\kappa\bar{\kappa})^\ell P_\ell(d,\hat{\kappa}\cdot\hat{\bar{\kappa}})\\
&=\nn_{d,\ell}|\mathbb{S}^{d-1}|\ \frac{1}{\ell !}\kappa_{<i_1}\kappa_{i_2}\ldots\kappa_{i_\ell>}\bar{\kappa}^{<i_1}\bar{\kappa}^{i_2}\ldots\bar{\kappa}^{i_\ell>}\ .
\end{split}\end{equation}
For example, the STF tensors corresponding to the orthonormal spherical harmonics have an inner product
given by
\begin{equation}\label{eq:STFOrtho}\begin{split}
\frac{\nn_{d,\ell}|\mathbb{S}^{d-1}|}{\ell !}
\mathscr{Y}_{(\ell\vec{m}')}^{\ast<i_1i_2\ldots i_\ell>} \mathscr{Y}^{(\ell\vec{m})}_{<i_1i_2\ldots i_\ell>}=
\delta^{\vec{m}}_{\vec{m}'}\ .
\end{split}\end{equation}
We recognise $\nn_{d,\ell}|\mathbb{S}^{d-1}|$ here as the conversion factor between the STF tensor inner product and the standard functional inner product between the spherical harmonics. The same factor
also appears in the statement of spherical harmonic addition theorem,  stated in terms of STF tensors:
\begin{equation}\label{eq:STFAddition}\begin{split}
\frac{\nn_{d,\ell}|\mathbb{S}^{d-1}|}{\ell!}\sum_{\vec{m}}
\mathscr{Y}_{(\ell\vec{m})}^{\ast <i_1i_2\ldots i_\ell>} \mathscr{Y}^{(\ell\vec{m})}_{<j_1j_2\ldots j_\ell>}=
 \Pi^{<i_1i_2\ldots i_\ell>}_{ <j_1j_2\ldots j_\ell>}\ .
\end{split}\end{equation}
This important relation can be proved in many ways: one way is to use Eq.\eqref{eq:LegSTFReln} to convert the standard addition theorem into STF tensors. Another ab initio derivation is to first argue that LHS should be proportional to RHS for symmetry reasons and then fix the normalisation by using the orthonormality relation Eq.\eqref{eq:STFOrtho}.

\subsection{Green functions in Minkowski spacetime}\label{app:FlatGreenFns}
We will begin by briefly reviewing the Green functions of the wave operator (i.e., the massless scalar operator) in $\mathbb{R}^{d,1}$. This theory is standard, although the notations and normalisations for Green functions in $d\neq 2,3$ are non-standard. Thus, this subsection mainly serves to establish our notation. We will state our results with an eye towards their generalisation to dS Green functions.

We begin with the unique spherically symmetric eigenfunction of the Laplacian in $\mathbb{R}^d$ with eigenvalue $-\w^2$ :
\begin{equation}
	\begin{split}
 J_0(d,\w r)&\equiv  {}_0F_1\left[\frac{d}{2},-\frac{\w^2r^2}{4}\right]\ .
	\end{split}
\end{equation}
We can construct a whole tower of descendants from this eigenfunction by taking an STF derivative
\begin{equation}
\begin{split}
\mathscr{Y}_\ell(-\vec{\nabla})J_0(d,\w r)&\equiv \w^\ell\  J_\ell(d,\w r)\ \mathscr{Y}_\ell(\vec{n})=\w^{\nu-\frac{d}{2}+1}\  J_\ell(d,\w r)\ \mathscr{Y}_\ell(\vec{n}) \ ,
\end{split}
\end{equation}
where we have defined (we remind the reader that $\nu\equiv \ell+\frac{d}{2}-1$)
\begin{equation}\label{eq:besselgen}
\begin{split}
J_\ell(d,\w r)&\equiv  \Gamma\left(\frac{d}{2}\right)\left(\frac{\w r}{2}\right)^{1-\frac{d}{2}}\  J_\nu(\w r)\equiv  \frac{\Gamma(d/2)}{\Gamma(1+\nu)}\left(\frac{\w r}{2}\right)^{\nu-\frac{d}{2}+1}\  {}_0F_1\left[1+\nu,-\frac{\w^2r^2}{4}\right]\ .
\end{split}
\end{equation}
The notation is motivated by the fact that the functions that appear here generalise the Bessel J functions in the $d=2$ version of the above problem. We can also define the functions analogous to Neumann and Hankel functions. We will define the \emph{Neumann Green function} via
\begin{equation}\label{eq:Neugen}
\begin{split}
&N_\ell(d,\w r)\equiv -\frac{1}{4}\frac{Y_\nu(\w r)}{(2\pi\w r)^{\frac{d}{2}-1} }\\
&= \frac{\Gamma(\nu)}{(4\pi)^{d/2}}\left(\frac{\w r}{2}\right)^{-\nu-\frac{d}{2}+1}\Biggl\{  {}_0F_1\left[1-\nu,-\frac{\w^2r^2}{4}\right]- \frac{\pi \cot \nu \pi}{\Gamma(\nu)\Gamma(1+\nu)}\left(\frac{\w r}{2}\right)^{2\nu}\  {}_0F_1\left[1+\nu,-\frac{\w^2r^2}{4}\right]\Biggr\}\ .
\end{split}
\end{equation}
In the above definition, for half-integer $\nu$ (i.e., for $d$ odd), we can set $\cot \nu \pi=0$, whereas for integer $\nu$, the divergence in the $\cot \nu \pi$ cancels the divergence in the first term and  this formula should be interpreted as a limit. Green functions are normalised such that
\begin{equation}
\begin{split}
-(\vec{\nabla}^2+\w^2)[\w^{\nu+\frac{d}{2}-1}N_\ell(d,\w r)\mathscr{Y}_\ell(\vec{n})]=\mathscr{Y}_\ell(-\vec{\nabla})\delta^d(\vec{r})\ .
\end{split}
\end{equation}
Since the RHS here is a multipole source, the combination $\w^{\nu+\frac{d}{2}-1}N_\ell(d,\w r)\mathscr{Y}_\ell(\vec{n})$ should then be interpreted as the amplitude of standing wave sourced by such a multipole source. We term this a \emph{standing wave} since it is an even function of frequency. In contrast, the 
outgoing/ingoing waves  are denoted by $\w^{\nu+\frac{d}{2}-1}H_\ell^\pm(d,\w r)$ respectively. We will refer to them as \emph{Hankel Green functions}. Given  a spherical harmonic $\mathscr{Y}_\ell(\vec{n})$ of degree $\ell$ on $\mathbb{S}^{d-1}$, both these Green functions satisfy 
\begin{equation}
\begin{split}
-(\vec{\nabla}^2+\w^2)[\w^{\nu+\frac{d}{2}-1}H_\ell^\pm(d,\w r)\mathscr{Y}_\ell(\vec{n})]=\mathscr{Y}_\ell(-\vec{\nabla})\delta^d(\vec{r})\ .
\end{split}
\end{equation}
The outgoing/ingoing conditions are imposed by taking $H_\ell^\pm(d,\w r)$ to be analytic  in the upper/lower half plane of complex frequency respectively. The notation here is again motivated by the fact that these functions generalise the Hankel functions in $d=2$ (up to normalisations). Their explicit forms are given by 
\begin{equation}\label{eq:Hankelgen}
\begin{split}
&H_\ell^\pm(d,\w r)\equiv \frac{\pm i}{4}\frac{H^{1,2}_\nu(\w r)}{(2\pi\w r)^{\frac{d}{2}-1} } \equiv 	N_\ell(d,\w r)\pm	\frac{i\pi}{\Gamma(d/2)(4\pi)^{d/2}}J_\ell(d,\w r)\\
&=\frac{\Gamma(\nu)}{(4\pi)^{d/2}}\left(\frac{\w r}{2}\right)^{-\nu-\frac{d}{2}+1}\Biggl\{ {}_0F_1\left[1-\nu,-\frac{\w^2r^2}{4}\right]\pm (1\pm i\cot \nu \pi)\frac{2\pi i}{\Gamma(\nu)^2}\frac{1}{2\nu}\left(\frac{\w r}{2}\right)^{2\nu}\  {}_0F_1\left[1+\nu,-\frac{\w^2r^2}{4}\right]\Biggr\}\ .
\end{split}
\end{equation}
As in the case of Neumann Green functions,
for half-integer $\nu$ (i.e., for $d$ odd), we can set $\cot \nu \pi=0$, whereas for integer $\nu$ the above expression is indeterminate and should be interpreted as a limit. 

As in the case of Bessel J functions, these Green functions could also be obtained by STF-differentiating their corresponding primary eigenfunction at $\ell=0$, viz.,
\begin{equation}
	\begin{split}
\mathscr{Y}_\ell(-\vec{\nabla})N_0(d,\w r)&=\w^\ell\  N_\ell(d,\w r)\ \mathscr{Y}_\ell(\vec{n})=\w^{\nu-\frac{d}{2}+1}\  N_\ell(d,\w r)\ \mathscr{Y}_\ell(\vec{n}) \ ,\\
\mathscr{Y}_\ell(-\vec{\nabla})H^\pm_0(d,\w r)&=\w^\ell\  H^\pm_\ell(d,\w r)\ \mathscr{Y}_\ell(\vec{n})=\w^{\nu-\frac{d}{2}+1}\  H^\pm_\ell(d,\w r)\ \mathscr{Y}_\ell(\vec{n}) \ .		
	\end{split}
\end{equation}
A related statement is the \emph{multipole-expansion} of these Green functions, which, in our normalisations, takes the following form: 
\begin{equation}
\begin{split}
&J_0(d,\w|\vec{r}-\vec{r}_0|)=
\sum_{\ell m} |\mathbb{S}^{d-1}|\  \mathscr{Y}_{\ell \vec{m}}(\hat{r})\mathscr{Y}_{\ell  \vec{m}}(\hat{r}_0)^\ast J_\ell(d,\w r) J_\ell(d,\w r_0)\ ,\\
&N_0(d,\w|\vec{r}-\vec{r}_0|)=
\sum_{\ell m} |\mathbb{S}^{d-1}|\  \mathscr{Y}_{\ell  \vec{m}}(\hat{r})\mathscr{Y}_{\ell  \vec{m}}(\hat{r}_0)^\ast\\
&\qquad\times\Bigl\{\Theta(r<r_0)J_\ell(d,\w r) N_\ell(d,\w r_0)+\Theta(r>r_0)J_\ell(d,\w r_0) N_\ell(d,\w r) \Bigr\}\ ,\\
&H_0^\pm(d,\w|\vec{r}-\vec{r}_0|)=
\sum_{\ell  \vec{m}} |\mathbb{S}^{d-1}|\  \mathscr{Y}_{\ell  \vec{m}}(\hat{r})\mathscr{Y}_{\ell  \vec{m}}(\hat{r}_0)^\ast\\
&\qquad\times\Bigl\{\Theta(r<r_0)J_\ell(d,\w r) H_\ell^\pm(d,\w r_0)+\Theta(r>r_0)J_\ell(d,\w r_0) H_\ell^\pm(d,\w r) \Bigr\}\ .
\end{split}
\end{equation}
Here, the set  of functions $\mathscr{Y}_{\ell  \vec{m}}(\hat{r})$ for different $ \vec{m}$ denote an orthonormal basis of $\mathbb{S}^{d-1}$ spherical harmonics of degree $\ell$. Further, in the equation above, the symbol 
\begin{equation}
	\begin{split}
|\mathbb{S}^{d-1}|\equiv \frac{2\pi^{\frac{d}{2}}}{\Gamma\left(\frac{d}{2}\right)}\ 
	\end{split}
\end{equation}
denotes the volume of the unit sphere. The argument for the above expansion is well-known within the theory of Green functions: we first expand the LHS in terms of eigenfunctions and then fix the coefficients by demanding continuity and a unit jump in the radial derivative. The jump can be readily evaluated using the Wronskian formulae\footnote{Our Wronskian convention is $\mathcal{W}[f(z),g(z)]\equiv f\partial_z g-g\partial_z f$.}
\begin{equation}\begin{split} 
\mathcal{W}[ N_{\ell}(d,z) ,J_{\ell}(d,z) ]= \mathcal{W}[ H^{\pm}_{\ell}(d,z) ,J_{\ell}(d,z) ] =\frac{1}{|\mathbb{S}^{d-1}|z^{d-1}}  \ .
\end{split}\end{equation}

We will be interested here in the multipole expansion of the retarded/outgoing Green function $\w^{d-2}H_0^+(d,\w|\vec{r}-\vec{r}_0|)$, which, using the relations quoted earlier, we can rewrite entirely in terms of ${}_0F_1$  functions: 
\begin{equation}\label{eq:HPlusDDW}
	\begin{split}
&\w^{d-2}H_0^+(d,\w|\vec{r}-\vec{r}_0|)=\w^{d-2}H_0^+(d,\w|\vec{r}_0-\vec{r}|)\\
&\ =\frac{i\pi}{2} \sum_{\ell  \vec{m}} \frac{(rr_0)^{\nu-\frac{d}{2}+1}}{\Gamma(1+\nu)^2}\left(\frac{\w }{2}\right)^{2\nu}\  \mathscr{Y}_{\ell  \vec{m}}(\hat{r})\mathscr{Y}_{\ell  \vec{m}}(\hat{r}_0)^\ast\  {}_0F_1\left[1+\nu,-\frac{\w^2r^2}{4}\right]{}_0F_1\left[1+\nu,-\frac{\w^2r_0^2}{4}\right]\\
&\quad+\sum_{\ell  \vec{m}}\frac{1}{2\nu}\frac{r_<^{\nu-\frac{d}{2}+1}}{r_>^{\nu+\frac{d}{2}-1}}\  \mathscr{Y}_{\ell  \vec{m}}(\hat{r})\mathscr{Y}_{\ell  \vec{m}}(\hat{r}_0)^\ast\  {}_0F_1\left[1+\nu,-\frac{\w^2r_<^2}{4}\right]\\
&\qquad\times \Biggl\{  {}_0F_1\left[1-\nu,-\frac{\w^2r_>^2}{4}\right]- \frac{\pi \cot \nu \pi}{\Gamma(\nu)\Gamma(1+\nu)}\left(\frac{\w r_>}{2}\right)^{2\nu}\  {}_0F_1\left[1+\nu,-\frac{\w^2r_>^2}{4}\right]\Biggr\}\ .
	\end{split}
\end{equation}
Here we have used a commonly used notation in such expansions, viz.,
\begin{equation}
\begin{split}
r_>\equiv \text{Max}(r,r_0)\ ,\quad	
r_<\equiv \text{Min}(r,r_0)\ .
\end{split}
\end{equation}
Further, we have also separated out the real and the imaginary parts  of the radial functions.

Consider double integrals of the form
\begin{equation}
	\begin{split}
		\int d^dr\ \rho_1(\vec{r},\w)\int d^dr_0\ \rho_2(\vec{r}_0,\w)\ f(d,\w|\vec{r}-\vec{r}_0|)
	\end{split}
\end{equation}
where $f$ could be any one of the functions discussed above. Using the multipole expansion, such a double integral can be decomposed into an infinite   sum of factorised integrals, one each for every spherical harmonic. For the practical computation of radiation reaction, it is then convenient to convert the spherical harmonic sum into an STF expression using Eq.\eqref{eq:STFAddition}.

Let us illustrate the above remarks by computing the flat spacetime scalar radiation reaction, which, in a slightly different notation, is explained in detail in references\cite{Birnholtz:2013ffa,Birnholtz:2013nta}.
Say we have an extended scalar source whose emissive part at frequency $\w$  is the average source $\rho_A(\w,\vec{r})$, and whose absorptive part is the difference source $\rho_D(\w,\vec{r})$. Ignoring all fluctuation effects,  the flat spacetime influence phase for this source, after integrating out the massless scalar field about vacuum, can be written down as
\begin{equation}\label{eq:flatRRbare}
	\begin{split}
		S_{RR}^{\text{bare}}=\int\frac{d\w}{2\pi} \int d^dr_0\int d^dr\  [\rho_D(\vec{r}_0,\w)]^\ast\ \rho_A(\vec{r},\w)\ \w^{d-2}H_0^+(d,\w|\vec{r}_0-\vec{r}|)\ .
	\end{split}
\end{equation}
Here $\w^{d-2}H_0^+(d,\w|\vec{r}_0-\vec{r}|)$ is the outgoing Green function  for the scalar field, and the superscript `bare' indicates that this expression is divergent and has to be counter-termed before it makes sense. We do not give here a derivation of the above influence phase except the heuristic that the above action describes causal propagation of a free scalar about the Minkowski vacuum. The above influence phase is also natural if one applies the original Feynman-Vernon argument in \cite{FEYNMAN1963118} for harmonic oscillators  to each  Minkowski mode of the scalar field and sums the result. A more proper derivation should involve a careful discussion of the fall-offs near space-like, time-like, and null asymptotia. We do not attempt such a discussion here because, as we shall see later, dS-SK geometry naturally incorporates such boundary conditions. Our dS answer in an appropriate limit will reduce to the above result.

We substitute the multipole expansion Eq.\eqref{eq:HPlusDDW} into the influence phase $S^{\text{bare}}_{RR}$. For simplicity, we will take the number of spatial dimensions (i.e., $d$) to be odd, so that $\nu\equiv\ell+\frac{d}{2}-1$ is a half-integer (and $\cot \nu\pi=0$). The above action then has two sets of terms:
the first set of terms, odd under time reversal, are 
\begin{equation}\label{eq:RRMinkn}
	\begin{split}
&\sum_{\ell\vec{m}}\int\frac{d\w}{2\pi} \frac{i\pi}{2} \left(\frac{\w }{2}\right)^{2\nu} \frac{1}{\Gamma(1+\nu)^2} \\
  &\times \int d^dr_0 \left\{\rho_D(\vec{r}_0,\w)r_0^{\nu-\frac{d}{2}+1}\mathscr{Y}_{\ell  \vec{m}}(\hat{r}_0)\ {}_0F_1\left[1+\nu,-\frac{\w^2r_0^2}{4}\right]\right\}^\ast\\
  &\qquad\times \int d^dr\ \left\{\rho_A(\vec{r},\w)r^{\nu-\frac{d}{2}+1}\mathscr{Y}_{\ell  \vec{m}}(\hat{r})\ {}_0F_1\left[1+\nu,-\frac{\w^2r^2}{4}\right]\right\}\ .
	\end{split}
\end{equation}
The combinations appearing in the second and the third line are the Bessel-smeared radiative multipole moments\footnote{The reader should compare this definition against electrostatic multipole moments defined in Eq.\eqref{eq:EstaticMp}, remembering $\nu\equiv\ell+\frac{d}{2}-1$. } of the sources, i.e.,
\begin{equation}\label{eq:jFlatdef}
	\begin{split}
\multj_A(\w,\ell,\vec{m})\equiv  \frac{1}{2\nu}\int d^dr\ \rho_A(\vec{r},\w)\ r^{\nu-\frac{d}{2}+1}\mathscr{Y}_{\ell  \vec{m}}(\hat{r})\ {}_0F_1\left[1+\nu,-\frac{\w^2r^2}{4}\right]\ ,\\
\multj_D(\w,\ell,\vec{m})\equiv  \frac{1}{2\nu}\int d^dr\ \rho_D(\vec{r},\w)\ r^{\nu-\frac{d}{2}+1}\mathscr{Y}_{\ell  \vec{m}}(\hat{r})\ {}_0F_1\left[1+\nu,-\frac{\w^2r^2}{4}\right] \ .
	\end{split}
\end{equation}
We can then write the time reversal odd terms in the form
\begin{equation}\label{eq:RRMinknII}
	\begin{split}
&\sum_{\ell\vec{m}}\int\frac{d\w}{2\pi} \frac{2\pi i}{\Gamma(\nu)^2} \left(\frac{\w }{2}\right)^{2\nu}  \multj_D^\ast(\w,\spL)\multj_A(\w,\spL)\ .
	\end{split}
\end{equation}

Given that 
${}_0F_1$ functions are completely regular when their first argument is positive (i.e., when $1+\nu=\ell+\frac{d}{2}>0$, we conclude that these multipole moments are finite, even for point-like sources. Hence, each term in Eq.\eqref{eq:RRMinkn} is finite. The reader should contrast this with the second set of terms, even under time reversal:
\begin{equation}\label{eq:RRMinknn}
	\begin{split}
&\sum_{\ell  \vec{m}}\int\frac{d\w}{2\pi} \int d^dr\int d^dr_0\  [\rho_D(\vec{r},\w)]^\ast\ \rho_A(\vec{r},\w)\\
&\qquad \times\frac{1}{2\nu}\frac{r_<^{\nu-\frac{d}{2}+1}}{r_>^{\nu+\frac{d}{2}-1}}\  \mathscr{Y}_{\ell  \vec{m}}(\hat{r})\mathscr{Y}_{\ell  \vec{m}}(\hat{r}_0)^\ast\  {}_0F_1\left[1+\nu,-\frac{\w^2r_<^2}{4}\right] {}_0F_1\left[1-\nu,-\frac{\w^2r_>^2}{4}\right] \ .
	\end{split}
\end{equation}
which are divergent due to the Green functions ${}_0F_1(1-\nu,\ldots)$. Fortunately, since these are all even under time reversal, one can counter-term away these terms. In other words, these terms in the influence phase serve to renormalise the non-dissipative terms already present in the action of the source.

Let us return to the terms in Eq.\eqref{eq:RRMinkn}: they are  odd in $\w$, and hence \emph{cannot} be countertermed or absorbed into the non-dissipative action. We can simplify these remaining terms by substituting the STF definition of spherical harmonics (see Eq.\eqref{eq:OrthoToSTF}) and invoking the STF addition theorem in Eq.\eqref{eq:STFAddition}. We then get the 
radiation-reaction influence phase as
\begin{equation}
	\begin{split}
S_{RR}^\text{Odd $d$}&=\sum_{\ell}\int\frac{d\w}{2\pi} \frac{i\pi}{2 \nn_{d,\ell}|\mathbb{S}^{d-1}|} \left(\frac{\w }{2}\right)^{2\nu} \frac{1}{\Gamma(1+\nu)^2} \frac{1}{\ell!}\Pi^{<i_1i_2\ldots i_\ell>}_{ <j_1j_2\ldots j_\ell>} \\
  &\times \int d^dr_0 \left\{\rho_D(\vec{r}_0,\w)x_0^{j_1}x_0^{j_2}\ldots x_0^{j_\ell}\ {}_0F_1\left[1+\nu,-\frac{\w^2r_0^2}{4}\right]\right\}^\ast\\
  &\qquad\times \int d^dr\ \left\{\rho_A(\vec{r},\w)x_{i_1}x_{i_2}\ldots x_{i_\ell}\ {}_0F_1\left[1+\nu,-\frac{\w^2r^2}{4}\right]\right\}\ .
\end{split}
\end{equation}
Here we recognise the STF multipole moments of the sources' absorptive and emissive parts. We will find it convenient to  define our STF multipole moments as 
\begin{equation}
	\begin{split}
\mathcal{Q}_{A,STF}^{i_1\ldots i_\ell}(\w) \equiv  \frac{1}{2\nu}\Pi^{<i_1i_2\ldots i_\ell>}_{ <j_1j_2\ldots j_\ell>}\int d^dr\  \rho_A(\vec{r},\w)x^{j_1}x^{j_2}\ldots x^{j_\ell}\ {}_0F_1\left[1+\nu,-\frac{\w^2r^2}{4}\right]\ ,\\
\mathcal{Q}_{D,STF}^{i_1\ldots i_\ell}(\w) \equiv  \frac{1}{2\nu}\Pi^{<i_1i_2\ldots i_\ell>}_{ <j_1j_2\ldots j_\ell>}\int d^dr\  \rho_D(\vec{r},\w)x^{j_1}x^{j_2}\ldots x^{j_\ell}\ {}_0F_1\left[1+\nu,-\frac{\w^2r^2}{4}\right]\ .
\end{split}
\end{equation}
In terms of these STF multipole moments, the action for radiation reaction takes the form
\begin{equation}
\label{eq:SRRMinkSTF}
	\begin{split}
S_{RR}^\text{Odd $d$}&=\sum_{\ell\vec{m}}\int\frac{d\w}{2\pi} \frac{2\pi i}{\Gamma(\nu)^2} \left(\frac{\w }{2}\right)^{2\nu}  \multj_D^\ast(\w,\spL)\multj_A(\w,\spL)\\
  &=\sum_{\ell}\int\frac{d\w}{2\pi} \frac{2\pi i}{ \Gamma(\nu)^2} \left(\frac{\w }{2}\right)^{2\nu} \frac{1}{\nn_{d,\ell}|\mathbb{S}^{d-1}|} \frac{1}{\ell!}\mathcal{Q}_{D,STF}^{\ast<i_1i_2\ldots i_\ell>}\mathcal{Q}^{A,STF}_{ <i_1i_2\ldots i_\ell>}\ .
\end{split}
\end{equation}
In the first line, we have quoted the answer in terms of the spherical multipole moments
for comparison. The multipole action above could also be derived entirely by using Cartesian STF harmonics from the very beginning (See \cite{Ross:2012fc} for a detailed derivation). Given the absence of Cartesian coordinates valid everywhere on the static patch, we will employ a judicious mix of  spherical harmonic and STF harmonic expansions
to compute the influence phase. The flat spacetime derivation we have given here closely mimics the strategy we will eventually use for dS.

Let us conclude this flat spacetime discussion by commenting on the case where $d$ is even and $\nu\in\mathbb{Z}$. We will tackle this case by a dimensional regularisation via analytic continuation in $\nu$. From our discussion of multipole expansion, it is clear that the time reversal even terms in Eq.\eqref{eq:RRMinknn} are the same for any $\nu$ and can be counter-termed away similarly.

The terms in  Eq.\eqref{eq:RRMinkn}, on the other hand, get multiplied by a factor of $(1+i\cot \pi \nu)$ for a general $\nu$: this can be seen, e.g., in  Eq.\eqref{eq:HPlusDDW}. The $\cot \pi \nu$ factor  leads to novel divergences as $\nu$ approaches an integer, necessitating further counter-terms. 

To compute the counter-terms as $\nu\to n\in \mathbb{Z}$, we need the following expansion:
\begin{equation}\label{eq:evenDimRegFlat}
(1+i\cot \pi \nu)\frac{2\pi i}{ \Gamma(\nu)^2} \left(\frac{\w }{2H}\right)^{2\nu}=\frac{1}{ \Gamma(n)^2} \left(\frac{\w }{2 H}\right)^{2n}\left\{\frac{2}{\nu-n}-4\psi^{(0)}(n) +\ln \left(\frac{\w}{2H}\right)^4+O(\nu-n)  \right\}\ . 
\end{equation}
Here $H$ is the characteristic scale for dimensional regularisation and $\psi^{(0)}(x)\equiv \frac{d}{dx}\ln \Gamma(x)$ is the di-gamma function. Using a version of modified minimal subtraction, we counter-term away the first two terms inside the bracket of RHS. Thus, the influence phase due to radiation reaction for even spatial dimensions is
\begin{equation}\label{eq:RRevenFlat}
\begin{split}
S_{RR}^\text{Even $d$}&=
\sum_{\ell}\int\frac{d\w}{2\pi} \frac{1}{ \Gamma(\nu)^2} \left(\frac{\w }{2}\right)^{2\nu} \ln \left(\frac{\w^4}{H^4}\right)\frac{1}{\nn_{d,\ell}|\mathbb{S}^{d-1}|} \frac{1}{\ell!}\mathcal{Q}_{D,STF}^{\ast<i_1i_2\ldots i_\ell>}\mathcal{Q}^{A,STF}_{ <i_1i_2\ldots i_\ell>}\ ,
\end{split}
\end{equation}
where we have reset $n$ again everywhere to the variable $\nu$. What we have here is a classical renormalisation group running of the multipole couplings present in the world line action, i.e., an RGE induced by the classical radiation reaction. Such classical RGE is, in fact, common in many radiation reaction problems (See e.g. discussions in \cite{Goldberger:2004jt,Goldberger:2007hy,Foffa:2011np,Birnholtz:2013nta}). We will see later how this non-local influence phase gets further modified in dS spacetime.

\section{Designer scalar in dS : Green functions, regularisation and renormalistion}\label{app:DesSc}
In this appendix, we aim to describe the scalar Green functions in dS spacetime in some amount of detail. Our focus will be on a point-like observer sitting on the south pole, and our Green functions are all hence `boundary-to-bulk' with the boundary being the world line at the south pole. The point-like nature necessitates a careful discussion of regularisation, counter-terms etc.: our discussion will closely parallel the flat spacetime discussion in the previous appendix as well as the dS discussion in \cite{Lopez-Ortega:2006aal,Anninos:2011af}. We will also confine ourselves to a single copy of the static patch in this appendix,  relegating the applications to dS-SK to the next appendix.

We will work with outgoing Eddington Finkelstein(EF) coordinates\cite{Spradlin:2001pw}  describing the static patch of dS spacetime dS${}_{d+1}$. This
spacetime is a solution of the Einstein equations with a positive cosmological constant
\begin{equation}
	\Lambda= \frac{1}{2}d(d-1)\ .
\end{equation}
We have chosen units where the Hubble constant is unity. The spacetime metric is
\begin{equation}\label{eq:dSEF}
	ds^2=-2\ du\ dr-(1-r^2)\ du^2+r^2 d\Om_{d-1}^2\ .
\end{equation}
Here $d\Om_{d-1}^2$ denotes the metric on a unit $\mathbb{S}^{d-1}$. The outgoing Eddington Finkelstein time $u$ is related
to the more commonly used time $t$ via $u=t-r_\ast$ where $r_\ast$ is the tortoise coordinate defined via
\begin{equation}\label{eq:tortoise}
	r_\ast\equiv -i \pi \zeta\equiv \int_0^r\frac{d\rho}{1-\rho^2}=\frac{1}{2}\ln\left(\frac{1+r}{1-r}\right)\ .
\end{equation}
The radial coordinate $r$ is centred around a static observer sitting at $r=0$. We will mostly work with the frequency domain
where the time dependence of fields\footnote{We note a slight  inconsistency in our definitions when compared to definitions in the  appendix \ref{app:FlatMult}. In appendix \ref{app:FlatMult}, we fourier-transformed with respect to standard time slices, whereas here in $dS$ we are fourier-transforming with respect to outgoing EF time $u$. Since in flat spacetime $u=t-r$, this means that all the flat space radial functions in appendix \ref{app:FlatMult} should be multiplied with a pre-factor of $e^{-i\w r}$ before they can be compared against the dS results described here.\label{ftnt:fourier}} is taken to be $\sim e^{-i\w u}$. Further, we will decompose everything into appropriate spherical
harmonics on $\mathbb{S}^{d-1}$. The spherical harmonics are labelled by the eigenvalue of the sphere Laplacian  $\nabla^2_{\mathbb{S}^{d-1}}$ which is $-\ell(\ell+d-2)$.

As described in the main text, we will consider a class of \emph{designer} scalar systems in dS with an action 
\begin{equation}\label{eq:ActPhi}
		S=-\frac{1}{2}\int d^{d+1}x \sqrt{-g}\ r^{\nn+1-d}\left\{\partial^\mu\Phi_{\nn}\,\partial_\mu\Phi_{\nn}+\frac{\Phi_{\nn}^2}{4r^2}\left[(d+\nn-3)(d-\nn-1)-r^2\left(4\mu^2-(\nn+1)^2\right)\right]\right\}
 	\end{equation}
After we strip out the harmonic dependence in time/angles, the above action results in a radial ODE of the form
\begin{equation}\label{Eq:phiRadODEInit}
	\begin{split}
		&\frac{1}{r^\nn}\Dp [r^\nn \Dp \phN] +\w^2\phN\\
		&\qquad+\frac{1-r^2}{4r^2}\Bigl\{(\nn-1)^2-(d+2\ell-2)^2+[4\mu^2-(\nn+1)^2]r^2\Bigr\}\phN=0\ .
	\end{split}
\end{equation}
Here $\phN(r,\w,\spL)$ is the radial part of the field, the derivative operators $D_\pm \equiv (1-r^2)\del_r\pm i\w$, and the equation depends on the parameters $\{\mu,\nn,\ell\}$ whose physical interpretation will be clear momentarily. 

The combination $(\nn+1)^2-4\mu^2$ can be interpreted as a mass term $4m^2$ for the scalar in Hubble units.
The exponent $\nn$ describes the auxiliary radial varying dilaton mentioned at the beginning of this appendix. The index $\ell$ is associated with the eigenvalue of the sphere laplacian.
The expressions involved simplify considerably if we use, instead of $\ell$, the following parameter:
\begin{align}
	\begin{split}
		\nu &\equiv \frac{d}{2}+\ell-1\ .
	\end{split}
\end{align}
For example, in terms of $\nu$, the eigenvalue of the sphere laplacian becomes $(\frac{d}{2}-1)^2-\nu^2$. Since we will be concerned with the cases where $d>2$ and $\ell\geq 0$, $\nu$ is a positive number. We can then rewrite the above ODE as
\begin{equation}\label{Eq:phiRadODE}
	\begin{split}
		&\frac{1}{r^\nn}\Dp [r^\nn \Dp \phN] +\w^2\phN\\
		&\qquad+\frac{1-r^2}{4r^2}\Bigl\{(\nn-1)^2-4\nu^2+[4\mu^2-(\nn+1)^2]r^2\Bigr\}\phN=0\ .
	\end{split}
\end{equation}
It is instructive to rewrite the above ODE in terms of a new field  $\psi\equiv r^{\frac{\nn}{2}}\phN$ as
\begin{equation}\label{Eq:psiRadODE}
	\begin{split}
		(\Dp^2+\w^2) \psi +\frac{1-r^2}{4r^2}\Bigl\{1-4\nu^2+[4\mu^2-1]r^2\Bigr\}\psi=0\ .
	\end{split}
\end{equation}
The absence of $\nn$ in this ODE shows that $\nn$ merely controls the overall pre-factor.
We also note a symmetry under $\nu\mapsto -\nu$ and $\mu\mapsto -\mu$: either of these sign changes should map one solution to the other.

\subsection{Outgoing Green function}
The above second-order radial ODE  can be exactly solved in terms of hypergeometric functions. The  worldline to bulk outgoing Green function is given by\cite{Bunch:1978yq,Lopez-Ortega:2006aal,Anninos:2011af} 
\begin{equation}\label{eq:GoutI}
	\begin{split}
		\gO(r,\w,\spL) &=r^{\nu-\frac{\nn}{2}}(1+r)^{-i\w}
		\\
		&\quad\times \frac{\Gamma\left(\frac{1+\nu-
				\mu-i\w}{2}\right)\Gamma\left(\frac{1+\nu+\mu-i\w}{2}\right)}{\Gamma(1-i\w)\Gamma\left(1+\nu \right)}\ {}_2F_1\left[\frac{1+\nu-
			\mu-i\w}{2},\frac{1+\nu+\mu-i\w}{2};1-i\w;1-r^2\right]\ .
	\end{split}
\end{equation}
Here we have fixed the overall normalisation by an appropriate boundary condition to be described below. We will devote this subsection to a detailed study of the above Green function.

We remind the reader that the hypergeometric function always has a nice series expansion around the point where its last argument vanishes. It then follows that  the above solution is manifestly regular at the future horizon $r=1$ without any branch cuts or poles. An alternate form for the same function that emphasises the small $r$ behaviour near the observer's worldline is
\begin{equation}\label{eq:GoutII}
	\begin{split}
		\gO &=r^{-\nu-\frac{1}{2}(\nn-1)}(1+r)^{-i\w}
		\\
		&\times\Bigl\{ {}_2F_1\left[\frac{1-\nu+\mu-i\w}{2},\frac{1-\nu-\mu-i\w}{2};1-\nu;r^2\right]\Bigr.\\
		&\Bigl.\quad\qquad -(1+i\cot\nu \pi)\khO\frac{r^{2\nu}}{2\nu}\ {}_2F_1\left[\frac{1+\nu-
			\mu-i\w}{2},\frac{1+\nu+\mu-i\w}{2};1+\nu;r^2\right] \Bigr\}\ .
	\end{split}
\end{equation}
Here $\khO$ is the worldline  retarded Green function given by the expression\cite{Lopez-Ortega:2006aal,Anninos:2011af}.
\begin{equation}\label{eq:KhOut}
	\begin{split}
		\khO(\w,\ell) &\equiv 2\frac{\Gamma\left(\frac{1+\nu-
				\mu-i\w}{2}\right)\Gamma\left(\frac{1+\nu+\mu-i\w}{2}\right)
			\Gamma\left(1-\nu\right)
		}{\Gamma\left(\frac{1-\nu+\mu-i\w}{2}\right)\Gamma\left(\frac{1-\nu-\mu-i\w}{2}\right)
			\Gamma\left(\nu\right)(1+i\cot\nu \pi)}\\
   &=-e^{i\nu\pi}\frac{2\pi i}{\Gamma(\nu)^2} \frac{\Gamma\left(\frac{1+\nu-\mu-i\w}{2}\right)\Gamma\left(\frac{1+\nu+\mu-i\w}{2}\right)}{\Gamma\left(\frac{1-\nu+\mu-i\w}{2}\right)\Gamma\left(\frac{1-\nu-\mu-i\w}{2}\right)} \ .
	\end{split}
\end{equation}
The reason for choosing the normalisation of $\khO$ this way will become clear eventually. The above equation is the dS analogue of the Hankel function decomposition into Neumann and Bessel functions. As in Eq.\eqref{eq:Hankelgen}, when $d$ is even and $\nu$ is an integer, the above expression should be understood as a limit, with the $\cot\nu\pi$ divergence exactly cancelling the divergence in the first term of Eq.\eqref{eq:GoutII}.

The hypergeometric identity used for the above decomposition is 
\begin{equation}
	\begin{split}
		&\frac{\Gamma(a)\Gamma(b)}{\Gamma(c)\Gamma(a+b-c)} z^{a+b-c} {}_2F_1(a,b;c;1-z)\\
		&={}_2F_1(c-a,c-b;1+c-a-b;z)\\
		&\quad+z^{a+b-c}\frac{\Gamma(a)\Gamma(b)}{\Gamma(c)\Gamma(c-a-b)} \frac{\Gamma(c)\Gamma(a+b-c)}  {\Gamma(c-a)\Gamma(c-b)}{}_2F_1(a,b;a+b-c+1;z)\ ,
	\end{split}
\end{equation}
where we have taken
\begin{equation}
	\begin{split}
		a=\frac{1+\nu-
			\mu-i\w}{2}\ ,\ b=\frac{1+\nu+\mu-i\w}{2}\ ,\
		c=1-i\w\ ,\ z=r^2\ .
	\end{split}
\end{equation}
In all these identities, we take the branch cuts of hypergeometric functions as well as $(1+r)^{-i\w}$ to be outside the open unit disk in the complex $r$ plane. Thus, all these functions are analytic within the open static patch and in turn, on the dS-SK contour.

With this new form for the outgoing Green function, it is straightforward to obtain  a near-origin expansion to all orders. The explicit expressions are given by 
\begin{equation}
	\begin{split}
		& {}_2F_1\left[\frac{1+\nu-
			\mu-i\w}{2},\frac{1+\nu+
			\mu-i\w}{2};1+\nu;r^2\right]\\
		&=\sum_{k=0}^\infty \frac{r^{2k}}{(2k)!}
		\frac{(\nu-\mu-i\w-1+2k)!!}{(\nu-\mu-i\w+1)!!}
		\frac{(\nu+\mu-i\w-1+2k)!!}{(\nu+\mu-i\w-1)!!} 
		\frac{(2\nu)!! (2k-1)!!}{(2\nu+2k)!!}\ ,
	\end{split}
\end{equation}
as well as
\begin{equation}
	\begin{split}
		& {}_2F_1\left[\frac{1-\nu+\mu-i\w}{2},\frac{1-\nu-\mu-i\w}{2};1-\nu;r^2\right]\\
		&=\sum_{k=0}^\infty \frac{(-r^2)^k}{(2k)!}
		\frac{(\nu-\mu+i\w-1)!!}{(\nu-\mu+i\w-1-2k)!!}
		\frac{(\nu+\mu+i\w-1)!!}{(\nu+\mu+i\w-1-2k)!!} 
		\frac{(2\nu-2-2k)!! (2k-1)!!}{(2\nu-2)!!}\ .
	\end{split}
\end{equation}
The second expansion can be interpreted literally only for $d$ odd (i.e., when $\nu\in \mathbb{Z}+\frac{1}{2}$). For $d$ even, the above expansion (and most of the discussion below) should be understood in a dimensionally regularised sense. 

The near-origin form of the outgoing Green function shows the normalisation
\begin{equation}\label{eq:PhiBC}
	\begin{split}
		\lim_{r\to 0}r^{\nu+\frac{\nn-1}{2}}\gO &=1\ .
	\end{split}
\end{equation}
This condition can be thought of as the dS analogue of the condition on the AdS boundary-to-bulk Green function. As in that case, the above condition along with outgoing property/analyticity at the future horizon  uniquely determines $\gO$. Extending this analogy to AdS, we can roughly read off the  retarded worldline Green function $\khO$  by looking at the ratio of coefficients of the sub-dominant solution to the dominant solution in the outgoing solution $\gO$. This is essentially the Son-Starinets prescription\cite{Son:2002sd} of AdS/CFT adapted to the present dS context. Such analogies have been noted before in \cite{Anninos:2011af}: our aim here is to give a more systematic derivation of these statements, taking into account the subtleties associated with divergences, regularisation, finite size effects, etc.   

To this end, let us begin with a physical interpretation of the outgoing Green function $\gO$.
If we are given that $\phN$ behaves at small $r$ near the worldline as
\begin{equation}\label{eq:PhiBndVal}
	\begin{split}
		\phN(r,\w,\spL)= \frac{\multj(\w,\spL)}{r^{\nu+\frac{1}{2}(\nn-1)}}+\ldots\ ,
	\end{split}
\end{equation}
where $\mathbb{S}_\spL$ is a spherical harmonic on $\mathbb{S}^{d-2}$, we then have a unique outgoing solution \[\phN(r,\w,\spL)=\gO(r,\w,\spL) \multj(\w,\spL) \] describing the field that is radiated out of the worldline. This is the dS analogue of the outgoing Hankel Green function in flat space.\footnote{More precisely, in outgoing EF coordinates the corresponding Green functions in flat space are the outgoing Hankel Green functions given in Eq.\eqref{eq:Hankelgen} multiplied with a prefactor of  $e^{-i\w r}$. See footnote \ref{ftnt:fourier} for an explanation for this pre-factor.} 

The alternate form we have written down above in Eq.\eqref{eq:GoutII}  is then the dS version of the familiar statement\footnote{We review this statement, for the benefit of the reader, around Eq.\eqref{eq:Hankelgen}.} that the outgoing Hankel Green function
can be written as the sum of a Neumann Green function (which diverges
near the origin) and a Bessel J function (which is regular at the origin). Such a decomposition
of the outgoing Green function into a singular Green function and a regular solution is a first step in Dirac's approach to the self-force\cite{Dirac:1938nz} (the curved space version is sometimes also termed as the Detweiler-Whiting decomposition\cite{Detweiler:2002mi}). We will later show in appendix\ref{app:DetWhitdS} that our answer matches in dS$_4$ with the regular part quoted in \cite{Burko:2002ge, Poisson:2011nh} using the rules of Deitweiler-Whiting decomposition.

\subsection{Renormalised conjugate field and \mathinhead{\kO}{Kout}}
We will now turn to the question of deriving the worldline Green function $\kO$ from the outgoing Green function $\gO$. As we will describe in detail below, the physics here is that of radiation reaction and the main subtlety is how to deal with divergences. Our main strategy here will be to define a renormalised conjugate field which reduces to $\kO$ near the source worldline. The idea here is philosophically similar to other radiation reaction computations in the literature\cite{Detweiler:2002mi,Goldberger:2004jt,Goldberger:2007hy,Porto:2016pyg,Levi:2018nxp,Barack:2018yvs} as well as the counter-term subtraction in AdS/CFT\cite{Balasubramanian:1999re}. The implementation is however sufficiently different that  we provide a detailed analysis below.

The radial ODE Eq.\eqref{Eq:phiRadODE} can be derived by extremising the action
\begin{equation}\label{Eq:phiAct}
	\begin{split}
		S&= -\frac{1}{2}\sum_\spL\int\frac{d\w}{2\pi}\oint  \frac{r^\nn dr}{1-r^2}\Bigl[(\Dp \phN)^* \Dp \phN-\w^2 \phN^* \phN\Bigr.\\
		&\Bigl.\qquad-\frac{1-r^2}{4r^2}\Bigl\{(\nn-1)^2-4\nu^2+[4\mu^2-(\nn+1)^2]r^2\Bigr\} \phN^* \phN\Bigr]+S_{ct}\ .
	\end{split}
\end{equation}
Here $S_{ct}$ denotes the counter-term action to be determined later. The integration over $r$ 
ranges over the regulated dS-SK contour (clockwise from the right static patch to the left static patch) and, in addition, we have indicated an integration over all frequencies and a  sum over spherical harmonics. The reality condition in the Fourier domain takes the form 
\begin{equation}
	\begin{split}
		\phN^* (r,w,\ell,\vec{m}) = \phN (r,-\w,\ell,-\vec{m}) \ .
	\end{split}
\end{equation}
Here $\vec{m}$ denotes the additional labels appearing in the spherical 
harmonic decomposition.

The canonical conjugate field for radial evolution is obtained by varying the above action with respect to $\partial_r \phN^*$ which yields $-r^\nn \Dp \phN$ after we take into account the fact that $\phN^*$ and $\phN$ are related by the reality condition quoted above. The minus sign in the canonical
conjugate is because we are looking at evolution along a space-like
direction. 

Taking into account the powers of $r$ multiplying the multipole moment $\multj$  in  Eq.\eqref{eq:PhiBndVal}, the canonical conjugate of $\multj$ should be defined with the opposite power, viz., we should consider instead
\begin{equation}
	\begin{split}
		-r^{-\nu-\frac{1}{2}(\nn-1)}\left[r^\nn \Dp \phN\right]\ .
	\end{split}
\end{equation}
The canonical conjugate field of the radial evolution at the two regulated boundaries is given by evaluating the above expressions at $r=r_c\pm i\varepsilon$. Naively the $r_c\to 0$ limit should then yield the required canonical conjugate that couples to the right/left point multipole source. This limit however does not work: on a generic solution, the $r_c\to 0$ limit is beset with divergences. Appropriate counter-terms need to be added to the above
bare expression before a sensible $r_c\to 0$ limit can be taken. The counter-terms arise from adding in a worldline counter-term action
\begin{equation}\label{Eq:PhiActCT}
	\begin{split}
		S_{ct}&= -\frac{1}{2}\sum_\spL\int\frac{d\w}{2\pi} r^{\nn-1} \ct_\nn(r,\w,\spL) \phN^* \phN|_{\text{Bnd}}\ .
	\end{split}
\end{equation}
Here $|_{\text{Bnd}}$ refers to the fact that we add such a contribution at every boundary. Being a boundary contribution, this addition does not change the equations of motion for the scalar field. If the original variational principle was defined with a Dirichlet boundary condition $\delta \phN|_{\text{Bnd}}=0$, the counterterm above does not change that boundary condition. The reader might expect, from the discussion at the end of appendix \S\ref{app:FlatMult} on flat spacetime, that  additional counter-terms will be required in even $d$ to deal with $\cot\pi\nu$ divergences. As we did there, we will first deal with singularities at the sources before solving the divergences peculiar to even $d$.

In the above expression, we should take  $\ct_\nn(r,\w,\spL)$ to be a real and even function of $\w$ to get a real counter-term action. Addition of this worldline action modifies the canonical conjugate evaluated at the radial boundaries to 
\begin{equation}\label{eq:piDef}
	\begin{split}
		r^{-\nu-\frac{1}{2}(\nn-1)}\piN\equiv -r^{-\nu-\frac{1}{2}(\nn-1)}\left[r^\nn \Dp+r^{\nn-1}\ct_\nn \right]\phN\ .
	\end{split}
\end{equation}
The $\ct_\nn$ should then be chosen such that this object evaluated at $r=r_c\pm i\varepsilon$ has a well-defined $r_c\to 0$ limit.

We will now determine $\ct_\nn$ by studying the outgoing Green function (the counter-terms determined using a generic enough solution should work for every other solution). As we shall
see, the boundary value of the \emph{renormalised} conjugate field in this case is  the boundary Green function $\kO$. Before going into the details of the computation, 
it might be useful to situate it in a familiar  physical context. 

In the case of electromagnetism, the worldline Green function $\kO$ for a charged particle encodes the radiation reaction or self-force due to the particle's EM fields acting on itself. While this statement is broadly true, it is clear that this idea has to be interpreted with some care. If we take the bare electric field produced by the point charge and try to compute the self-force on it naively, the calculation will be dominated by the Coulomb divergence at the origin yielding an infinite answer. 

A little bit of thought however reveals that these divergences merely serve to relate the bare properties (e.g., mass) of the fictitious charge-free particle to the properties of the actual physical particle. What we should do instead is to compute the renormalised electric field felt by the particle after adding counter-terms which shift the mass to the experimentally measured value. This renormalised field associated with the radiation is determined from the near field by imposing the outgoing boundary condition and can then be used to compute the self-force of the particle. 

With this physical example in mind, we can interpret the first term in Eq.\eqref{eq:GoutII} as analogous to the Coulomb field in the near region whose divergent contributions need to be removed by using counter-terms. It is only after this is done that we can extract $\khO$ as the renormalised worldline Green function.

We will now  demand that the renormalised conjugate field computed over the first term in Eq.\eqref{eq:GoutII} vanish. This fixes the counter-term function $\ct_\nn$ to be 
\begin{equation}\label{eq:ctnn}
	\begin{split}
		\frac{\ct_\nn}{1-r^2}&\equiv
		-r\frac{d}{dr}\ln\left\{ r^{-\nu -\frac{1}{2}
			(\nn-1)} (1-r^2)^{-\frac{i\w}{2}} {}_2F_1\left[\frac{1-\nu+\mu-i\w}{2},\frac{1-\nu-\mu-i\w}{2};1-\nu;r^2\right] \right\}\ .
	\end{split}
\end{equation}
Here we take the branch cut of $(1-r^2)^{-\frac{i\w}{2}}$ to be away from the open unit disc $|r|<1$ in the complex $r$ plane and, with this choice, $\ct_\nn$ is  analytic everywhere inside each copy of the static patch, and has no discontinuity across the dS-SK branch-cut.
While it is not obvious from the expression above, we can invoke the  Euler transformation formula for the hyper-geometric function which states that
\begin{equation}\label{eq:EulerRev}
	\begin{split}
		{}_2F_1&\left[\frac{1\pm \nu+\mu+i\w}{2},\frac{1\pm \nu-\mu+i\w}{2};1\pm \nu;r^2\right]\\ &=(1-r^2)^{-i\w}{}_2F_1\left[\frac{1\pm\nu+\mu-i\w}{2},\frac{1\pm\nu-\mu-i\w}{2};1\pm \nu;r^2\right] \ ,
	\end{split}
\end{equation}
to conclude that $\ct_\nn$  is  a real and even function of $\w$. Here we have taken the function to be analytic in static patch again and hence  $\ct_\nn$ has a well-behaved small $r$ expansion. The first few terms in this expansion are given by 
\begin{equation}
	\begin{split}
		\ct_\nn&=(1-r^2)\left(\nu+\frac{1}{2}
		(\nn-1)\right) + r^2 \frac{(\nu-\mu-1)(\nu+\mu-1)-\w^2}{2\nu-2}\\
		&+r^4\frac{
			[(\nu-\mu-1)^2+\w^2][(\nu+\mu-1)^2+\w^2]}
		{(2\nu-2)^2(2\nu-4)}\\
		&+r^6\frac{[(\nu-\mu-1)^2+\w^2][(\nu+\mu-1)^2+\w^2]}
		{(2\nu-2)^3(2\nu-4)(2\nu-6)}\\
		&\quad \times[(2\nu-2)(2\nu-4)-2(\nu-\mu-1)(\nu+\mu-1)+2\w^2]\ +\ldots .
	\end{split}
\end{equation}
Note  that all terms in the above expansion are indeed real and even functions of $\w$ as claimed. Note that all the $r$ and $\w$ factors appear in the numerator implying that this counter-term is local in time/radial direction. 

Now that we have the expression for the counter-term, it is straightforward to compute
the renormalised conjugate field evaluated over the outgoing Green function. We obtain the following answer
\begin{equation}\label{eq:piOut}
	\begin{split}
		\piN^{\text{Out}}&\equiv -\left[r^\nn \Dp+r^{\nn-1}\ct_\nn \right]\gO\\
		&= (1+i\cot \pi \nu)\khO \mathscr{Z}_\nn(r,\w)r^{\nu+\frac{1}{2}(\nn-1)}(1+r)^{-i\w}
		{}_2F_1\left[\frac{1+\nu-\mu-i\w}{2},\frac{1+\nu+\mu-i\w}{2};1+\nu;r^2\right]\ ,
	\end{split}
\end{equation}
where $\mathscr{Z}_\nn(r,\w)$ is a function given by the expression
\begin{equation}
	\begin{split}
		\frac{\mathscr{Z}_\nn}{1-r^2}&\equiv 1-\frac{r}{2\nu}\frac{d}{dr}\ln\left\{ 
		\frac{{}_2F_1\left[\frac{1-\nu-\mu-i\w}{2},\frac{1-\nu+\mu-i\w}{2};1-\nu;r^2\right]} {{}_2F_1\left[\frac{1+\nu+\mu-i\w}{2},\frac{1+\nu-\mu-i\w}{2};1+\nu;r^2\right]} \right\}\ .
	\end{split}
\end{equation}
This is also a real and even function of $\w$ with a well-behaved series expansion near the origin. We thus see that the renormalised conjugate field of the outgoing wave is essentially its regular part, obtained after dropping its singular part and then renormalised by a factor of $\mathscr{Z}_\nn$. Taking the $r\to 0$ limit yields
\begin{equation}
	\begin{split}
	\lim_{r\to 0}r^{-\nu-\frac{1}{2}(\nn-1)}\piN^{\text{Out}}\equiv 	-\lim_{r\to 0}r^{-\nu-\frac{1}{2}(\nn-1)}\left[r^\nn \Dp+r^{\nn-1}\ct_\nn \right]\gO= (1+i\cot \pi \nu)\khO \ .
	\end{split}
\end{equation}
This then justifies our original definition for $\khO$.

If $d$ is odd and $\nu\equiv \frac{d}{2}+\ell-1\in\mathbb{Z}+\frac{1}{2}$, we can set $\nu$ to its actual value everywhere (i.e., remove dim-reg.) in our result: the value of the renormalised conjugate field at the world line (which we shall henceforth refer to by the symbol $\kO$) is then finite. We can then write
\begin{equation}\label{eq:KOutOdd}
\begin{split}
		\kO|_{\text{Odd d}} &= (1+i\cot\pi\nu)\khO|_{\text{Odd d}}
   =-e^{i\nu\pi}\frac{2\pi i}{\Gamma(\nu)^2} \frac{\Gamma\left(\frac{1+\nu-\mu-i\w}{2}\right)\Gamma\left(\frac{1+\nu+\mu-i\w}{2}\right)}{\Gamma\left(\frac{1-\nu+\mu-i\w}{2}\right)\Gamma\left(\frac{1-\nu-\mu-i\w}{2}\right)} \ .
	\end{split}
\end{equation}
For the massless case in odd $d$, we have $\mu,\nu\in\mathbb{Z}+\frac{1}{2}$ for all values of interest given in table \ref{table:Nmuvalues}. If we further assume that $\mu\neq 1+\nu\equiv \frac{d}{2}+\ell$, the above expression is, in fact, an odd polynomial of $i\w$ with degree $2\nu$ (see table \ref{tab:KoutMark} for an illustration). An interesting example is  that of a conformally coupled scalar in odd $d$, where we have a closed-form expression
\begin{equation}
\kO\Big|_{\mu=\frac{1}{2}}=\frac{(-1)^{\nu-\frac{1}{2}}}{(2\nu-2)!!^2}\prod\limits_{k=1}^{2\nu}\left[\nu+\frac{1}{2}-k-i\w\right] 
\end{equation}
In all such cases, for every multipole moment, the Hubble corrections for the radiation correction terminate. Hence, we get a completely Markovian influence phase with no memory/tail terms. Further, as we shall explain in detail in the appendix \ref{app:RadReact}, for an arbitrarily moving point-like source, all the multipole  contributions add up nicely into a local generally covariant expression for the radiation reaction force.

\begin{table}[H]
	\centering
		\caption{$\frac{\kO}{-i\w}$ for $\mu\in\left\{\frac{d}{2}-1,\frac{d}{2}-2\right\}$ (gauge/gravity scalar/vector sectors) }
		\setlength{\extrarowheight}{2pt}
	\begin{tabular}{|c|c|c|c|}
		\hline
		$\mu=\frac{d}{2}-1$&$\ell=0$&$\ell=1$&$\ell=2$\\[0.5ex]\hline
		$d=3$& $1$&$ \w^2+1 $ & $\frac{ \w
			^4}{9}+\frac{5 \w^2}{9}+\frac{4 }{9}$  \\[0.5ex] $d=5$ & $
		\w^2+4$ & $\frac{\w^4}{9}+\frac{10
			\w^2}{9}+1 $ & $ \frac{\w
			^6}{225}+\frac{7 \w^4}{75}+\frac{28 \w
			^2}{75}+\frac{64 }{225} $\\[0.5ex]
		$d=7$&$\frac{\w^4}{9}+\frac{20 \w^2}{9}+\frac{64 }{9} $ &$\frac{\w^6}{225}+\frac{7 \w ^4}{45}+\frac{259 \w^2}{225}+1$ &$ \frac{ \w^8}{11025}+\frac{19 \w^6}{3675}+\frac{8 \w ^4}{105}+\frac{3088 \w^2}{11025}+\frac{256
			}{1225} $\\[0.5ex] \hline
		\hline
		$\mu=\frac{d}{2}-2$&$\ell=0$&$\ell=1$&$\ell=2$\\[0.5ex]\hline
		$d=3$& $1$&$ \w^2+1 $ & $\frac{ \w
			^4}{9}+\frac{5 \w^2}{9}+\frac{4 }{9}$  \\ $d=5$ & $\w^2+1$ & $\frac{\w^4}{9}+\frac{5 \w ^2}{9}+\frac{4 }{9} $ & $\frac{ \w
			^6}{225}+\frac{14 \w^4}{225}+\frac{49 \w ^2}{225}+\frac{4 }{25} $\\
		$d=7$&$\frac{\w^4}{9}+\frac{10 \w^2}{9}+1 $ &$\frac{\w^6}{225}+\frac{7 \w
			^4}{75}+\frac{28 \w^2}{75}+\frac{64
		}{225}$ &$\frac{ \w^8}{11025}+\frac{13 \w^6}{3675}+\frac{19 \w^4}{525}+\frac{1261  \w
		^2}{11025}+\frac{4 }{49} $\\[0.5ex] \hline
	\end{tabular}\label{tab:KoutMark}
\end{table}

For the minimally coupled massless scalar ($\mu=\frac{d}{2}$), we still obtain a polynomial $\kO$ for all multipoles except the monopole ($\ell=0$) contribution. The monopole has an extra $1/\w$ correction in addition to the polynomial terms odd in $\w$ (See tables \ref{tab:KoutNMark0} and \ref{tab:KoutNMark12}). An explicit expression for $\ell=0$ contribution is given by 
\begin{equation}
	\kO|_{\mu=1+\nu=\frac{d}{2}}=\frac{(d-2)^2}{i\w}\cosh\frac{\pi\w}{2}\  \frac{\Gamma\left(\frac{d-i\w}{2}\right)\Gamma\left(\frac{d+i\w}{2}\right)}{\Gamma\left(\frac{d}{2}\right)^2}
\end{equation}
The inverse omega that appears in the front of this expression suggests that the correct variable for a low frequency expansion in this case is the time integral of the scalar source rather than the source itself. Such a mild non-Markovianity for minimally coupled scalars in dS has been noted before\cite{Burko:2002ge,Burko:2002gf}, and we will review its physical interpretation in appendix \ref{app:RadReact}.

\begin{table}[H]
	\centering
	\setlength{\extrarowheight}{2pt}
	\caption{$i\w \kO$}
\begin{tabular}{|c|c|}
	\hline
	$\mu=\frac{d}{2}$&$\ell=0$ \\[0.5ex]\hline $d=3$&$  \w ^2+1  $ \\[0.5ex] $d=5$ & $
 \w ^4+10\w ^2  +9$\\[0.5ex]
	$d=7$&$\frac{\w^6}{9}+\frac{35 \w^4}{9}+\frac{259 \w ^2}{9}+25$ \\[0.5ex] \hline
\end{tabular}\label{tab:KoutNMark0}
\end{table}

\begin{table}[H]
	\centering
	\setlength{\extrarowheight}{2pt}
	\caption{$\frac{\kO}{-i\w}$}
	\begin{tabular}{|c|c|c|}
		\hline
		$\mu=\frac{d}{2}$&$\ell=1$&$\ell=2$\\[0.5ex]\hline
		$d=3$&$  \w ^2+4  $ & $\frac{ \w ^4}{9}+\frac{10  \w ^2}{9}+1 $  \\[0.5ex] $d=5$ & $\frac{\w ^4}{9}+\frac{20 \w ^2}{9}+\frac{64  
		}{9}  $ & $\frac{\w^6}{225}+\frac{7  \w^4}{45}+\frac{259 \w^2}{225}+1$\\[0.5ex]
		$d=7$ &$\frac{\w ^6}{225}+\frac{56 \w^4}{225}+\frac{784  \w^2}{225}+\frac{256 1 }{25}$ &$ \frac{ \w^8}{11025}+\frac{4 \w^6}{525}+\frac{94  \w ^4}{525}+\frac{12916 \w^2}{11025}+1$\\[0.5ex] \hline
	\end{tabular}\label{tab:KoutNMark12}
\end{table}

For generic values of $(\mu,\nu)$, a small $\w$
expansion of $\kO$ is easy to obtain by expanding out the gamma functions in terms of polygamma functions. We get
\begin{equation}\label{eq:Koutwexp}
	\begin{split}
		&\kO|_\text{Odd $d$} =-e^{i\nu\pi}\frac{2\pi i}{\Gamma(\nu)^2} \frac{\Gamma\left(\frac{1+\nu-\mu}{2}\right)\Gamma\left(\frac{1+\nu+\mu}{2}\right)}{\Gamma\left(\frac{1-\nu+\mu}{2}\right)\Gamma\left(\frac{1-\nu-\mu}{2}\right)}\\
		&\times \exp\left\{\sum\limits_{k=0}^{\infty}\frac{\left(\frac{-i\w}{2}\right)^{k+1}}{(k+1)!}\left[\psi^{(k)}\left(\frac{1+\nu-\mu}{2}\right)+\psi^{(k)}\left(\frac{1+\nu+\mu}{2}\right)-\psi^{(k)}\left(\frac{1-\nu-\mu}{2}\right)-\psi^{(k)}\left(\frac{1-\nu+\mu}{2}\right)\right]\right\}\ ,
	\end{split}
\end{equation}
where $\psi^{(k)}(z)\equiv \frac{d^{k+1}}{dz^{k+1}}\ln \Gamma(z)$ is the polygamma function. When  both $\nu+\mu$ or $\nu-\mu$ are non-negative integers, the terms in the above expressions become indeterminate and should instead be interpreted as a limit. In such cases, explicit computations show that the above exponential terminates yielding an odd polynomial in $\w$ when $\nu$ is half-integer.

We will now comment on the even $d$/integer $\nu$ case. The $\cot \pi \nu$ diverges in this limit, and  we need the analogue of Eq.\eqref{eq:evenDimRegFlat} to figure out the counter-terms needed to remove this divergence. The analogous expansion is given by
\begin{equation}\label{eq:evenDimRegdSOut}
\begin{split}
   (1+i \cot (\pi\nu))\khO&=
  \frac{(-)^n}{\Gamma(n)^2} \frac{\Gamma\left(\frac{1+n-\mu-i\w}{2}\right)\Gamma\left(\frac{1+n+\mu-i\w}{2}\right)}{\Gamma\left(\frac{1-n+\mu-i\w}{2}\right)\Gamma\left(\frac{1-n-\mu-i\w}{2}\right)}\left[\frac{2}{\nu-n}\right.\\ &\left.+\psi^{(0)}\left(\frac{1+n-\mu-i\w}{2}\right)+\psi^{(0)}\left(\frac{1+n+\mu-i\w}{2}\right)\right.\\ &\left.+\psi^{(0)}\left(\frac{1-n-\mu-i\w}{2}\right)+\psi^{(0)}\left(\frac{1-n+\mu-i\w}{2}\right)-4\psi^{(0)}(n)+O(\nu-n)\right]\ .
\end{split}
\end{equation} 
As in the flat spacetime, we can counter-term away the first two terms, and change the $n$ back to $\nu$. This yields  the renormalised worldline Green function as\cite{Lopez-Ortega:2006aal,Anninos:2011af}
\begin{equation}\label{eq:KoutEven}
	\begin{split}
		\kO|_\text{Even $d$} &=\Delta_\nn(\nu,\mu,\w)\left[\psi^{(0)}\left(\frac{1+\nu-\mu-i\w}{2}\right)+\psi^{(0)}\left(\frac{1+\nu+\mu-i\w}{2}\right)\right.\\ &\left.+\psi^{(0)}\left(\frac{1-\nu-\mu-i\w}{2}\right)+\psi^{(0)}\left(\frac{1-\nu+\mu-i\w}{2}\right)-4\psi^{(0)}(\nu)\right]\ ,
	\end{split}
\end{equation}
where the function $\Delta_\nn$ is defined below in Eq.\eqref{eq:ctEvenH}. To get this answer, we add to counterterm in Eq.\eqref{Eq:PhiActCT}, further terms of the form
\begin{equation}\label{Eq:PhiActCTnu}
	\begin{split}
		S_{ct,\text{Even}}&= \sum_\spL\frac{1}{\nu-n}\int\frac{d\w}{2\pi} r^{\nn-1+2n}\Delta_\nn(n,\mu,\w)\phN^* \phN|_{r_c}\ ,
	\end{split}
\end{equation}
where $n=\ell+\frac{d}{2}-1$ and we have defined 
\begin{equation}\label{eq:ctEvenH}
\begin{split}
   \Delta_\nn(n,\mu,\w)&\equiv
  \frac{(-)^n}{\Gamma(n)^2} \frac{\Gamma\left(\frac{1+n-\mu-i\w}{2}\right)\Gamma\left(\frac{1+n+\mu-i\w}{2}\right)}{\Gamma\left(\frac{1-n+\mu-i\w}{2}\right)\Gamma\left(\frac{1-n-\mu-i\w}{2}\right)}=\frac{1}{\Gamma(n)^2}\prod\limits_{k=1}^{n}\left[\frac{\w^2}{4}+\frac{1}{4}(\mu-n+2k-1)^2\right]\\
  &=\Delta_\nn^\ast(n,\mu,\w)\ .
\end{split}
\end{equation} 
Note that the explicit product form we give above is valid for $n\in\mathbb{Z}_+$. This form shows that  $\Delta_\nn$ is a real and even  function of $\w$, which is an essential condition for such a counterterm to be admissible. With this counterterm, Eq.\eqref{eq:KoutEven} is the dS generalisation of the radiation reaction influence phase in flat spacetime described by Eq.\eqref{eq:RRevenFlat}.
The simple logarithmic running in flat spacetime is now replaced by a more complicated RGE with the Hubble constant playing the role of the IR cutoff. A low frequency expansion
$\kO$ can be obtained by using the polygamma series expansion
\begin{align}
\begin{split}
   \psi^{(0)}&\left(\frac{1+\nu-\mu-i\w}{2}\right)+\psi^{(0)}\left(\frac{1+\nu+\mu-i\w}{2}\right)+\psi^{(0)}\left(\frac{1-\nu-\mu-i\w}{2}\right)+\psi^{(0)}\left(\frac{1-\nu+\mu-i\w}{2}\right)\\ &= \sum\limits_{k=0}^{\infty}\frac{\left(\frac{-i\w}{2}\right)^{k+1}}{(k+1)!}\left[\psi^{(k)}\left(\frac{1+\nu-\mu}{2}\right)+\psi^{(k)}\left(\frac{1+\nu-\mu}{2}\right)\right.\\ &\hspace{4 cm}\left.+\psi^{(k)}\left(\frac{1-\nu-\mu}{2}\right)+\psi^{(k)}\left(\frac{1-\nu+\mu}{2}\right)\right]\ ,
   \end{split}
\end{align}
which is well-defined except when any one of the polygamma arguments is a negative integer. These results agree with the dS expressions derived in \cite{Anninos:2011af}.

We will now add a remark about the poles of $\kO$ in the complex frequency plane. These are sometimes termed `de-Sitter quasi-normal modes' although we think this is a misleading terminology for the following reason. The adjective `quasi-normal' is usually applied to poles of Green functions that have a real as well as an imaginary part: as the name suggests, these are `almost' normal modes that characterise the physics of ring-down. The dS horizon does not ring-down and has no quasi-normal modes in this sense. 

\begin{table}
	\centering
		\caption{Residues of $\kO$ in even $d$ for $\mu\in\left\{\frac{d}{2},\frac{d}{2}-1,\frac{d}{2}-2\right\}$ (gauge/gravity scalar/vector/tensor sectors) \\ at $\w=-i(\mu+\nu+1)$. }
		\setlength{\extrarowheight}{2pt}
	\begin{tabular}{|c|c|c|c|c|c|}
		\hline
		$\mu=\frac{d}{2}$&$\ell=0$&$\ell=1$&$\ell=2$&$\ell=3$&$\ell=4$\\[0.5ex]\hline
		$d=4$& $24i$&$ -192i $ & $720i$ & $-1920i$ & $4200i$ \\[0.5ex] $d=6$& $-320i$&$ 1440i $ & $-4480i$ & $11200i$ & $-24192i$ \\[0.5ex]
		$d=8$& $2520i$&$ -8960i $ & $25200i$ & $-60480i$ & $129360i$ \\[0.5ex] \hline
		\hline
		$\mu=\frac{d}{2}-1$&$\ell=0$&$\ell=1$&$\ell=2$&$\ell=3$&$\ell=4$\\[0.5ex]\hline
		$d=4$& $16i$&$ -96i $ & $288i$ & $-640i$ & $1200i$ \\[0.5ex] $d=6$& $ -192i $ & $720i$ & $-1920i$ & $4200i$ & $-8064i$\\[0.5ex]
		$d=8$&$1440i$ & $-4480i$ & $11200i$ & $-24192i$ & $47040i$\\[0.5ex] \hline
  \hline
		$\mu=\frac{d}{2}-2$&$\ell=0$&$\ell=1$&$\ell=2$&$\ell=3$&$\ell=4$\\[0.5ex]\hline
		$d=4$& $8i$&$ -32i $ & $72i$ & $-128i$ & $200i$ \\[0.5ex] $d=6$& $-96i$&$ 288i $ & $-640i$ & $-1200i$ & $-2016i$ \\[0.5ex]
		$d=8$& $720i$&$ -1920i $ & $4200i$ & $-8064i$ & $14112i$ \\[0.5ex] \hline
	\end{tabular}\label{tab:EvendQNM1}
\end{table}

\begin{table}
	\centering
		\caption{Residues of $\kO$ in even $d$ for $\mu\in\left\{\frac{d}{2},\frac{d}{2}-1,\frac{d}{2}-2\right\}$ (gauge/gravity scalar/vector/tensor sectors) \\ at $\w=-i(5+\nu-\mu)$. }
		\setlength{\extrarowheight}{2pt}
	\begin{tabular}{|c|c|c|c|c|c|}
		\hline
		$\mu=\frac{d}{2}$&$\ell=0$&$\ell=1$&$\ell=2$&$\ell=3$&$\ell=4$\\[0.5ex]\hline
		$d=4$& $24i$&$ -192i $ & $720i$ & $-1920i$ & $4200i$ \\[0.5ex] $d=6$& $0$&$ 0 $ & $0$ & $0$ & $0$ \\[0.5ex]
		$d=8$& $0$&$0 $ & $0$ & $0$ & $0$ \\[0.5ex] \hline
		\hline
		$\mu=\frac{d}{2}-1$&$\ell=0$&$\ell=1$&$\ell=2$&$\ell=3$&$\ell=4$\\[0.5ex]\hline
		$d=4$& $48i$&$ -576i $ & $2880i$ & $-9600i$ & $25200i$ \\[0.5ex] $d=6$& $ -192i $ & $720i$ & $-1920i$ & $4200i$ & $-8064i$\\[0.5ex]
		$d=8$&$0$ & $0$ & $0$ & $0$ & $0$\\[0.5ex] \hline
  \hline
		$\mu=\frac{d}{2}-2$&$\ell=0$&$\ell=1$&$\ell=2$&$\ell=3$&$\ell=4$\\[0.5ex]\hline
		$d=4$& $72i$&$ -1152i $ & $7200i$ & $-28800i$ & $88200i$ \\[0.5ex] $d=6$& $-576i$&$ 2880i $ & $-9600i$ & $25200i$ & $-56448i$ \\[0.5ex]
		$d=8$& $720i$&$ -1920i $ & $4200i$ & $-8064i$ & $14112i$ \\[0.5ex] \hline
	\end{tabular}\label{tab:EvendQNM2}
\end{table}

\begin{table}
	\centering
		\caption{Residues of $\kO$ in even $d$ for $\mu\in\left\{\frac{d}{2},\frac{d}{2}-1,\frac{d}{2}-2\right\}$ (gauge/gravity scalar/vector/tensor sectors) \\ at $\w=-i(7-\nu+\mu)$. }
		\setlength{\extrarowheight}{2pt}
	\begin{tabular}{|c|c|c|c|c|c|}
		\hline
		$\mu=\frac{d}{2}$&$\ell=0$&$\ell=1$&$\ell=2$&$\ell=3$&$\ell=4$\\[0.5ex]\hline
		$d=4$& $120i$&$ -960i $ & $720i$ & $0$ & $0$ \\[0.5ex] $d=6$& $-1440i$&$1440i $ & $0$ & $0$ & $0$ \\[0.5ex]
		$d=8$& $2520i$&$0 $ & $0$ & $0$ & $0$ \\[0.5ex] \hline
		\hline
		$\mu=\frac{d}{2}-1$&$\ell=0$&$\ell=1$&$\ell=2$&$\ell=3$&$\ell=4$\\[0.5ex]\hline
		$d=4$& $96i$&$ -576i $ & $288i$ & $0$ & $0$ \\[0.5ex] $d=6$& $-960i$&$720i $ & $0$ & $0$ & $0$ \\[0.5ex]
		$d=8$& $1440i$&$0 $ & $0$ & $0$ & $0$ \\[0.5ex] \hline
  \hline
		$\mu=\frac{d}{2}-2$&$\ell=0$&$\ell=1$&$\ell=2$&$\ell=3$&$\ell=4$\\[0.5ex]\hline
		$d=4$& $72i$&$ -288i $ & $72i$ & $0$ & $0$ \\[0.5ex] $d=6$& $-576i$&$288i $ & $0$ & $0$ & $0$ \\[0.5ex]
		$d=8$& $720i$&$0 $ & $0$ & $0$ & $0$ \\[0.5ex] \hline
	\end{tabular}\label{tab:EvendQNM3}
\end{table}

\begin{table}
	\centering
		\caption{Residues of $\kO$ in even $d$ for $\mu\in\left\{\frac{d}{2},\frac{d}{2}-1,\frac{d}{2}-2\right\}$ (gauge/gravity scalar/vector/tensor sectors) \\ at $\w=-i(11-\nu-\mu)$. }
		\setlength{\extrarowheight}{2pt}
	\begin{tabular}{|c|c|c|c|c|c|}
		\hline
		$\mu=\frac{d}{2}$&$\ell=0$&$\ell=1$&$\ell=2$&$\ell=3$&$\ell=4$\\[0.5ex]\hline
		$d=4$& $120i$&$ -960i $ & $720i$ & $0$ & $0$ \\[0.5ex] $d=6$& $-320i$&$0 $ & $0$ & $0$ & $0$ \\[0.5ex]
		$d=8$& $0$&$0 $ & $0$ & $0$ & $0$ \\[0.5ex] \hline
		\hline
		$\mu=\frac{d}{2}-1$&$\ell=0$&$\ell=1$&$\ell=2$&$\ell=3$&$\ell=4$\\[0.5ex]\hline
		$d=4$& $160i$&$ -1920i $ & $2880i$ & $-640i$ & $0$ \\[0.5ex] $d=6$& $-960i$&$720i $ & $0$ & $0$ & $0$ \\[0.5ex]
		$d=8$& $0$&$0 $ & $0$ & $0$ & $0$ \\[0.5ex] \hline
  \hline
		$\mu=\frac{d}{2}-2$&$\ell=0$&$\ell=1$&$\ell=2$&$\ell=3$&$\ell=4$\\[0.5ex]\hline
		$d=4$& $200i$&$ -3200i $ & $7200i$ & $-3200i$ & $200i$ \\[0.5ex] $d=6$& $-1920i$&$2880i $ & $-640i$ & $0$ & $0$ \\[0.5ex]
		$d=8$& $720i$&$0 $ & $0$ & $0$ & $0$ \\[0.5ex] \hline
	\end{tabular}\label{tab:EvendQNM4}
\end{table}

The poles of $\kO$, when present, are more akin to Matsubara modes of thermal Green functions in that they  lie along the imaginary axis in the complex frequency plane. As is evident from Tables~\ref{tab:KoutMark},\ref{tab:KoutNMark0} and \ref{tab:KoutNMark12}, for $d$ odd and $\mu\in\left\{\frac{d}{2},\frac{d}{2}-1,\frac{d}{2}-2\right\}$, $\kO$ is a polynomial function of $\w$ and has no poles whatsoever. For $d$ even, the polygamma functions appearing in  Eq.\eqref{eq:KoutEven} have simple poles when their arguments become negative integers (this happens only along the negative imaginary axis in the complex frequency plane). The kernel $\kO$ inherits these poles except when they get cancelled by the zeroes of $\Delta_\nn(n,\mu,\w)$ given in Eq.\eqref{eq:ctEvenH}. We exhibit the residues of some of these  poles in the table~\ref{tab:EvendQNM1},\ref{tab:EvendQNM2},\ref{tab:EvendQNM3} and \ref{tab:EvendQNM4}: the ones with zero residues correspond to cancelled poles. The presence of these poles indicates that the small-frequency Langevin description might fail beyond a certain cut-off frequency.

We will conclude this section with a comment on the flat spacetime limit of the expressions derived in this appendix. 
Intuitively, we expect that the high-frequency modes  with $\w\gg 1$ would be insensitive to the cosmological constant, and would behave like Minkowski modes. This intuition can indeed be made precise by examining the high frequency expansion of $\kO$. Using Stirling approximation for the Gamma functions, we can indeed check the following statement valid for $\w\gg 1$:
\begin{align}
\begin{split}
\kO \approx \left\{\begin{array}{cc} \frac{2\pi i}{\Gamma(\nu)^2}\left(\frac{\w}{2}\right)^{2\nu} & \text{for $d$  odd ,} \\
\frac{1}{ \Gamma(\nu)^2} \left(\frac{\w }{2}\right)^{2\nu} \ln \left(\frac{\w^4}{H^4}\right) & \text{for $d$  even .} \\
\end{array}\right.
\end{split}
\end{align}
Comparing these limits with Eqs.\eqref{eq:SRRMinkSTF} and \eqref{eq:RRevenFlat}, we conclude $\kO$ is indeed the dS generalisation of the radiation reaction kernel.

\section{SK Green functions and the cosmological influence phase}\label{app:Inf}
We now turn to the problem of constructing the solution on the dS-SK spacetime contour. The construction here closely parallels corresponding derivation in AdS\cite{Skenderis:2008dg,Skenderis:2008dh,Glorioso:2018mmw,Chakrabarty:2019aeu,Jana:2020vyx} and we include a concise summary here mainly for completeness. The reader is encouraged to see these references for a more extensive discussion and interpretation of the expressions quoted below.

Our discussion in this appendix is structured as follows: we begin by extending our discussion of counter-terms etc. to the \emph{incoming} Green functions. Physically, such Green functions
are relevant while describing the effect of a distant source in the past of the observer.
As will be derived below, even if there are no sources present, an observer in dS spacetime sees cosmic background radiation at the dS temperature. We will need the incoming Green function to describe these 
waves.

\subsection{Time reversal, incoming waves and their branch-cut}

We would now like to argue that the renormalised conjugate field continues to be finite for the Green function  describing incoming waves. The incoming Green function can be computed from the answers we already have by using the time reversal isometry of the dS spacetime. The only non-trivial step involved is to realise  how the time reversal  isometry acts on EF coordinates.

The action of time reversal is achieved by the diffeomorphism
\begin{equation}
	\begin{split}
		u\mapsto 2\pi i \zeta-u\ ,\ \w\mapsto -\w\ ,
	\end{split}
\end{equation} 
where $\zeta$ is the mock tortoise coordinate introduced in Eq.\eqref{eq:zeta-def}. One can check that this diffeomorphism preserves the metric in Eq.\eqref{eq:dSEF} and is hence an isometry. The map $\w\mapsto -\w$ is necessary maintain the $\sim e^{-i\w u}$ factor in  Fourier domain. The time reversal is hence achieved by reversing $\w$ and then multiplying all  Fourier domain functions by a factor $e^{-2\pi\w \zeta}$.

Using the time reversal isometry the bulk to worldline Green function with incoming boundary condition takes the form
\begin{equation}\label{eq:GinI}
	\gI(r,\w,\spL) =e^{-2\pi\w\zeta}\gs(r,\w,\spL)
\end{equation}
Unlike $\gO$, the Green function $\gI$ has a branch-cut on the dS-SK contour, taking different values in the left vs. right static patches. The near origin expansion of $\gI$ can be obtained by using  the Euler transformation in Eq.\eqref{eq:EulerRev}:
\begin{equation}\label{eq:GinII}
	\begin{split}
		\gI(r,\w,\spL) &\equiv e^{-2\pi\w\zeta}\gs=e^{-2\pi\w\zeta}\left(\frac{1-r}{1+r}\right)^{-i\w}\times  r^{-\nu-\frac{1}{2}(\nn-1)}(1+r)^{-i\w}
		\\
		&\times\Bigl\{ {}_2F_1\left[\frac{1-\nu+\mu-i\w}{2},\frac{1-\nu-\mu-i\w}{2};1-\nu;r^2\right]\Bigr.\\
		&\Bigl.\quad\qquad -(1-i\cot\pi\nu)\khI \frac{r^{2\nu}}{2\nu}\ {}_2F_1\left[\frac{1+\nu-
			\mu-i\w}{2},\frac{1+\nu+\mu-i\w}{2};1+\nu;r^2\right] \Bigr\}\ .
	\end{split}
\end{equation}
The branch cuts of the explicit $(1\pm r)^{-i\w}$ are chosen to lie outside the open unit disc in the complex $r$ plane and a careful evaluation of the pre-factor  above yields
\begin{equation}\label{eq:LRExpId}
	\begin{split}
		e^{-2\pi\w\zeta}\left(\frac{1-r}{1+r}\right)^{-i\w}=
		\begin{cases}
			1 \qquad &\text{L contour}\\
			e^{-2\pi\w} \qquad &\text{R contour .}\ 
		\end{cases}
	\end{split}
\end{equation}
This shows explicitly the branch-cut and  jump in the incoming Green function. In the above equation, the symbol $\khI$ denotes the worldline  advanced Green function given by the expression
\begin{equation}\label{eq:KhIn}
	\begin{split}
		\khI(\w,\ell) &\equiv [\khO(\w,\ell)]^\ast=-e^{-2\pi i\nu}\Khout(-\w,\ell)\\
   &=e^{-i\nu\pi}\frac{2\pi i}{\Gamma(\nu)^2} \frac{\Gamma\left(\frac{1+\nu-\mu+i\w}{2}\right)\Gamma\left(\frac{1+\nu+\mu+i\w}{2}\right)}{\Gamma\left(\frac{1-\nu+\mu+i\w}{2}\right)\Gamma\left(\frac{1-\nu-\mu+i\w}{2}\right)} \ .
\end{split}
\end{equation}
The comments made in the context of $\khO$ below Eq.\eqref{eq:KhOut} apply also in this case. The decomposition in Eq.\eqref{eq:GinII}

Given the above definition of $\gI$, it is now straightforward to compute the renormalised
conjugate field. Since the incoming mode has a branch cut, it behaves differently at the two boundaries. Adding in the counterterm in Eq.\eqref{eq:ctnn}, we get the renormalised conjugate field as
\begin{equation}\label{eq:DmToDp}
	\begin{split}
		\piN^{\text{In}}&\equiv -\left[r^\nn \Dp+r^{\nn-1}\ct_\nn \right]\gI \\
&= -e^{-2\pi\w\zeta}\left[r^\nn \Dm+r^{\nn-1}\ct_\nn \right]\gs\\
&= e^{-2\pi\w\zeta}\ps .
	\end{split}
\end{equation}
Here we have used $D_\pm \equiv (1-r^2)\del_r\pm i\w$ as well as the property that $\Dp[e^{-2\pi\w\zeta}\#]=e^{-2\pi\w\zeta}\Dm[\#]$. Using Eq.\eqref{eq:piOut}, we obtain
\begin{equation}\label{eq:piIn}
	\begin{split}
		\piN^{\text{In}}
		&=(1-i\cot\pi\nu)\khI e^{-2\pi\w\zeta}\left(\frac{1-r}{1+r}\right)^{-i\w}\mathscr{Z}_\nn(r,\w)\\
		&\quad\times r^{\nu+\frac{1}{2}(\nn-1)}(1+r)^{-i\w}
		{}_2F_1\left[\frac{1+\nu-\mu-i\w}{2},\frac{1+\nu+\mu-i\w}{2};1+\nu;r^2\right]\ .
	\end{split}
\end{equation}
As in the case of outgoing waves, we see again that the renormalised conjugate field is the regular
part of the incoming waves renormalised with the same factor $\mathscr{Z}_\nn(r,\w)$. 
We can then take $r\to 0$ limit  above and below the branch cut to get
\begin{equation}\label{eq:piIn3}
	\begin{split}
		\lim_{r\to 0}r^{-\nu-\frac{1}{2}(\nn-1)}\piN^{\text{In}}=
		\begin{cases}
			(1-i\cot\pi\nu)\khI\qquad &\text{L boundary}\ ,\\
			e^{-2\pi\w}(1-i\cot\pi\nu)\khI \qquad &\text{R boundary}\  .
		\end{cases}
	\end{split}
\end{equation}
This shows that the counter-term we derived also works for the incoming waves. When $d$ is odd and $\cot\pi\nu=0$, we can remove the dimensional regularisation without any further counterterms.
The analogue of Eq.\eqref{eq:KOutOdd} for the incoming waves is 
\begin{equation}\label{eq:KOutIn}
\begin{split}
		\kI|_{\text{Odd d}} &= (1-i\cot\pi\nu)\khI|_{\text{Odd d}}=(\kO)^\ast|_{\text{Odd d}}
   =e^{-i\nu\pi}\frac{2\pi i}{\Gamma(\nu)^2} \frac{\Gamma\left(\frac{1+\nu-\mu+i\w}{2}\right)\Gamma\left(\frac{1+\nu+\mu+i\w}{2}\right)}{\Gamma\left(\frac{1-\nu+\mu+i\w}{2}\right)\Gamma\left(\frac{1-\nu-\mu+i\w}{2}\right)} \ .
	\end{split}
\end{equation}
All our statements about $\kO$ in odd $d$ apply mutatis mutandis to $\kI$.

When $d$ is even and $\nu$ approaches an integer, there are additional divergences due to $\cot\pi\nu$. We already encountered such divergences and countertermed them away for outgoing waves. We have to check now that the  counterterms in Eq.\eqref{Eq:PhiActCTnu} added to cancel such divergences out of outgoing waves, work also for the incoming waves. To see this, we examine the expansion
\begin{equation}\label{eq:evenDimRegdSIn}
\begin{split}
   (1-i \cot\pi\nu)\khI(\nu)&=
  \Delta_\nn(n,\mu,\w)\left[\frac{2}{\nu-n}\right.\\ &\left.+\psi^{(0)}\left(\frac{1+n-\mu+i\w}{2}\right)+\psi^{(0)}\left(\frac{1+n+\mu+i\w}{2}\right)\right.\\ &\left.+\psi^{(0)}\left(\frac{1-n-\mu+i\w}{2}\right)+\psi^{(0)}\left(\frac{1-n+\mu+i\w}{2}\right)-4\psi^{(0)}(n)+O(\nu-n)\right]\ ,
\end{split}
\end{equation} 
where $\Delta_\nn(n,\mu,\w)$ is given by Eq.\eqref{eq:ctEvenH}. Here, we have used crucially the fact that  $\Delta_\nn$ is a real, even function of $\w$. 

From the above expression, we can see  that the incoming  conjugate field in Eq.\eqref{eq:piIn} is also rendered finite by the same counterterms as before. Crucially, the monodromy factors of $e^{-2\pi\w\zeta}$ work out correctly to cancel the divergences near both the left/right world lines. We get the final renormalised advanced worldline Green function as
\begin{equation}\label{eq:KinEven}
\begin{split}
		\kI|_\text{Even $d$} &=\Delta_\nn(\nu,\mu,\w)\left[\psi^{(0)}\left(\frac{1+\nu-\mu+i\w}{2}\right)+\psi^{(0)}\left(\frac{1+\nu+\mu+i\w}{2}\right)\right.\\ &\left.+\psi^{(0)}\left(\frac{1-\nu-\mu+i\w}{2}\right)+\psi^{(0)}\left(\frac{1-\nu+\mu+i\w}{2}\right)-4\psi^{(0)}(\nu)\right]\ ,
\end{split}
\end{equation}

To conclude, we have demonstrated a set of counterterms which result in finite answers for conjugate fields evaluated over both outgoing as well as incoming waves. The final renormalised conjugate field is given by
\begin{equation}\label{eq:piIn4}
	\begin{split}
		\lim_{r\to 0}r^{-\nu-\frac{1}{2}(\nn-1)}\piN^{\text{In}}=
		\begin{cases}
			\kI\qquad &\text{L boundary}\ ,\\
			e^{-2\pi\w}\kI \qquad &\text{R boundary}\  .
		\end{cases}
	\end{split}
\end{equation}
Since the  most general solution on the dS-SK geometry is a linear combination of outgoing/incoming waves, it follows that our counterterm prescription will yield a finite answer for the cosmological influence phase.

\subsection{Point-like sources and Green functions on dS-SK contour}
In this subsection, we solve for the unique combination of outgoing and incoming waves  corresponding to a point source placed at the centre(s) of left/right static patches in dS-SK geometry. As we will describe subsequently, with some more effort, arbitrary extended sources on the dS-SK background can also be dealt with. 

We  describe the point source problem first to introduce, within a simpler setting, the ingredients needed for the extended sources. As we shall see, in analogy with AdS, we can think of the problem of point sources placed at the centre of the static patch as one involving boundary-to-bulk Green functions. In contrast, the problem of extended sources is that of bulk-to-bulk Green functions, and it is hence fairly more involved. 

As described in the main text, the  solution for the bulk field produced by point-like sources is given by Eq.\eqref{eq:phN_FPbasis}. Using
Eq.\eqref{eq:jPjF}, we then have
\begin{equation}\label{eq:dsSKsolnRL}
	\begin{split}
		\phN=g_R\multj_R-g_L\multj_L\ ,
	\end{split}
\end{equation}
where we have defined 
\begin{equation}\label{eq:gLgRdef}
	\begin{split}
		g_L&\equiv n_\w\Bigl(\gO-e^{2\pi\w(1-\zeta)}\gs\Bigr)\ ,\\
		g_R&\equiv (1+n_\w)\Bigl(\gO-e^{-2\pi\w\zeta}\gs\Bigr)\ .
	\end{split}
\end{equation}
These are the dS analogues of the left/right \emph{boundary-to-bulk} Green functions which tell us how left and right sources affect the solution on the dS-SK geometry. They obey the Kubo-Martin-Schwinger (KMS) relation  $g_R(\zeta)=e^{2\pi\w}g_L(1+\zeta)$ as well as the following boundary conditions on the dS-SK contour: 
\begin{equation}\label{eq:gLRbc}
	\begin{split}
		\lim_{\zeta\to 0}r^{\nu+\frac{\nn-1}{2}}g_L =-1\ ,&\quad \lim_{\zeta\to 0}r^{\nu+\frac{\nn-1}{2}}g_R =0\ ,\\
	\lim_{\zeta\to 1}r^{\nu+\frac{\nn-1}{2}}g_L =0\ ,&\quad \lim_{\zeta\to 1}r^{\nu+\frac{\nn-1}{2}}g_R =1.
	\end{split}
\end{equation}
This result can be derived directly from the boundary condition in Eq.\eqref{eq:PhiBC}. The above
conditions imply that the Green function $g_{L,R}$ are two different smooth interpolations between the homogeneous solution regular at the origin on one side and a Green function with a source singularity on the other side. Thus, $g_R$ is regular near the left boundary whereas $g_L$ is regular near the right boundary.

The Green functions $g_{L,R}$ can be written down explicitly. Substituting Eqs.\eqref{eq:GoutII} and \eqref{eq:GinII} into Eq.\eqref{eq:gLgRdef}, we get the following  expressions:
\begin{equation}\label{eq:gLgRexp2A}
	\begin{split}
		g_L&=
 n_\w r^{-\nu-\frac{1}{2}(\nn-1)}(1+r)^{-i\w}
		\\
		&\times\Bigl\{ 	\left[1-	e^{2\pi\w(1-\zeta)}\left(\frac{1-r}{1+r}\right)^{-i\w}\right]\times {}_2F_1\left[\frac{1-\nu+\mu-i\w}{2},\frac{1-\nu-\mu-i\w}{2};1-\nu;r^2\right]\Bigr.\\
  &\quad -	i\cot\pi\nu\left[\khO+	e^{2\pi\w(1-\zeta)}\left(\frac{1-r}{1+r}\right)^{-i\w}\khI\right]\times  \frac{r^{2\nu}}{2\nu}\ {}_2F_1\left[\frac{1+\nu-
			\mu-i\w}{2},\frac{1+\nu+\mu-i\w}{2};1+\nu;r^2\right] \\
		&\Bigl.\quad -	\left[\khO-	e^{2\pi\w(1-\zeta)}\left(\frac{1-r}{1+r}\right)^{-i\w}\khI\right]\times  \frac{r^{2\nu}}{2\nu}\ {}_2F_1\left[\frac{1+\nu-
			\mu-i\w}{2},\frac{1+\nu+\mu-i\w}{2};1+\nu;r^2\right] \Bigr\}\ ,\\
  	\end{split}
\end{equation}
and
\begin{equation}\label{eq:gLgRexp2B}
	\begin{split} 
		g_R&=(1+n_\w) r^{-\nu-\frac{1}{2}(\nn-1)}(1+r)^{-i\w}
		\\
		&\times\Bigl\{ 	\left[1-	e^{-2\pi\w\zeta}\left(\frac{1-r}{1+r}\right)^{-i\w}\right]\times {}_2F_1\left[\frac{1-\nu+\mu-i\w}{2},\frac{1-\nu-\mu-i\w}{2};1-\nu;r^2\right]\Bigr.\\
    &\quad -	i\cot\pi\nu\left[\khO+	e^{-2\pi\w\zeta}\left(\frac{1-r}{1+r}\right)^{-i\w}\khI\right]\times  \frac{r^{2\nu}}{2\nu}\ {}_2F_1\left[\frac{1+\nu-
			\mu-i\w}{2},\frac{1+\nu+\mu-i\w}{2};1+\nu;r^2\right] \\
		&\Bigl.\quad -	\left[\khO-	e^{-2\pi\w \zeta}\left(\frac{1-r}{1+r}\right)^{-i\w}\khI\right]\times  \frac{r^{2\nu}}{2\nu}\ {}_2F_1\left[\frac{1+\nu-
			\mu-i\w}{2},\frac{1+\nu+\mu-i\w}{2};1+\nu;r^2\right] \Bigr\}\ .
	\end{split}
\end{equation}
These equations describe the Dirac-Deitweiler-Whiting\cite{Dirac:1938nz,Detweiler:2002mi} type decomposition of the left/right Green functions into a singular solution which does not contribute to the radiation reaction, and a regular solution (the terms in the last line of each equation) which contribute to the  finite influence phase.

Having said that, the reader should note that the expressions above are fairly complicated,  with an elaborate branch cut structure that cannot be easily guessed a priori without the dS-SK prescription. These formulae are  more complicated by the fact that we are forced to work with dimensional regularisation for even $d$. We will simplify the expressions for these dS boundary-to-bulk propagators in the next subsection when we describe extended sources. For present purposes, it is, however, sufficient to note the following: despite the complexity of expressions, given that we have  a counterterm procedure that works both for outgoing and incoming waves, we are guaranteed a finite renormalised conjugate field. 

To see this explicitly, we construct the corresponding renormalised conjugate field 
\begin{equation}
	\begin{split}
		\piN(\zeta,\w,\ell)=-\pO(r,\w,\ell)\JFb+e^{2\pi\w(1-\zeta)}\ps(r,\w,\ell)\JPb=\pi_R(\zeta,\w,\ell)\multj_R-\pi_L(\zeta,\w,\ell)\multj_L\ ,
	\end{split}
\end{equation}
with the left/right boundary-to-bulk Green functions for the conjugate field defined by
\begin{align}\label{eq:piLRdef}
	\begin{split}
		\pi_L(\zeta,\w,\ell)&\equiv-\left[r^\nn \Dp+r^{\nn-1}\ct_\nn \right]g_L(\zeta,\w,\ell) = n_\w\Bigl(\pO(r,\w,\ell)-e^{2\pi\w(1-\zeta)}\ps(r,\w,\ell)\Bigr)\ ,\\
		\pi_R(\zeta,\w,\ell)&\equiv-\left[r^\nn \Dp+r^{\nn-1}\ct_\nn \right]g_R(\zeta,\w,\ell) =(1+n_\w)\Bigl(\pO(r,\w,\ell)-e^{-2\pi\w\zeta}\ps(r,\w,\ell)\Bigr)\ .
	\end{split}
\end{align}
The equality here follows from a logic similar to that used in Eq.\eqref{eq:DmToDp}. The explicit forms of $\pO$ and $e^{-2\pi\w\zeta}\ps$ are given in Eqs.\eqref{eq:piOut} and \eqref{eq:piIn} respectively. Substituting them in, we get
\begin{equation}\label{eq:piLRexp}
	\begin{split}
		\pi_L&=
		n_\w \left[(1+i\cot{\pi\nu})\khO-	e^{2\pi\w(1-\zeta)}\left(\frac{1-r}{1+r}\right)^{-i\w}(1-i\cot{\pi\nu})\khI\right] \mathscr{Z}_\nn(\w,r)\\
		&\quad\times r^{\nu+\frac{1}{2}(\nn-1)}(1+r)^{-i\w}{}_2F_1\left[\frac{1+\nu-
			\mu-i\w}{2},\frac{1+\nu+\mu-i\w}{2};1+\nu;r^2\right] \ ,\\
		\pi_R&=(1+n_\w) \left[(1+i\cot{\pi\nu})\khO-	e^{-2\pi\w \zeta}\left(\frac{1-r}{1+r}\right)^{-i\w}(1-i\cot{\pi\nu})\khI\right]\mathscr{Z}_\nn(\w,r)\\
		&\quad\times r^{\nu+\frac{1}{2}(\nn-1)}(1+r)^{-i\w}{}_2F_1\left[\frac{1+\nu-
			\mu-i\w}{2},\frac{1+\nu+\mu-i\w}{2};1+\nu;r^2\right]\ .
	\end{split}
\end{equation}
This picks out the regular part of the solution on dS-SK contour renormalised by $\mathscr{Z}_\nn(\w,r)$, as expected. 

For $\nu\in\mathbb{Z}$, we should subtract the $\cot\pi\nu$ divergences using further counterterms in Eq.\eqref{Eq:PhiActCTnu}: once this is done, we can relax the dimensional regularisation and effectively replace \[(1+i\cot\pi \nu)\khO\to \kO\ ,\quad (1-i\cot\pi \nu)\khI\to \kI\ .\] After this is done, we can  take $r\to 0$ limit on both sides of the dS-SK contour to get
\begin{equation}
	\begin{split}
		\lim_{r\to 0}r^{-\nu-\frac{1}{2}(\nn-1)}\piN=
		\begin{cases}
			K_{LR}\multj_R-K_{LL}\multj_L \qquad &\text{L boundary}\ ,\\
			K_{RR}\multj_R-K_{RL}\multj_L \qquad &\text{R boundary}\ ,
		\end{cases}
	\end{split}
\end{equation}
where we have defined the Schwinger-Keldysh worldline Green functions defined via
\begin{equation}\label{eq:Ksk}
	\begin{split}
		K_{LL}\equiv n_\w \kO-(1+n_\w)\kI\ ,&\quad K_{LR}\equiv(1+n_\w)\Bigl( \kO-\kI\Bigr)\ ,\\
		K_{RL}\equiv n_\w\Bigl( \kO-\kI\Bigr)\ ,&\quad K_{RR}\equiv(1+n_\w) \kO-n_\w\kI\ .
	\end{split}
\end{equation}
These are exactly the expressions for the Schwinger-Keldysh two-point functions of a bosonic system coupled to a thermal bath\cite{Schwinger:1960qe,FEYNMAN1963118,Caldeira:1982iu}.

Now that we have the near origin values of both the generalised free scalar system as well as its renormalised conjugate field, we are ready to compute the influence phase of the observer in the saddle point approximation by evaluating the on-shell action. We want to compute the action  given in Eq.\eqref{Eq:phiAct} along with the counter-term in Eqs.\eqref{Eq:PhiActCT} and \eqref{Eq:PhiActCTnu} over the dS-SK solution we found in Eq.\eqref{eq:dsSKsolnRL}. We begin with the full action
\begin{equation}
	\begin{split}
		S&= -\frac{1}{2}\sum_\spL\int\frac{d\w}{2\pi}\oint  \frac{r^\nn dr}{1-r^2}\Bigl[(\Dp \phN)^* \Dp \phN-\w^2(1-r^2) \phN^* \phN\Bigr.\\
		&\Bigl.\qquad-\frac{1}{4r^2}\Bigl\{(\nn-1)^2-4\nu^2+[4\mu^2-(\nn+1)^2]r^2\Bigr\} \phN^* \phN\Bigr] +S_{ct}\ ,
	\end{split}
\end{equation}
integrate by parts over the bulk terms and then use the equation of motion in Eq.\eqref{Eq:phiRadODE}. This results in an on-shell action written purely in terms of boundary terms:
\begin{equation}
	\begin{split}
		S_{\text{On-Shell}}&=  \frac{1}{2}\sum_\spL\int\frac{d\w}{2\pi} \phN^* \piN|_{\text{Bnd}}= - \frac{1}{2}\sum_\spL\int\frac{d\w}{2\pi} \Bigl\{
		\multj_R^*[K_{RR}\multj_R-K_{RL}\multj_L]-\multj_L^*[K_{LR}\multj_R-K_{LL}\multj_L]\Bigr\}\ ,
	\end{split}
\end{equation}
where $\piN$ is the renormalised conjugate field defined in Eq.\eqref{eq:piDef}. Here we have used the fact that the integrand in the first step can be written as a product
\begin{equation}
	\begin{split}
	[r^{\nu+\frac{1}{2}(\nn-1)}\phN]^*	r^{-\nu-\frac{1}{2}(\nn-1)}\piN\ ,
	\end{split}
\end{equation}
and each factor in this product has a finite limit as we remove the regulator at the boundary (i.e. take $r_c\to 0$ limit). The dS-SK contour integral $\oint$ runs clockwise from the right static patch to the left static patch, thus resulting in the sign of the final expression above.

We can further simplify the above expression using the reality properties of the multipole sources as well as $1+n_\w+n_{-\w}=0$. The cosmological influence phase of the point-like dS observer can then be written in the form given in Eq.\eqref{eq:SCIPpt}.


\subsection{Extended sources on dS-SK contour I : bulk-to-bulk propagator}
 In this section, we will describe the problem of a finite size observer within dS spacetime. One motivation for such an exercise is to give a more physical version of the regularisation, counter-terms and renormalisation described in the previous sub-sections. We will see that indeed a finite size observer has a renormalised cosmological influence phase, which, as its size is reduced, approaches the result for a point-like observer. Apart from this formal motivation, we are also interested in checking whether the conjectured dS-SK saddle point correctly reproduces the finite size physics in dS. As we shall see, this is also a way to naturally generalise our construction to a non-co-moving observer with a peculiar velocity as well as  to describe observers made of multiple worldlines (or equivalently the case of a string or a membrane in dS).
 
 The main physics in all the above cases is that of relative time-delays: for an extended  source, its effective radiative multipole moments have to be computed by adding up source strengths at various points with different time-delays. This is necessary because the emitted wave takes a finite  amount of time to cross an extended source, and this wave-crossing time has to be accounted for when adding up emissions from two farther ends of the source. For spherical sources in flat space, this translates to modulating the source with an appropriate Bessel J function in frequency domain. We will see below that an analogous statement in dS emerges naturally out of dS-SK saddle-point geometry.
 
 Let us begin by describing our setup. Consider an extended source of the generalised/designer scalar field in dS spacetime. This means modifying the radial ODE in  Eq.\eqref{Eq:phiRadODE} 
 by a source term of the form
 \begin{equation}\label{Eq:phiRadODEs}
 	\begin{split}
 		&\frac{1}{r^\nn}\Dp [r^\nn \Dp \phN] +\w^2\phN\\
 		&\qquad+\frac{1-r^2}{4r^2}\Bigl\{(\nn-1)^2-4\nu^2+[4\mu^2-(\nn+1)^2]r^2\Bigr\}\phN+(1-r^2)\varrho_\nn(\zeta,\w,\spL)=0\ .
 	\end{split}
 \end{equation}  
In the context of dS-SK contour, we will let $\varrho_\nn$ be a general function  over the saddle-point geometry, allowing it to even take completely different values in the two copies of the static patch (i.e., as a function of complex $r$, it is allowed to have a branch-cut along the static patch). The $(\w,\spL)$ arguments of $\varrho_\nn$ imply that we also allow the most general time/angle dependence.

The solution for the above ODE can then be written in terms of an appropriate dS-SK contour-ordered, bulk-to-bulk Green function:
\begin{equation}\label{eq:dSSKBlkF}
	\begin{split}
		\phN(\zeta,\w,\spL)&=\oint  r_0^\nn dr_0\ \mathbb{G}(\zeta|\zeta_0,\w,\spL)\varrho_{_\nn}(\zeta_0,\w,\spL)\ .
	\end{split}
\end{equation}
Here $\oint$ refers to the integral over clockwise  dS-SK contour and $\mathbb{G}$ is the radial Green function satisfying the appropriate boundary conditions (which we will detail below). 

According to our proposal in this note, the influence phase of the extended source can be computed by solving the above ODE everywhere on dS-SK and then substituting the solution into the 
action corresponding to the above ODE, viz., by evaluating 
\begin{equation}\label{Eq:phiActs}
	\begin{split}
		S&= -\frac{1}{2}\sum_\spL\int\frac{d\w}{2\pi}\oint  \frac{r^\nn dr}{1-r^2}\Bigl[(\Dp \phN)^* \Dp \phN-\w^2\phN^* \phN\Bigr.\\
		&\Bigl.\qquad-\frac{1-r^2}{4r^2}\Bigl\{(\nn-1)^2-4\nu^2+[4\mu^2-(\nn+1)^2]r^2\Bigr\} \phN^* \phN\Bigr]\\
		&\qquad + \sum_\spL\int\frac{d\w}{2\pi}\oint  r^\nn dr\ \phN^*\varrho_{_\nn} +S_{ct}[\varrho_{_\nn}]\ 
	\end{split}
\end{equation}
on the Green function solution above. The last line in the action above gives the source term and the counter-term parts of the action.\footnote{The reader should note that the counterterms used here for extended sources need not (and, indeed will not) match with the counterterms used for point sources in the previous subsections. } For a truly extended source, counter-terms are not necessary for finiteness and their job is to provide the finite renormalisation of the conservative part of the action. 

Using the radial ODE above,  on-shell action can be reduced to the following simple form  
\begin{equation}
\label{eq:SOnShellBlkSrc}
	\begin{split}
		S|_{\textbf{On-shell}}&= \frac{1}{2} \sum_\spL\int\frac{d\w}{2\pi}\oint  r^\nn dr\ \varrho_{_\nn}^*\phN|_{\textbf{On-shell}}+S_{ct}[\varrho_{_\nn}]\\
		&= \frac{1}{2} \sum_\spL\int\frac{d\w}{2\pi}\oint  r^\nn dr\  \oint  r_0^\nn dr_0\ [\varrho_{_\nn}(\zeta,\w,\spL)]^*\mathbb{G}(\zeta|\zeta_0,\w,\spL)\varrho_{_\nn}(\zeta_0,\w,\spL) +S_{ct}[\varrho_{_\nn}]\ .
	\end{split}
\end{equation}
Thus, once we solve for the bulk-to-bulk Green function $\mathbb{G}$, we can substitute  it into the above expression to obtain the dS-SK saddle point answer for  cosmological influence phase $\SCIP$. While it is not immediately evident, we will demonstrate in the next subsection that \emph{the dissipative part of the influence phase for the extended sources computed from the expression above, when written in terms of appropriate multipole moments, takes a form identical to that for a point source derived before.} In addition to this radiation reaction, for extended sources, we also expect conservative interactions between its different internal parts.

Let us now derive an explicit expression for the bulk-to-bulk Green function $\mathbb{G}$.
The construction here is analogous to the one in vacuum AdS\cite{Witten:1998qj}, as well as the contour-ordered bulk-to-bulk Green function in the SK contour corresponding to planar AdS black holes\cite{Loganayagam:2022zmq,OpenEFT}. We will demand that this Green function be regular at the edges of dS-SK contour, viz., we require that
\begin{equation}\label{eq:Gbc}
	\begin{split}
		\lim_{\zeta\to 0}r^{\nu+\frac{\nn-1}{2}}\mathbb{G} =\lim_{\zeta\to 1}r^{\nu+\frac{\nn-1}{2}}\mathbb{G} =0.
	\end{split}
\end{equation}
Further, to be a Green function, it should obey the ODE
\begin{equation}
	\begin{split}
		&\frac{1}{r^\nn}\Dp [r^\nn \Dp \mathbb{G}] +\w^2\mathbb{G}\\
		&\qquad+\frac{1-r^2}{4r^2}\Bigl\{(\nn-1)^2-4\nu^2+[4\mu^2-(\nn+1)^2]r^2\Bigr\}\mathbb{G}+\frac{1}{r^\nn}(1-r^2)\delta_c(r-r_0)=0\ .
	\end{split}
\end{equation}
Here $\delta_c(r-r_0)$ is the contour-ordered delta function on the dS-SK contour. The above ODE implies that $\mathbb{G}$ is  a solution of the homogeneous radial ODE for  $\zeta\neq\zeta_0$ with a unit discontinuity in the conjugate field at  $\zeta=\zeta_0$.
We have already solved the homogeneous radial ODE for point sources to construct the left and right boundary-to-bulk Green functions in Eq.\eqref{eq:gLgRdef}. These are solutions characterised by the boundary conditions specified in Eq.\eqref{eq:gLRbc}. 

Looking at Eq.\eqref{eq:gLRbc}, we conclude that, we should take  $\mathbb{G}\propto g_R$ near the left boundary and $\mathbb{G}\propto g_L$ near the right boundary since these are the solutions that satisfy the necessary regularity conditions in Eq.\eqref{eq:Gbc}. Demanding continuity, we surmise that
\begin{equation}\label{eq:BlkBlkdS}
	\begin{split}
	 \mathbb{G}(\zeta|\zeta_0,\w,\spL)&=\frac{1}{W_{LR}(\zeta_0,\w,\spL)} g_R(\zeta_\succ,\w,\spL)g_L(\zeta_\prec,\w,\spL)\\
  &\equiv\frac{1}{W_{LR}(\zeta_0,\w,\spL)}\begin{cases}
	 	 g_R(\zeta,\w,\spL)g_L(\zeta_0,\w,\spL) &\quad \text{if}\; \zeta\succ\zeta_0\\
         g_L(\zeta,\w,\spL)g_R(\zeta_0,\w,\spL) &\quad \text{if}\; \zeta\prec\zeta_0
	 \end{cases}	\ .
	\end{split}
\end{equation}
Here the symbols $\succ$ and $\prec$ denote  comparison using the radial contour ordering of dS-SK contour. The unit discontinuity condition on the conjugate field fixes the function $W_{RL}$ to be the Wronskian between  right and left boundary-to-bulk Green functions, viz.,
\begin{equation}\label{eq:WronskRL}
	\begin{split}
		W_{RL}(\zeta,\w,\spL) &\equiv g_L \pi_R-g_R \pi_L=(1+n_\w)e^{-2\pi\w\zeta}\Bigl(\gO \ps-\gs\pO\Bigr)\\
  &=(1+n_\w)e^{-2\pi\w\zeta}\left[	(1-i\cot\pi\nu)\khI-(1+i\cot\pi\nu)\khO\right]\ .
	\end{split}
\end{equation}
Here, the equality in the first line follows from Eqs.\eqref{eq:gLgRdef} and \eqref{eq:piLRdef}. The last equality follows by substituting the expressions for $\gO$
and $\pO$ from Eqs.\eqref{eq:GoutII} and \eqref{eq:piOut}, and then invoking the following hypergeometric Wronskian identity
\begin{equation}
	\begin{split}\label{eq:ZnnAlt}
 \mathscr{Z}_\nn(r,\w)& r^{\nu+\frac{1}{2}(\nn-1)}(1+r)^{-i\w} {}_2F_1\left[\frac{1+\nu-
 	\mu-i\w}{2},\frac{1+\nu+\mu-i\w}{2};1+\nu;r^2\right]\\
 &= \left(\frac{1-r}{1+r}\right)^{i\w}\Biggl\{ r^{-\nu-\frac{1}{2}(\nn-1)}(1+r)^{-i\w}
{}_2F_1\left[\frac{1-\nu+\mu-i\w}{2},\frac{1-\nu-\mu-i\w}{2};1-\nu;r^2\right]\Biggr\}^{-1}\ .
	\end{split}
\end{equation}
This identity expresses a combination of the derivatives of hypergeometric functions in terms of the hypergeometric functions, and such an identity  can be derived from a Wronskian like argument associated with the corresponding radial ODE.

The reader should note an important subtlety in the statement above: the Wronskian  here is \emph{not} a constant function along radial direction, but rather varies as we traverse the dS-SK contour. A similar subtlety was already noted in the AdS context by \cite{Loganayagam:2022zmq}. As we shall eventually see, the extra $e^{-2\pi\w\zeta}$ factor is here for a good physical reason: it  ensures that multipole moments that enter into cosmological influence phase are computed using source distributions in standard time-slices, instead of source distributions along Eddington-Finkelstein null time-slices.

To proceed further, we should now substitute the explicit forms of dS-SK boundary-to-bulk propagators given in Eqs.\eqref{eq:gLgRexp2A} and \eqref{eq:gLgRexp2B} into the expression for the bulk-to-bulk propagator in Eq.\eqref{eq:BlkBlkdS}, and then perform the dS-SK contour integral in Eq.\eqref{eq:SOnShellBlkSrc}. To this end, we first regroup the  expressions for $g_L$ and $g_R$ into somewhat more tractable expressions with clear branch cut structures. For what follows, we will find it convenient to separate out the solutions into a singular (non-normalisable) part $\Xi_{nn}$ vs a regular (normalisable) part $\Xi_{n}$, \emph{using the renormalised world line Green functions instead of the bare ones from the start}. The adjectives singular/regular refer here to their behaviour near the  worldline (i.e., near $r=0$). To this end, let us begin by defining two functions $\Xi_{nn},\Xi_n$   implicitly via
\begin{equation}\label{eq:GXiformula}
\begin{split}
\left(\frac{1-r}{1+r}\right)^{-\frac{i\w}{2}}\gO(r,\w,\spL) &\equiv   \Xi_{nn}(r,\w,\spL)-\kO\ \Xi_n(r,\w,\spL)\ ,\\
\left(\frac{1-r}{1+r}\right)^{\frac{i\w}{2}}\gs(r,\w,\spL) &\equiv \Xi_{nn}(r,\w,\spL)-\kI\ \Xi_n(r,\w,\spL) \ ,
\end{split}
\end{equation}
where $\kO$ and $\kI$ are the final renormalised world line Green functions. The above equality  should be thought of as defining  the functions $\Xi_n(r,\w,\spL)$ and $\Xi_{nn}(r,\w,\spL)$ as analytic functions on the open static patch $0< r<1$, viz., in the equations above, we align all the potential branch cuts  away from the unit disc in complex radius plane. The above equations can be inverted to give a direct definition of these functions
\begin{equation}
\begin{split}
(\kI-\kO)\ \Xi_n &\equiv  \left(\frac{1-r}{1+r}\right)^{-\frac{i\w}{2}}\gO- \left(\frac{1-r}{1+r}\right)^{\frac{i\w}{2}}\gs \ ,\\
(\kI-\kO)\ \Xi_{nn} &\equiv  \left(\frac{1-r}{1+r}\right)^{-\frac{i\w}{2}}\kI\gO- \left(\frac{1-r}{1+r}\right)^{\frac{i\w}{2}}\kO\gs\ .
\end{split}
\end{equation}
Since $\kI(\w,
\ell)=\kO(-\w,\ell)$ and $\gs(\w,\ell)=\gO(-\w,\ell)$, the above expressions imply that both $\Xi_n$ and $\Xi_{nn}$ are even functions of $\w$. Explicit expressions can be written down for these two functions using Eq.\eqref{eq:GoutII}. We have
\begin{equation}\label{eq:XinExp}
\begin{split}
 \Xi_n &\equiv \frac{1}{2\nu}r^{\nu-\frac{1}{2}(\nn-1)}(1-r^2)^{-\frac{i\w}{2}}
		 \frac{(1+i\cot\nu \pi)\khO-	(1-i\cot\nu \pi)\khI}{\kO-\kI}\\
  &\qquad\times {}_2F_1\left[\frac{1+\nu-
			\mu-i\w}{2},\frac{1+\nu+\mu-i\w}{2};1+\nu;r^2\right] \ ,\\
\end{split}
\end{equation}
for the normalisable/regular mode and 
\begin{equation}\label{eq:XinnExp}
	\begin{split}   
 \Xi_{nn} &\equiv  r^{-\nu-\frac{1}{2}(\nn-1)}(1-r^2)^{-\frac{i\w}{2}}\Bigl\{  {}_2F_1\left[\frac{1-\nu+\mu-i\w}{2},\frac{1-\nu-\mu-i\w}{2};1-\nu;r^2\right]\Bigr.\\
& -\ \frac{\kI(1+i\cot\nu \pi)\khO-	\kO(1-i\cot\nu \pi)\khI}{\kI-\kO}\\\
&\Bigl.\qquad\times\frac{r^{2\nu}}{2\nu}{}_2F_1\left[\frac{1+\nu-
			\mu-i\w}{2},\frac{1+\nu+\mu-i\w}{2};1+\nu;r^2\right] \Bigr\}\ 
\end{split}
\end{equation}
for the non-normalisable/singular mode. One advantage of working with such renormalised functions is that we can safely remove the dimensional regularisation in the above expressions resulting in a finite limit. When $d$ is odd and $\nu\equiv \ell+\frac{d}{2}-1\in \mathbb{Z}+\frac{1}{2}$, we can simply set $\cot \nu\pi =0$ and take $\khO\to \kO$ and $\khI\to \kI$. All the $K$'s then drop out of the above expression, and $\Xi_{nn}$ and $\Xi_n$ become proportional to single hypergeometric functions. 

When $d$ is even and $\nu\to n\in \mathbb{Z}$, we can use 
Eqs.\eqref{eq:evenDimRegdSOut} and \eqref{eq:evenDimRegdSIn} to write 
\begin{equation}\label{eq:KhExpnsEven}
	\begin{split}   
(1+i\cot\nu \pi)\khO&=\frac{2}{\nu-n}\Delta_\nn(n,\mu,\w)+\kO + O(\nu-n)\ ,\\ 
(1-i\cot\nu \pi)\kI&=\frac{2}{\nu-n}\Delta_\nn(n,\mu,\w)+\kI + O(\nu-n)\ .  
\end{split}
\end{equation}
Using these expansions, the $K$s cancel out again  and we are left with the following limits:
\begin{equation}\label{eq:EvenXi}
\begin{split}
 \Xi_n|_\text{Even d} &\equiv \lim_{\nu\to n}\frac{1}{2\nu}r^{\nu-\frac{1}{2}(\nn-1)}(1-r^2)^{-\frac{i\w}{2}} {}_2F_1\left[\frac{1+\nu-
			\mu-i\w}{2},\frac{1+\nu+\mu-i\w}{2};1+\nu;r^2\right] \ ,\\
 \Xi_{nn}|_\text{Even d} &\equiv  \lim_{\nu\to n} r^{-\nu-\frac{1}{2}(\nn-1)}(1-r^2)^{-\frac{i\w}{2}}\Biggl\{  {}_2F_1\left[\frac{1-\nu+\mu-i\w}{2},\frac{1-\nu-\mu-i\w}{2};1-\nu;r^2\right]\Biggr.\\
&\Biggl.\qquad -\ \frac{r^{2\nu}}{\nu(\nu-n)}\Delta_\nn(n,\mu,\w)\ {}_2F_1\left[\frac{1+\nu-
			\mu-i\w}{2},\frac{1+\nu+\mu-i\w}{2};1+\nu;r^2\right] \Biggr\}\  .  
\end{split}
\end{equation}
One can explicitly check that these limits exist and result in finite expressions for both regular/singular modes when $d$ is even. To summarise, Eq.\eqref{eq:GXiformula} decomposes the outgoing/incoming Green functions into renormalised pieces in any $d$. 

We will now rewrite the full bulk-to-bulk propagator in Eq.\eqref{eq:BlkBlkdS} in terms of these renormalised modes. We begin by rewriting the boundary-to-bulk propagators: using Eq.\eqref{eq:gLgRdef}, we obtain
\begin{equation}\label{eq:gLgRXiexp}
	\begin{split}   
g_L&= n_\w\left(\frac{1-r}{1+r}\right)^{\frac{i\w}{2}}\Biggl\{\left[1-e^{2\pi\w(1-\zeta)} \left(\frac{1-r}{1+r}\right)^{-i\w}\right]\Xi_{nn} \Biggr.\\
& \qquad\Biggl.\qquad -\left[\kO-\kI e^{2\pi\w(1-\zeta)} \left(\frac{1-r}{1+r}\right)^{-i\w}\right]\Xi_{n}\Biggr\}\ ,\\
g_R&= (1+n_\w)\left(\frac{1-r}{1+r}\right)^{\frac{i\w}{2}}\Biggl\{\left[1-e^{-2\pi\w\zeta} \left(\frac{1-r}{1+r}\right)^{-i\w}\right]\Xi_{nn} \Biggr.\\
&\qquad \Biggl.\qquad -\left[\kO-\kI e^{-2\pi\w\zeta} \left(\frac{1-r}{1+r}\right)^{-i\w}\right]\Xi_{n}\Biggr\}\ .
\end{split}
\end{equation}
Substituting them back into Eq.\eqref{eq:BlkBlkdS}, we get an explicit expression for the bulk-to-bulk propagator of the form
\begin{equation}\label{eq:BlkBlkdSFull}
	\begin{split}
	 \mathbb{G}(\zeta|\zeta_0,\w,\spL)&=\frac{1}{W_{LR}(\zeta_0,\w,\spL)} g_R(\zeta_\succ,\w,\spL)g_L(\zeta_\prec,\w,\spL)\\
  &=\frac{n_\w e^{2\pi\w\zeta_0}}{\kI-\kO}\left(\frac{1-r}{1+r}\right)^{\frac{i\w}{2}}\left(\frac{1-r_0}{1+r_0}\right)^{\frac{i\w}{2}}\\
  &\times \Biggl\{\left[1-e^{-2\pi\w\zeta_\succ} \left(\frac{1-r_\succ}{1+r_\succ}\right)^{-i\w}\right]\Xi_{nn}(r_\succ) \Biggr.\\
&\qquad \Biggl.\qquad -\left[ \kO-\kI e^{-2\pi\w\zeta_\succ} \left(\frac{1-r_>}{1+r_\succ}\right)^{-i\w}\right]\Xi_{n}(r_\succ)\Biggr\}\\
&\times\Biggl\{\left[1-e^{2\pi\w(1-\zeta_\prec)} \left(\frac{1-r_\prec}{1+r_\prec}\right)^{-i\w}\right]\Xi_{nn}(r_\prec) \Biggr.\\
& \qquad\Biggl.\qquad -\left[\kO-\kI e^{2\pi\w(1-\zeta_\prec)} \left(\frac{1-r_\prec}{1+r_\prec}\right)^{-i\w}\right]\Xi_{n}(r_\prec)\Biggr\}\ .
	\end{split}
\end{equation}
Here, since all quantities are already renormalised, we have removed the dimensional regularisation\footnote{For odd $d$, we set $\cot\nu \pi=0$ and remove the hats on $K$s. For  even $d$,  we use Eq.\eqref{eq:KhExpnsEven}.} in the Wronskian given in Eq.\eqref{eq:WronskRL}. To conclude, given an arbitrary extended source on the dS-SK geometry, the above bulk-to-bulk propagator, we can get the bulk field by substituting the above bulk-to-bulk Green function into Eq.\eqref{eq:dSSKBlkF}. Further, we can also compute the on-shell action Eq.\eqref{eq:SOnShellBlkSrc}, which, according to our prescription, should yield the influence phase of that extended source.

\subsection{Extended sources on dS-SK contour II: Radiative multipoles }
In this subsection, we would like to evaluate both the field and the influence of an extended source. We will find it convenient to discretise the source into a set of spherical shells around the centre of the right/left  static patches. Let $\zeta=1+\zeta_i$ characterise the radial position of the $i^{th}$ spherical shell in the right patch, the same radial position in the left patch is then characterised by $\zeta=\zeta_i$. We will let the $i$ vary over $1$ to $N_s$, where $N_s$
is the number of shells in each copy of the static patch. We will take the strength of the scalar source on these spherical shells to be
\begin{equation}
	\begin{split}
r^\nn\varrho_{_\nn}(\zeta,\w,\spL)=\sum_i\sigma^R_i(\w,\spL)\ \delta_c(\zeta|1+\zeta_i)-\sum_i\sigma^L_i(\w,\spL)\ \delta_c(\zeta|\zeta_i)\ .
	\end{split}
\end{equation}
Here, as before, we work in frequency domain/orthonormal spherical harmonic basis, and allow arbitrary time/angle dependence. Any arbitrary source distribution confined within the open static patch can be approximated to any desired accuracy as being built from such spherical shell sources. As we shall see, such a discrete model regularises the divergences associated with the self-interactions.

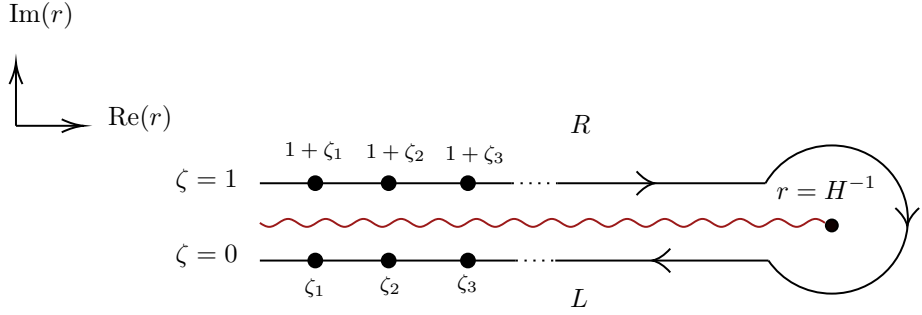
\begin{figure}[H]\label{fig:3regionsSKApp}
		\centering

\tikzset{every picture/.style={line width=0.75pt}} 

\begin{tikzpicture}[x=0.75pt,y=0.75pt,yscale=-1,xscale=1]

\draw  [draw opacity=0] (440,114) .. controls (446.57,102.6) and (459.1,94.91) .. (473.47,94.91) .. controls (494.66,94.91) and (511.83,111.65) .. (511.83,132.31) .. controls (511.83,152.96) and (494.66,169.71) .. (473.47,169.71) .. controls (460.13,169.71) and (448.39,163.08) .. (441.51,153.02) -- (473.47,132.31) -- cycle ; \draw   (440,114) .. controls (446.57,102.6) and (459.1,94.91) .. (473.47,94.91) .. controls (494.66,94.91) and (511.83,111.65) .. (511.83,132.31) .. controls (511.83,152.96) and (494.66,169.71) .. (473.47,169.71) .. controls (460.13,169.71) and (448.39,163.08) .. (441.51,153.02) ;  
\draw  [fill={rgb, 255:red, 15; green, 1; blue, 1 }  ,fill opacity=1 ] (470.47,135.31) .. controls (470.47,133.65) and (471.81,132.31) .. (473.47,132.31) .. controls (475.12,132.31) and (476.47,133.65) .. (476.47,135.31) .. controls (476.47,136.96) and (475.12,138.31) .. (473.47,138.31) .. controls (471.81,138.31) and (470.47,136.96) .. (470.47,135.31) -- cycle ;
\draw   (392,159) .. controls (389,155.67) and (386,153.67) .. (383,153) .. controls (386,152.33) and (389,150.33) .. (392,147) ;
\draw   (375,108) .. controls (378,111.33) and (381,113.33) .. (384,114) .. controls (381,114.67) and (378,116.67) .. (375,120) ;
\draw   (517.5,126.5) .. controls (514.17,129.5) and (512.17,132.5) .. (511.5,135.5) .. controls (510.83,132.5) and (508.83,129.5) .. (505.5,126.5) ;
\draw  [fill={rgb, 255:red, 0; green, 0; blue, 0 }  ,fill opacity=1 ] (209.5,113.96) .. controls (209.5,112.05) and (211.05,110.5) .. (212.96,110.5) .. controls (214.87,110.5) and (216.42,112.05) .. (216.42,113.96) .. controls (216.42,115.87) and (214.87,117.42) .. (212.96,117.42) .. controls (211.05,117.42) and (209.5,115.87) .. (209.5,113.96) -- cycle ;
\draw  [fill={rgb, 255:red, 0; green, 0; blue, 0 }  ,fill opacity=1 ] (209.5,153) .. controls (209.5,151.09) and (211.05,149.54) .. (212.96,149.54) .. controls (214.87,149.54) and (216.42,151.09) .. (216.42,153) .. controls (216.42,154.91) and (214.87,156.46) .. (212.96,156.46) .. controls (211.05,156.46) and (209.5,154.91) .. (209.5,153) -- cycle ;
\draw  [fill={rgb, 255:red, 0; green, 0; blue, 0 }  ,fill opacity=1 ] (246.5,113.96) .. controls (246.5,112.05) and (248.05,110.5) .. (249.96,110.5) .. controls (251.87,110.5) and (253.42,112.05) .. (253.42,113.96) .. controls (253.42,115.87) and (251.87,117.42) .. (249.96,117.42) .. controls (248.05,117.42) and (246.5,115.87) .. (246.5,113.96) -- cycle ;
\draw  [fill={rgb, 255:red, 0; green, 0; blue, 0 }  ,fill opacity=1 ] (286.5,113.96) .. controls (286.5,112.05) and (288.05,110.5) .. (289.96,110.5) .. controls (291.87,110.5) and (293.42,112.05) .. (293.42,113.96) .. controls (293.42,115.87) and (291.87,117.42) .. (289.96,117.42) .. controls (288.05,117.42) and (286.5,115.87) .. (286.5,113.96) -- cycle ;
\draw    (185,114) -- (312,114) ;
\draw    (335,114) -- (440,114) ;
\draw  [dash pattern={on 0.84pt off 2.51pt}]  (312,114) -- (335,114) ;
\draw    (185,153.02) -- (314,153.02) ;
\draw    (334,153.02) -- (441.51,153.02) ;
\draw  [dash pattern={on 0.84pt off 2.51pt}]  (314,153.02) -- (334,153.02) ;
\draw  [fill={rgb, 255:red, 0; green, 0; blue, 0 }  ,fill opacity=1 ] (246.5,153) .. controls (246.5,151.09) and (248.05,149.54) .. (249.96,149.54) .. controls (251.87,149.54) and (253.42,151.09) .. (253.42,153) .. controls (253.42,154.91) and (251.87,156.46) .. (249.96,156.46) .. controls (248.05,156.46) and (246.5,154.91) .. (246.5,153) -- cycle ;
\draw  [fill={rgb, 255:red, 0; green, 0; blue, 0 }  ,fill opacity=1 ] (286.5,153) .. controls (286.5,151.09) and (288.05,149.54) .. (289.96,149.54) .. controls (291.87,149.54) and (293.42,151.09) .. (293.42,153) .. controls (293.42,154.91) and (291.87,156.46) .. (289.96,156.46) .. controls (288.05,156.46) and (286.5,154.91) .. (286.5,153) -- cycle ;
\draw    (62,85) -- (62,55) ;
\draw [shift={(62,53)}, rotate = 90] [color={rgb, 255:red, 0; green, 0; blue, 0 }  ][line width=0.75]    (10.93,-3.29) .. controls (6.95,-1.4) and (3.31,-0.3) .. (0,0) .. controls (3.31,0.3) and (6.95,1.4) .. (10.93,3.29)   ;
\draw    (62,85) -- (94,85) ;
\draw [shift={(96,85)}, rotate = 180] [color={rgb, 255:red, 0; green, 0; blue, 0 }  ][line width=0.75]    (10.93,-3.29) .. controls (6.95,-1.4) and (3.31,-0.3) .. (0,0) .. controls (3.31,0.3) and (6.95,1.4) .. (10.93,3.29)   ;
\draw  [color={rgb, 255:red, 155; green, 25; blue, 25 }  ,draw opacity=1 ] (185,134) .. controls (186.55,135.02) and (188.03,136) .. (189.75,136) .. controls (191.47,136) and (192.95,135.02) .. (194.5,134) .. controls (196.05,132.98) and (197.53,132) .. (199.25,132) .. controls (200.97,132) and (202.45,132.98) .. (204,134) .. controls (205.55,135.02) and (207.03,136) .. (208.75,136) .. controls (210.47,136) and (211.95,135.02) .. (213.5,134) .. controls (215.05,132.98) and (216.53,132) .. (218.25,132) .. controls (219.97,132) and (221.45,132.98) .. (223,134) .. controls (224.55,135.02) and (226.03,136) .. (227.75,136) .. controls (229.47,136) and (230.95,135.02) .. (232.5,134) .. controls (234.05,132.98) and (235.53,132) .. (237.25,132) .. controls (238.97,132) and (240.45,132.98) .. (242,134) .. controls (243.55,135.02) and (245.03,136) .. (246.75,136) .. controls (248.47,136) and (249.95,135.02) .. (251.5,134) .. controls (253.05,132.98) and (254.53,132) .. (256.25,132) .. controls (257.97,132) and (259.45,132.98) .. (261,134) .. controls (262.55,135.02) and (264.03,136) .. (265.75,136) .. controls (267.47,136) and (268.95,135.02) .. (270.5,134) .. controls (272.05,132.98) and (273.53,132) .. (275.25,132) .. controls (276.97,132) and (278.45,132.98) .. (280,134) .. controls (281.55,135.02) and (283.03,136) .. (284.75,136) .. controls (286.47,136) and (287.95,135.02) .. (289.5,134) .. controls (291.05,132.98) and (292.53,132) .. (294.25,132) .. controls (295.97,132) and (297.45,132.98) .. (299,134) .. controls (300.55,135.02) and (302.03,136) .. (303.75,136) .. controls (305.47,136) and (306.95,135.02) .. (308.5,134) .. controls (310.05,132.98) and (311.53,132) .. (313.25,132) .. controls (314.97,132) and (316.45,132.98) .. (318,134) .. controls (319.55,135.02) and (321.03,136) .. (322.75,136) .. controls (324.47,136) and (325.95,135.02) .. (327.5,134) .. controls (329.05,132.98) and (330.53,132) .. (332.25,132) .. controls (333.97,132) and (335.45,132.98) .. (337,134) .. controls (338.55,135.02) and (340.03,136) .. (341.75,136) .. controls (343.47,136) and (344.95,135.02) .. (346.5,134) .. controls (348.05,132.98) and (349.53,132) .. (351.25,132) .. controls (352.97,132) and (354.45,132.98) .. (356,134) .. controls (357.55,135.02) and (359.03,136) .. (360.75,136) .. controls (362.47,136) and (363.95,135.02) .. (365.5,134) .. controls (367.05,132.98) and (368.53,132) .. (370.25,132) .. controls (371.97,132) and (373.45,132.98) .. (375,134) .. controls (376.55,135.02) and (378.03,136) .. (379.75,136) .. controls (381.47,136) and (382.95,135.02) .. (384.5,134) .. controls (386.05,132.98) and (387.53,132) .. (389.25,132) .. controls (390.97,132) and (392.45,132.98) .. (394,134) .. controls (395.55,135.02) and (397.03,136) .. (398.75,136) .. controls (400.47,136) and (401.95,135.02) .. (403.5,134) .. controls (405.05,132.98) and (406.53,132) .. (408.25,132) .. controls (409.97,132) and (411.45,132.98) .. (413,134) .. controls (414.55,135.02) and (416.03,136) .. (417.75,136) .. controls (419.47,136) and (420.95,135.02) .. (422.5,134) .. controls (424.05,132.98) and (425.53,132) .. (427.25,132) .. controls (428.97,132) and (430.45,132.98) .. (432,134) .. controls (433.55,135.02) and (435.03,136) .. (436.75,136) .. controls (438.47,136) and (439.95,135.02) .. (441.5,134) .. controls (443.05,132.98) and (444.53,132) .. (446.25,132) .. controls (447.97,132) and (449.45,132.98) .. (451,134) .. controls (452.55,135.02) and (454.03,136) .. (455.75,136) .. controls (457.47,136) and (458.95,135.02) .. (460.5,134) .. controls (462.05,132.98) and (463.53,132) .. (465.25,132) .. controls (466.97,132) and (468.45,132.98) .. (470,134) ;

\draw (141,143) node [anchor=north west][inner sep=0.75pt]   [align=left] {$\displaystyle \zeta =0$};
\draw (141,105) node [anchor=north west][inner sep=0.75pt]   [align=left] {$\displaystyle \zeta =1$};
\draw (196,91.4) node [anchor=north west][inner sep=0.75pt]  [font=\footnotesize]  {$1+\zeta _{1}$};
\draw (206,160.4) node [anchor=north west][inner sep=0.75pt]  [font=\footnotesize]  {$\zeta _{1}$};
\draw (277,93.4) node [anchor=north west][inner sep=0.75pt]  [font=\footnotesize]  {$1+\zeta _{3}$};
\draw (237,92.4) node [anchor=north west][inner sep=0.75pt]  [font=\footnotesize]  {$1+\zeta _{2}$};
\draw (244,159.4) node [anchor=north west][inner sep=0.75pt]  [font=\footnotesize]  {$\zeta _{2}$};
\draw (283,158.4) node [anchor=north west][inner sep=0.75pt]  [font=\footnotesize]  {$\zeta _{3}$};
\draw (107,73) node [anchor=north west][inner sep=0.75pt]   [align=left] {Re($\displaystyle r$)};
\draw (57,21) node [anchor=north west][inner sep=0.75pt]   [align=left] {Im($\displaystyle r$)};
\draw (444,109) node [anchor=north west][inner sep=0.75pt]   [align=left] {$\displaystyle r=H^{-1}$};
\draw (340,78) node [anchor=north west][inner sep=0.75pt]   [align=left] {$\displaystyle R$};
\draw (340,166) node [anchor=north west][inner sep=0.75pt]   [align=left] {$\displaystyle L$};

\end{tikzpicture}
\caption{Spherical shell sources centred around the right/left static patches shown in the complex r plane. Their positions on the $L$ contour are related to their position on the $R$ contour by the branch cut discontinuity in $\zeta$. }
	\end{figure}

We will begin by writing down the bulk field due to the spherical shell sources described above. We have, using Eq.\eqref{eq:dSSKBlkF}, a superposition of fields produced by each shell source, i.e.,
\begin{equation}
	\begin{split}
		\phN(\zeta,\w,\spL)&=\oint  r_0^\nn dr_0\ \mathbb{G}(\zeta|\zeta_0,\w,\spL)\varrho_{_\nn}(\zeta_0,\w,\spL)\\
  &=\sum_i\frac{1}{W_{LR}(\zeta_i,\w,\spL)}
	\begin{cases}
        e^{2\pi\w}g_L(\zeta,\w,\spL)\ \Bigl[g_R(1+\zeta_i,\w,\spL)\ \sigma^R_i-g_L(1+\zeta_i,\w,\spL)\ \sigma^L_i\Bigr]&\quad \text{if}\; \zeta\prec 1+\zeta_i\ ,\\ & \\
	\quad g_R(\zeta_i,\w,\spL)\ \Bigl[g_R(\zeta,\w,\spL)\ \sigma^R_i-g_L(\zeta,\w,\spL)\ \sigma^L_i\Bigr] &\quad \text{if}\; 1+\zeta_i\prec\zeta\prec \zeta_i\	, \\  & \\
  	\quad g_R(\zeta,\w,\spL)\ \Bigl[g_R(\zeta_i,\w,\spL)\ \sigma^R_i-g_L(\zeta_i,\w,\spL)\ \sigma^L_i\Bigr] &\quad \text{if}\;  \zeta\succ \zeta_i\ .
	\end{cases}		
	\end{split}
\end{equation}
We remind the reader that $\prec$ and $\succ$ are comparisons using the radial contour-ordering of the dS-Sk contour. We also remind the reader that $\zeta$ changes from $1$ to $0$, as we traverse the clockwise dS-SK contour, starting from the right static patch (See Fig.\ref{fig:3regionsSKApp}). The reader should note that the above superposition of fields is continuous everywhere, but its derivative (and hence the conjugate field) is discontinuous at each spherical shell, with the discontinuity being determined by the strength of the scalar source at that shell. This is expected since the bulk-to-bulk Green function was constructed in the last subsection with precisely these boundary conditions in mind.  

Given the above field, computing the on-shell action is straightforward.
We use Eq.\eqref{eq:SOnShellBlkSrc} to write
\begin{equation}
	\begin{split}
		S|_{\textbf{On-shell}}&= \frac{1}{2} \sum_\spL\int\frac{d\w}{2\pi}\oint  r^\nn dr\ \varrho_{_\nn}^*\phN|_{\textbf{On-shell}}\\
  &=\frac{1}{2} \sum_{ij\spL}\int\frac{d\w}{2\pi} \frac{g_R(\zeta_i,\w,\spL)}{W_{LR}(\zeta_i,\w,\spL)}\Bigl\{ \sigma^{R\ast}_j \ \Bigl[g_R(1+\zeta_j,\w,\spL)\sigma^R_i-g_L(1+\zeta_j,\w,\spL)\sigma^L_i\Bigr]\Bigr.\\
  &\qquad \Bigl.\qquad -\sigma^{L\ast}_j\ \Bigl[g_R(\zeta_j,\w,\spL)\sigma^R_i-g_L(\zeta_j,\w,\spL)\sigma^L_i\Bigr] \Bigr\} \ .
	\end{split}
\end{equation}
Even though we are working with distributional sources/fields, given the continuity of $\phN$, the computation above is unambiguous. Next, we substitute explicit forms of the boundary-to-bulk Green functions as well as the Wronskian in terms of renormalised quantities. We have, using Eqs.\eqref{eq:gLgRXiexp} and \eqref{eq:LRExpId}, the following set of equalities:
\begin{equation}
	\begin{split}   
W_{LR}(\zeta_i,\w,\spL) &=- (1+n_\w)\left(\frac{1-r_i}{1+r_i}\right)^{i\w} [\kO-\kI]\ ,\\
\frac{g_R(\zeta_i,\w,\spL)}{W_{LR}(\zeta_i,\w,\spL)} &=\left(\frac{1-r_i}{1+r_i}\right)^{-\frac{i\w}{2}} \Xi_{n}(r_i,\w,\spL)\ ,\\
g_L(\zeta_i,\w,\spL)&= -\left(\frac{1-r_i}{1+r_i}\right)^{\frac{i\w}{2}}\Biggl\{\Xi_{nn}(r_i,\w,\spL) +\left[n_\w \kO-(1+n_\w)\kI  \right]\Xi_{n}(r_i,\w,\spL)\Biggr\}\ ,\\
g_L(1+\zeta_i,\w,\spL)&= -n_\w\left(\frac{1-r_i}{1+r_i}\right)^{\frac{i\w}{2}}\left[ \kO-\kI  \right]\Xi_{n}(r_i,\w,\spL)\ ,\\
g_R(\zeta_i,\w,\spL)&= -(1+n_\w)\left(\frac{1-r_i}{1+r_i}\right)^{\frac{i\w}{2}}\left[\kO-\kI \right]\Xi_{n}(r_i,\w,\spL)\ ,\\
g_R(1+\zeta_i,\w,\spL)&= \left(\frac{1-r_i}{1+r_i}\right)^{\frac{i\w}{2}}\Biggl\{\Xi_{nn}(r_i,\w,\spL)  -\left[(1+n_\w)\kO-n_\w \kI \right]\Xi_{n}(r_i,\w,\spL)\Biggr\}\ .
\end{split}
\end{equation}
Substituting these expressions back into the on-shell action yields the following double sum:
\begin{equation}\label{Eq:shellDoubleSum}
	\begin{split}
		S|_{\textbf{On-shell}}
  &=\frac{1}{2}\sum_{ij\spL}\int\frac{d\w}{2\pi}\left(\frac{1-r_i}{1+r_i}\right)^{-\frac{i\w}{2}}\left(\frac{1-r_j}{1+r_j}\right)^{\frac{i\w}{2}}\\
  &\qquad\times\Bigl\{ 
  \Xi_{n}(r_i,\w,\spL)\ \Xi_{nn}(r_j,\w,\spL)\ [\sigma^{R\ast}_j\sigma^R_i-\sigma^{L\ast}_j\sigma^L_i]\Bigr.\\
  &\quad\qquad-\Xi_{n}(r_i,\w,\spL)\ \Xi_{n}(r_j,\w,\spL)\ \kO(\sigma^R_j-\sigma^L_j)^\ast[(1+n_\w)\sigma^R_i-n_\w\sigma^L_i]\\
  &\quad\qquad\Bigl.-\Xi_{n}(r_i,\w,\spL)\ \Xi_{n}(r_j,\w,\spL)\ \kI(\sigma^R_i-\sigma^L_i)[(1+n_{-\w})\sigma^{R\ast}_j-n_{-\w}\sigma^{L\ast}_j]\Bigr]\ .
	\end{split}
\end{equation}

Let us begin by interpreting the terms in the above double sum. We first note that the last two lines in the above expression are related by the relabelling $\w\to-\w$ and are hence equal. The physical meaning of the last two lines is clarified by defining the \emph{radiative multipole moments}: 
\begin{equation}\label{eq:ShellMultDef}
	\begin{split}
	\multj_R(\w,\spL)\equiv \sum_i \left(\frac{1-r_i}{1+r_i}\right)^{-\frac{i\w}{2}} \Xi_n(r_i,\w,\spL)\ \sigma_i^R \equiv \int_R dr\  r^\nn \Xi_n(r,\w,\spL)\ \left(\frac{1-r}{1+r}\right)^{-\frac{i\w}{2}}\varrho_{_\nn}(\zeta,\w,\spL)  \ ,\\
 \multj_L(\w,\spL) \equiv \sum_i \left(\frac{1-r_i}{1+r_i}\right)^{-\frac{i\w}{2}} \Xi_n(r_i,\w,\spL)\ \sigma_i^L\equiv -\int_L dr\ r^\nn\Xi_n(r,\w,\spL)\ \left(\frac{1-r}{1+r}\right)^{-\frac{i\w}{2}}\varrho_{_\nn}(\zeta,\w,\spL) \  .
	\end{split}
\end{equation}
The integrals here are performed over right/left \emph{half} of the dS-SK contour respectively. We will also find it convenient to define the average/difference multipole moments via 
\[
 \multj_A(\w,\spL)\equiv\frac{1}{2}[\multj_R(\w,\spL)+\multj_L(\w,\spL)]\ ,\] and \[
 \multj_D(\w,\spL)\equiv \multj_R(\w,\spL)-\multj_L(\w,\spL) \]
Here we deliberately use the same notation as we did for multipole moments in flat spacetime (see Eq.\eqref{eq:jFlatdef}) and for  point-like dS sources (See Eq.\eqref{eq:jdSPtdef}). One reason for this is as follows: the last two lines of Eq.\eqref{Eq:shellDoubleSum} can be recast in terms of the above definitions, into the cosmological influence phase of a point-source
\begin{equation}
    \label{eq:SCIPpt2}
	\begin{split}
		\SCIP^\text{Pt}
  &\equiv  -\sum_{\spL}\int\frac{d\w}{2\pi} \kO(\multj_R-\multj_L)^\ast[(1+n_\w)\multj_R-n_\w\multj_L]=-\sum_\spL\int\frac{d\w}{2\pi} \kO\  \multj_D^*\ \Bigl[\multj_A+\left(n_\w+\frac{1}{2}\right)\multj_D\Bigr]\ .
	\end{split}
\end{equation}
We recognise here the exact influence phase derived for a point-like dS observer in Eq.\eqref{eq:SCIPpt}, using a detailed counterterm procedure. More evidence for this identification  will be presented in appendix \ref{app:RadReact}, where we describe how these multipole moments correctly reproduce the flat space answers with Hubble corrections.

For now, we turn our attention to the remaining terms, viz., the first double sum in Eq.\eqref{Eq:shellDoubleSum}. The presence of the singular Green solution $\Xi_{nn}$, as well as the right/left factorised form of this sum, indicates that these terms incorporate non-dissipative/conservative \emph{self-energy} corrections of the extended source.
The final on-shell action can then be written as $	S|_{\textbf{On-shell}}=\SCIP^\text{Pt}+S_{\text{Int}}$, where $S_{\text{Int}}$ denotes
the internal potential energy of the spherical shells:
\begin{equation}
	\begin{split}
		S_{\text{Int}}
  &\equiv  \frac{1}{2}\sum_{i\neq j\spL}\int\frac{d\w}{2\pi}\left(\frac{1-r_i}{1+r_i}\right)^{-\frac{i\w}{2}}\left(\frac{1-r_j}{1+r_j}\right)^{\frac{i\w}{2}}
  \Xi_{n}(r_i,\w,\spL)\ \Xi_{nn}(r_j,\w,\spL)\ [\sigma^{R\ast}_j\sigma^R_i-\sigma^{L\ast}_j\sigma^L_i]\ .
	\end{split}
\end{equation}
Another instructive way to rewrite this potential energy contribution is to define radially averaged mean fields on the right/left static patch via
\begin{equation}
	\begin{split}
	\overline{\varphi}_{R,\text{Int}}(\w,\spL)\equiv \sum_i \left(\frac{1-r_i}{1+r_i}\right)^{-\frac{i\w}{2}} \Xi_{nn}(r_i,\w,\spL)\ \sigma_i^R \equiv \int_R dr\  r^\nn \Xi_{nn}(r,\w,\spL)\ \left(\frac{1-r}{1+r}\right)^{-\frac{i\w}{2}}\varrho_{_\nn}(\zeta,\w,\spL)  \ ,\\
 \overline{\varphi}_{L,\text{Int}}(\w,\spL) \equiv \sum_i \left(\frac{1-r_i}{1+r_i}\right)^{-\frac{i\w}{2}} \Xi_{nn}(r_i,\w,\spL)\ \sigma_i^L\equiv -\int_L dr\ r^\nn\Xi_{nn}(r,\w,\spL)\ \left(\frac{1-r}{1+r}\right)^{-\frac{i\w}{2}}\varrho_{_\nn}(\zeta,\w,\spL) \  .
	\end{split}
\end{equation}
We can then rewrite the potential energy as that of  multipole moments placed in such an average field, viz.,
\begin{equation}
\begin{split}
S_{\text{Int}}
  &= \frac{1}{2}\sum_{\spL}\int\frac{d\w}{2\pi} [\multj_R^\ast\overline{\varphi}_{R,\text{Int}}-\multj_L^\ast\overline{\varphi}_{L,\text{Int}}]\ .
	\end{split}
\end{equation}

\subsection{Relation to Detweiler-Whiting decomposition}\label{app:DetWhitdS}

We will begin by writing down the non-normalisable solution $\Xi_{nn}$ and the normalisable solution $\Xi_{n}$ in the Schwarzschild time valid for odd $d$ ( restoring all factors of $H$):
\begin{equation}\label{eq:XiKH}
\begin{split}
\Xi_{nn}&\equiv r^{-\nu-\frac{d}{2}+\frac{1}{2}(d-1-\nn)}(1-H^2r^2)^{-\frac{i\w}{2H}}\\
&\quad\times{}_2F_1\left[ \frac{1}{2}\left(1+\mu-\nu-\frac{i\w}{H}\right),\frac{1}{2}\left(1-\mu-\nu-\frac{i\w}{H}\right),1-\nu,H^2r^2\right]\ ,\\
\Xi_n&\equiv \frac{1}{2\nu} r^{\nu-\frac{d}{2}+1+\frac{1}{2}(d-1-\nn)}(1-H^2r^2)^{-\frac{i\w}{2H}}\\
&\quad \times{}_2F_1\left[ \frac{1}{2}\left(1+\mu+\nu-\frac{i\w}{H}\right),\frac{1}{2}\left(1-\mu+\nu-\frac{i\w}{H}\right),1+\nu,H^2r^2\right]\ ,\\
\kO &\equiv 2\frac{\Gamma\left(\frac{1+\nu-
				\mu-\frac{i\w}{H}}{2}\right)\Gamma\left(\frac{1+\nu+\mu-\frac{i\w}{H}}{2}\right)
			\Gamma\left(1-\nu\right)
		}{\Gamma\left(\frac{1-\nu+\mu-\frac{i\w}{H}}{2}\right)\Gamma\left(\frac{1-\nu-\mu-\frac{i\w}{H}}{2}\right)
			\Gamma\left(\nu\right)} \ .
\end{split}
\end{equation}
We have also quoted above the retarded two-point function on the world line. The outgoing Green function  can then be decomposed into  $\kO\ \Xi_n $ and $\Xi_{nn}$: we will now argue that this 
should be thought of as the regular/singular Green functions ala Deitweiler-Whiting(DW)\cite{Detweiler:2002mi} corresponding to dS spacetime. 

The relation to DW decomposition is not prima facie clear, since  DW formulated their rules for  general curved spacetimes in time domain, whereas the above expressions are quoted in frequency domain. So, to substantiate our assertion, we need to Fourier transform the complicated expressions above into time domain, and then show that the DW axioms are satisfied. Rather than do that exercise in general, we will content ourselves with showing how this works in the particular example of a massless scalar field in $dS_4$, whose DW decomposition is described in \cite{Burko:2002ge,Poisson:2011nh}. 

The regular term for DW decomposition in this case was calculated by the authors of \cite{Burko:2002ge} in FLRW-like coordinates as
\begin{equation}\begin{split}
G_R &=\frac{\eta\eta'}{2|\mathbf{x}-\mathbf{x}'|}\left[\delta(\eta-\eta'-|\mathbf{x}-\mathbf{x}'|)-\delta(\eta-\eta'+|\mathbf{x}-\mathbf{x}'|)\right]\\
&+\frac{1}{2}\left[\theta(\eta-\eta'-|\mathbf{x}-\mathbf{x}'|)+\theta(\eta-\eta'+|\mathbf{x}-\mathbf{x}'|)\right]\ ,
\end{split} \end{equation}
To check this expression against $\kO\Xi_n$, we will convert it into static coordinates and then Fourier transform the result to frequency domain. 

The coordinate transformation between static and FLRW coordinates is given by
\begin{equation}
	\begin{split}
		\eta &= -\frac{e^{-Ht}}{\sqrt{1-r^2 H^2}}\ ,\quad 
		\rho = \frac{re^{-Ht}}{\sqrt{1-r^2 H^2}}\ .
	\end{split}
\end{equation} 
We will assume the source to be at the origin $\rho'=0$, so that only the $\ell=0$ term survives
by spherical symmetry. With this choice, $G_R$ becomes
\begin{equation}
	\begin{split}
	G_R&=\frac{e^{H(t-t')}}{2r}\\
 &\times\left[\sqrt{\frac{1-Hr}{1+Hr}}\delta\left(t'-t-\frac{1}{H}\ln\left(\sqrt{\frac{1-Hr}{1+Hr}}\right)\right)-\sqrt{\frac{1+Hr}{1-Hr}}\delta\left(t'-t-\frac{1}{H}\ln\left(\sqrt{\frac{1+Hr}{1-Hr}}\right)\right)\right]\\
& + \frac{1}{2}\left[\theta\left(t'-t-\frac{1}{H}\ln\left(\sqrt{\frac{1-Hr}{1+Hr}}\right)\right)+\theta\left(t'-t-\frac{1}{H}\ln\left(\sqrt{\frac{1+Hr}{1-Hr}}\right)\right)\right]\ .
	\end{split}
\end{equation} 
This expression can be readily Fourier transformed with respect to $t-t'$ yielding
\begin{equation}\label{eq:GtDWReg}
	\widetilde{G}_R=	\frac{1}{2r}\left[\left(\frac{1-Hr}{1+Hr}\right)^{-\frac{i \w}{2H}}-\left(\frac{1+Hr}{1-Hr}\right)^{-\frac{i \w}{2H}}\right]-\frac{H^2}{2i\w}\left[\left(\frac{1-Hr}{1+Hr}\right)^{-\frac{i \w}{2H}}+\left(\frac{1+Hr}{1-Hr}\right)^{-\frac{i \w}{2H}}\right]\ .
\end{equation}
Regularity near the origin is manifest in frequency domain. Further, the above expression is also an odd function of the frequency $\w$, signalling that these terms encode the dissipation due to radiation reaction.

Similarly, we can consider the singular Green's function quoted in \cite{Burko:2002ge}:
\begin{align}
	\begin{split}
		G_S=\frac{\eta\eta'}{2|\mathbf{x}-\mathbf{x}'|}\left[\delta(\eta-\eta'-|\mathbf{x}-\mathbf{x}'|)+\delta(\eta-\eta'+|\mathbf{x}-\mathbf{x}'|)\right]\\+\frac{1}{2}\left[\theta(\eta-\eta'-|\mathbf{x}-\mathbf{x}'|)-\theta(\eta-\eta'+|\mathbf{x}-\mathbf{x}'|)\right]
	\end{split}
\end{align}	
whose Fourier transform at $\rho'=0$ is
\begin{equation}\label{eq:GtDWSing}
	\tilde{G}_S=	\frac{1}{2r}\left[\left(\frac{1-Hr}{1+Hr}\right)^{-\frac{i \w}{2H}}+\left(\frac{1+Hr}{1-Hr}\right)^{-\frac{i \w}{2H}}\right]-\frac{H^2}{2i\w}\left[\left(\frac{1-Hr}{1+Hr}\right)^{-\frac{i \w}{2H}}-\left(\frac{1+Hr}{1-Hr}\right)^{-\frac{i \w}{2H}}\right]\ .
\end{equation}
This expression has a $\sim \frac{1}{r}$ behaviour near the origin and is an even function of $\w$. The expressions in Eq.\eqref{eq:GtDWReg} and Eq.\eqref{eq:GtDWSing} can then be matched against $\kO\ \Xi_n $ and $\Xi_{nn}$ respectively. This is done by taking Eq.\eqref{eq:XiKH}, setting $\nn=d-1, \mu=\frac{d}{2}, \nu=\ell+\frac{d}{2}-1$, and then taking the limit $d=3$ and $\ell=0$.

\section{Radiation reaction due to light scalar fields}\label{app:RadReact}
In this section, we will evaluate the radiation reaction force on a dS point particle coupled to a scalar field. We will do this  in a small curvature approximation, i.e., we begin with the leading order result in  flat spacetime\cite{Birnholtz:2013ffa,Birnholtz:2013nta} and then systematically correct it for curvature effects. In dS, Hubble constant $H$ parametrises the deviation from flat spacetime, so the small curvature expansion is an expansion in $H$. We will also work within a non-relativistic expansion and a multipole expansion, and then eventually covariantise the final answer for the radiation reaction(RR).

To this end, consider a point-like source moving along a time-like  worldline $x(\tau)$ in dS, where $\tau$ denotes the proper time of the source. We will assume that the particle trajectory is close to the south pole ($rH\ll 1$) and the radiation wavelength is taken to be much larger than the length scale of the particle trajectory ($\w r\ll 1$), but much smaller than the curvature length scale ($\w\gg H$). Further, we work in a non-relativistic limit ($v\ll 1$). Thus, we consider the following hierarchy of scales (See Fig.\ref{fig:nearflatexp}):
\begin{align}
	\begin{split}
H \ll \w \ll 1/r\ .
	\end{split}
\end{align}
In analogy with flat spacetime, we will refer to this expansion as the  post-Newtonian(PN) expansion in dS.

The source density for a moving source in dS is given by 
\begin{align}
\label{eq:ParticleSrc}
	\begin{split}
		\widetilde{\rho}(x')=\int \delta^{d+1}(x-x')d\tau=&\sqrt{1-H^2 r^2-\frac{\dot{r}^2}{1-H^2 r^2}+\dot{r}^2-v^2}\ \delta^d(\vec{x}-\vec{x}') \ ,
	\end{split}
\end{align}
where we have defined $v=\sqrt{\sum_{i=1}^{d}\dot{x}_i^{2}}$. Here  the dots denote  the derivative with respect to the standard time $t$. The time-dilation/length-contraction factor for the particle worldline can then be expanded as follows:
\begin{equation}
\label{eq:SrcFlatExp}
	\begin{split}
&\sqrt{1-H^2 r^2-\frac{\dot{r}^2}{1-H^2 r^2}+\dot{r}^2-v^2}\\
&=-\sum_{n,s,k} \binom{n}{k} \frac{(2s+2n+2k-5)!!(2s+2k-5)!!}{2^{n+s-1}(2s+4k-5)!!\  n!(s-1)!}(Hr)^{2n}\dot{r}^{2k}v^{2s-2}\ .
	\end{split}
\end{equation}
Here the sum is over all non-negative integers, and the binomial coefficient vanishes for all values of $k$ outside the range $0\leq k\leq n$. This expansion describes the red-shift of the particle within dS spacetime in a slow-motion approximation, assuming 
that both the cosmological redshift $Hr$ as well as the Doppler red-shifts due to peculiar motions (proportional to $\dot{r}$ and $v$) are small.
Our strategy below will be to use the above expansion to compute the symmetric trace-free(STF) multipole moments of the source, which can then be fed into the cosmological influence phase to compute the RR force.

We caution the reader that the source form given in Eq.\eqref{eq:ParticleSrc} is specific to the KG scalar case with $\nn=d-1$. This is \emph{not} the correct form of the source for the scalar/vector/tensor sectors of EM field and gravity. For such cases, the explicit form of sources involves extra velocity/time-dilation factors, e.g.,  EM vector sector source is the electric current carried by the particle which has additional velocity dependence not captured by Eq.\eqref{eq:ParticleSrc}. Another related comment on EM/gravity sources is that the RR force coming from just one of the sectors is not expected to be covariant\cite{Birnholtz:2013nta}: one should add in the contributions from all sectors to derive a covariant force expression. To do this for EM/linearised gravity, we need a theory of vector/tensor STF expansions (i.e., a formalism analogous to the one described in appendix\ \ref{app:FlatMult}). We will derive such a formalism elsewhere\cite{dSSKvec}: in this note, we will limit our RR force analysis to KG scalars. 

In the dS-SK geometry the above source will be doubled to a $\widetilde{\rho}_L$ and a $\widetilde{\rho}_R$ coming from left/right trajectories $x_L(\tau_L)$ and $x_R(\tau_R)$. The degrees of freedom of our open system are thus two copies of the position of the  particle and its time derivatives: $\{x_L, x_R, \dot{x}_L, \dot{x}_R, \ddot{x}_L, \ddot{x}_R,\dots \}$. The scalar ALD force and its post-Newtonian corrections only require expressions linear in $x_D, \dot{x}_D, \ddot{x}_D,\dots$ which are the difference in the positions and their derivatives. We will also keep terms up to cubic powers of $x_D$. In this approximation, the average and difference functions of the sources can be written in a simple way. Consider for illustration, the average and difference functions of just the position: 
\begin{align}
	\frac{1}{2}\left[\mathfrak{f}\left(x_A+\frac{x_D}{2}\right)+\mathfrak{f}\left(x_A-\frac{x_D}{2}\right)\right]=\mathfrak{f}(x_A)+\frac{x_D^2}{4}\frac{\partial^2 \mathfrak{f}}{\partial x_A^2}+O(x_D^4)\\
	\mathfrak{f}\left(x_A+\frac{x_D}{2}\right)-\mathfrak{f}\left(x_A-\frac{x_D}{2}\right)=x_D\frac{\partial\,\mathfrak{f}}{\partial x_A}+\frac{x_D^3}{24}\frac{\partial^3\,\mathfrak{f}}{\partial x_A^3}+O(x_D^4)
\end{align}
In general, the sources will be functions not only of the positions but also their time derivatives: in such cases, the above formula should be interpreted as a multi-variable Madhava-Taylor expansion.

We will now substitute the particle source Eq.\eqref{eq:ParticleSrc} into the multipole moments defined in Eq.\eqref{eq:ShellMultDef} and obtain the Lagrangian for RR force in  PN expansion. We begin by expressing the influence phase in terms of STF moments of the particle density: we proceed similarly to how we rewrote the RR influence  phase in flat spacetime (Eq.\eqref{eq:RRMinknII}) in terms of STF multipole moments (Eq.\eqref{eq:SRRMinkSTF}). Using the STF addition theorem in Eq.\eqref{eq:STFAddition}, we can rewrite the dissipative part of Eq.\eqref{eq:SCIPpt2} in time-domain  as:
\begin{equation}
	\begin{split}
S_{RR}^\text{Odd $d$}&=\sum_{\ell}\int\frac{d\w}{2\pi}  \frac{\kO}{\nn_{d,\ell}|\mathbb{S}^{d-1}|} \frac{1}{\ell!}\mathcal{Q}_{D,STF}^{\ast<i_1i_2\ldots i_\ell>}\mathcal{Q}^{A,STF}_{ <i_1i_2\ldots i_\ell>}\ .
\end{split}
\end{equation}
where we have defined the time-domain STF multipole moments in dS as 
\begin{equation}
	\begin{split}
\mathcal{Q}_{A,STF}^{i_1\ldots i_\ell}(t) \equiv  \Pi^{<i_1i_2\ldots i_\ell>}_{ <j_1j_2\ldots j_\ell>}\int r^{d-1} dr\  \hat{r}^{j_1}\hat{r}^{j_2}\ldots \hat{r}^{j_\ell}\ \Xi_n(i\partial_t, r)\widetilde{\rho}_A(t,\vec{r})\ ,\\
\mathcal{Q}_{D,STF}^{i_1\ldots i_\ell}(t) \equiv \Pi^{<i_1i_2\ldots i_\ell>}_{ <j_1j_2\ldots j_\ell>}\int r^{d-1} dr\  \hat{r}^{j_1}\hat{r}^{j_2}\ldots \hat{r}^{j_\ell}\ \Xi_n(i\partial_t,r)\widetilde{\rho}_D(t,\vec{r})\ .
\end{split}
\end{equation}
We will now use the PN expansion for $\widetilde{\rho}_D$ in 
Eq.\eqref{eq:SrcFlatExp}, the expansion for $\kO$ from  
Eq.\eqref{eq:Koutflat} and the expansion for $\Xi_n$ from 
Eq.\eqref{eq:XiExp} respectively. Keeping all terms in the action up to quartic order in amplitudes (i.e. in position $x$) and quartic order in  Hubble constant $H$, we get an effective Lagrangian of the form
\begin{equation}
	\begin{split}
|\mathbb{S}^{d-1}|(d-2)!!\times(-1)^{\frac{d-1}{2}}L=&-[x_i]_D\mathbb{D}_1[x^i]_A+\frac{1}{2}\left[x_ix_j-\frac{x^2}{d}\delta_{ij}\right]_D\mathbb{D}_2\left[x^i x^j-\frac{x^2}{d}\delta_{ij}\right]_A\\
		&-\left\{\frac{1}{2}(x_i)_D\mathbb{D}_{1}^{X}[x^ix^2]_A+\frac{1}{2}(x^2x_i)_D\mathbb{D}_{1}^{X}[x^i]_A\right\}\\
		&+\left\{\frac{1}{2}(x_i)_D\mathbb{D}_{1}^{V}[x^iv^2]_A+\frac{1}{2}(v^2x_i)_D\mathbb{D}_{1}^{V}[x^i]_A\ \right\}\\
		&+\frac{1}{2d}[x^2]_D\mathbb{D}_0^{XX}[x^2]_A+\frac{1}{4}[v^2]_D  \mathbb{D}_{0}^{VV}[v^2]_A\\
		&-\frac{1}{4}[x^2]_D  \mathbb{D}_{0}^{XV}[v^2]_A -\frac{1}{4}[v^2]_D  \mathbb{D}_{0}^{XV}[x^2]_A\ .
	\end{split}
\end{equation}
Here, we have the seven differential operators, each built out of a finite number of time-derivatives with constant coefficients, and labelled by $\{\mathbb{D}_1,\mathbb{D}_2,\mathbb{D}_1^X,\mathbb{D}_1^V,\mathbb{D}_0^{XX},\mathbb{D}_0^{VV},\mathbb{D}_0^{XV}\}$. We use the subscripts of these operators to denote the multipole number, whereas the superscripts are used to distinguish between different structures occurring at the same multipole number. The explicit form of these operators is tabulated in table \ref{tab:RRoperators}. We note that terms beyond quadrupole moment do not contribute to the quartic influence phase.

We can expand out the average and the difference multipole moments in terms of the average/difference in the particle position by using the following identities:
\begin{equation}
	\begin{split}
[Z]_d[Y^3]_a &=Z_dY_a^3+\frac{1}{4}Z_d\ (3Y_d^2 Y_a)\\
[Z^2]_d[Y^2]_a &=(2Z_dZ_a)Y_a^2+\frac{1}{4}(2Z_dZ_a) Y_d^2\\
[Z^3]_d [Y]_a&= (3Z_a^2Z_d)Y_a	+\frac{1}{4}Z_d^3 Y_a \ . 
\end{split}
\end{equation}
After  integration by parts, the above Lagrangian can be cast into the form:
\begin{equation}
L=\frac{(-1)^{\frac{d-1 }{2}}}{|\mathbb{S}^{d-1}|(d-2)!!}\left[F_i(x_A)x^i_D+\frac{1}{4}N_i(x_D)x^i_A\right]
\end{equation}
where $F^i$ are the Euler-Lagrange derivatives of the terms linear in $x_D$ with respect to $x^i_D$. Similarly, $N^i$ are the Euler-Lagrange derivatives of the terms linear in $x_A$ with respect to $x^i_A$.

The terms in the Lagrangian which are cubic in $x_D$ give rise to noise terms $N^i$. These terms are different from pure noise terms, i.e. those quartic in $x_D$. The $N^i$ contribute to the radiation reaction with a $x^i_A$ dependent term. It should be noted that this noise is not thermal in origin. Rather, the origin of this noise can be understood as follows: Despite the scalar field coupling linearly to the multipole moments, the moments themselves are non-linear functions of the positions. Hence, the open system described in terms of position has  extra noise terms.

Both the $F^i$ and the $N^i$ can be written in terms of the operators given in table \ref{tab:RRoperators} as:
\begin{equation}
\begin{split}
F^i=&-\mathbb{D}_1[x^i]+x_j\mathbb{D}_2[x^i x^j]-\frac{x^i}{d}\mathbb{D}_2[x^2]\\
&-\left\{\frac{1}{2}\mathbb{D}_{1}^{X}[x^ix^2]+\frac{1}{2}x^2\mathbb{D}_{1}^{X}[x^i]+x^i x^j \mathbb{D}_{1}^{X}[x_j]\right\}+\left\{\frac{1}{2}\mathbb{D}_{1}^{V}[x^iv^2]+\frac{1}{2}v^2\mathbb{D}_{1}^{V}[x^i]-\partial_t\left(v^i x^j \mathbb{D}_{1}^{V}[x_j]\ \right)\right\}\\
&+\frac{x^i}{d}\mathbb{D}_0^{XX}[x^2]-\frac{1}{2}\partial_t\left(v^i  \mathbb{D}_{0}^{VV}[v^2]\ \right)+\left\{\frac{1}{2}\partial_t\left(v^i  \mathbb{D}_{0}^{XV}[x^2]\ \right)-\frac{1}{2}x^i\mathbb{D}_{0}^{XV}[v^2]\right\}\ ,
\end{split}
\end{equation}
as well as
\begin{equation}
	\begin{split}
		N^i(x)=&x_j\mathbb{D}_2[x^i x^j]-\frac{x^i}{d}\mathbb{D}_2[x^2]\\
		&+\left\{\frac{1}{2}\mathbb{D}_{1}^{X}[x^ix^2]+\frac{1}{2}x^2\mathbb{D}_{1}^{X}[x^i]+x^i x^j \mathbb{D}_{1}^{X}[x_j]\right\}-\left\{\frac{1}{2}\mathbb{D}_{1}^{V}[x^iv^2]+\frac{1}{2}v^2\mathbb{D}_{1}^{V}[x^i]-\partial_t\left(v^i x^j \mathbb{D}_{1}^{V}[x_j]\ \right)\right\}\\
		&+\frac{x^i}{2d}\mathbb{D}_0^{XX}[x^2]-\frac{1}{2}\partial_t\left(v^i  \mathbb{D}_{0}^{VV}[v^2]\ \right)+\left\{\frac{1}{2}\partial_t\left(v^i  \mathbb{D}_{0}^{XV}[x^2]\ \right)-\frac{1}{2}x^i\mathbb{D}_{0}^{XV}[v^2]\right\}
	\end{split}
\end{equation}
The RR force $F^i$ then covariantises to the  expression given in Eq.\eqref{eq:RRrecursion}. Beyond the expressions quoted in the main text, we have also covariantised RR force expressions in $d=9,11$. they are given by
\begin{equation}
\begin{split}
{}^0f^\mu_9&\equiv \frac{P^{\mu\nu}}{9!!}\left\{-a_\nu^{(7)}+30\ (a\cdot a)\  a_\nu^{(5)}+210\ (a\cdot a^{(1)})\ a_\nu^{(4)}+378\ (a\cdot a^{(2)})\ a_\nu^{(3)}\right.\\
&\left.\qquad\qquad\quad+420\ (a\cdot a^{(3)})\ a_\nu^{(2)}+300\ (a\cdot a^{(4)})\ a_\nu^{(1)}+108\ (a\cdot a^{(5)})\ a_\nu\right.\\
&\left.\qquad\qquad+336\ (a^{(1)}\cdot a^{(1)})\  a_\nu^{(3)}+1050\ (a^{(1)}\cdot a^{(2)})\  a_\nu^{(2)}+960\ (a^{(1)}\cdot a^{(3)})\  a_\nu^{(1)}+420\ (a^{(1)}\cdot a^{(4)})\ a_\nu\right.\\
&\left.\qquad\qquad+675\ (a^{(2)}\cdot a^{(2)})\  a_\nu^{(1)}+756\ (a^{(2)}\cdot a^{(3)})\  a_\nu+O(a^5)\right\}\\
&-H^2\frac{P^{\mu\nu}}{9!!} \left\{a_\nu^{(5)}+97\ (a\cdot a)\  a_\nu^{(3)}+433\ (a\cdot a^{(1)})\ a_\nu^{(2)}+408\ (a\cdot a^{(2)})\ a_\nu^{(1)}+199\ (a\cdot a^{(3)})\ a_\nu\right.\\
&\left.\qquad\qquad+339\ (a^{(1)}\cdot a^{(1)})\  a_\nu^{(1)}+448\ (a^{(1)}\cdot a^{(2)})\ a_\nu+O(a^5)\right\}\\
&+H^4\frac{P^{\mu\nu}}{9!!}\left\{-a_\nu^{(3)}+157\ (a\cdot a)\  a_\nu^{(1)}+296\ (a\cdot a^{(1)})\ a_\nu\right\}+O(H^6)\ ,
\end{split}
\end{equation}
as well as
\begin{equation}
\begin{split}
{}^0f^\mu_{11}&\equiv \frac{P^{\mu\nu}}{11!!}\left\{-a_\nu^{(9)}+55\ (a\cdot a)\  a_\nu^{(7)}+495\ (a\cdot a^{(1)})\ a_\nu^{(6)}+1188\ (a\cdot a^{(2)})\ a_\nu^{(5)}\right.\\
&\left.\qquad\qquad\quad+1848\ (a\cdot a^{(3)})\ a_\nu^{(4)}+1980\ (a\cdot a^{(4)})\ a_\nu^{(3)}\right.\\
&\left.\qquad\qquad\quad+1485\ (a\cdot a^{(5)})\ a_\nu^{(2)}+770\ (a\cdot a^{(6)})\ a_\nu^{(1)}+220\ (a\cdot a^{(7)})\ a_\nu\right.\\
&\left.\qquad\qquad+1056\ (a^{(1)}\cdot a^{(1)})\  a_\nu^{(5)}+4620\ (a^{(1)}\cdot a^{(2)})\  a_\nu^{(4)}\right.\\
&\left.\qquad\qquad\quad+6336\ (a^{(1)}\cdot a^{(3)})\  a_\nu^{(3)}+5775\ (a^{(1)}\cdot a^{(4)})\  a_\nu^{(2)}+3520\ (a^{(1)}\cdot a^{(5)})\  a_\nu^{(1)}+1155\ (a^{(1)}\cdot a^{(6)})\ a_\nu\right.\\
&\left.\qquad\qquad+4455\ (a^{(2)}\cdot a^{(2)})\  a_\nu^{(3)}+10395\ (a^{(2)}\cdot a^{(3)})\  a_\nu^{(2)}+7700\ (a^{(2)}\cdot a^{(4)})\  a_\nu^{(1)}+2970\ (a^{(2)}\cdot a^{(5)})\  a_\nu\right.\\
&\left.\qquad\qquad+4928\ (a^{(3)}\cdot a^{(3)})\  a_\nu^{(1)}+4620\ (a^{(3)}\cdot a^{(4)})\  a_\nu\right\}\\
&-H^2\frac{P^{\mu\nu}}{11!}\left\{a_\nu^{(7)}+342\ (a\cdot a)\  a_\nu^{(5)}+2294\ (a\cdot a^{(1)})\ a_\nu^{(4)}+3826\ (a\cdot a^{(2)})\ a_\nu^{(3)}\right.\\
&\left.\qquad\qquad\quad+3737\ (a\cdot a^{(3)})\ a_\nu^{(2)}+2066\ (a\cdot a^{(4)})\ a_\nu^{(1)}+622\ (a\cdot a^{(5)})\ a_\nu\right.\\
&\left.\qquad\qquad+3231\ (a^{(1)}\cdot a^{(1)})\  a_\nu^{(3)}+8490\ (a^{(1)}\cdot a^{(2)})\  a_\nu^{(2)}+5663\ (a^{(1)}\cdot a^{(3)})\  a_\nu^{(1)}+1974\ (a^{(1)}\cdot a^{(4)})\ a_\nu\right.\\
&\left.\qquad\qquad+3785\ (a^{(2)}\cdot a^{(2)})\  a_\nu^{(1)}+3210\ (a^{(2)}\cdot a^{(3)})\  a_\nu+O(a^5)\right\}\ ,\\
&-H^4\frac{P^{\mu\nu}}{11!!} \left\{a_\nu^{(5)}+1340\ (a\cdot a)\  a_\nu^{(3)}+6108\ (a\cdot a^{(1)})\ a_\nu^{(2)}+6148\ (a\cdot a^{(2)})\ a_\nu^{(1)}+2599\ (a\cdot a^{(3)})\ a_\nu\right.\\
&\left.\qquad\qquad+5182\ (a^{(1)}\cdot a^{(1)})\  a_\nu^{(1)}+5876\ (a^{(1)}\cdot a^{(2)})\ a_\nu+O(a^5)\right\}\ .
\end{split}
\end{equation}
We have not yet succeeded in finding similar covariant expressions for the $N^i$. 

\subsection{Near flat expansions for odd $d$}\label{app:NearFlatExp}
In this subsection, we will describe how normalisable modes of the generalised scalar equation in dS can be thought of as perturbations of the  corresponding Bessel J modes in the flat spacetime. In the context of radiation reaction problems, these modes are essential in  defining the radiative multipole moments: their role is to  appropriately smear the sources to take into account time-delay effects. Once such an expansion is obtained, it is easy to find the flat space expansion of the non-normalisable mode for even dimensional dS just by analytical continuation. 

The solution of the generalised scalar wave equation, regular at $r=0$, is given by 
\begin{equation}
	\begin{split}
\Xi_n&\equiv \frac{1}{2\nu} r^{\nu-\frac{d}{2}+1+\frac{1}{2}(d-1-\nn)}(1-H^2r^2)^{-\frac{i\w}{2H}}\\
&\quad \times{}_2F_1\left[ \frac{1}{2}\left(1+\mu+\nu-\frac{i\w}{H}\right),\frac{1}{2}\left(1-\mu+\nu-\frac{i\w}{H}\right),1+\nu,H^2r^2\right]\ .
	\end{split}
\end{equation}
Here we have made all $H$ factors explicit so that the $H\to 0$ limit can readily be taken. In such a limit, the above expression reduces to a Bessel $J$ function. More explicitly, we will find it convenient to define a sequence of scaled Bessel $J$ functions of the form
\begin{equation}
	\begin{split}
\mathfrak{B}_k&\equiv\frac{r^{\nu-\frac{d}{2}+1+\frac{1}{2}(d-1-\nn)+2k}}{2\nu(\nu+1)\ldots(\nu+k)}\  {}_0F_1\left[1+k+\nu,-\frac{\w^2r^2}{4}\right]=\frac{\Gamma(\nu)\ r^{\frac{1}{2}(1-\nn)+k}}{2(\w/2)^{k+\nu}}J_{k+\nu}(\w r)\ 
	\end{split}
\end{equation}
in flat spacetime. In terms of these functions, the dS wavefunction $\Xi_n$ has a small $H$ expansion
\begin{equation}\label{eq:XiExp}
	\begin{split}
	\Xi_n =\sum_{k=0}^\infty \mathfrak{p}_{k}(\nu,H^2,\w^2)
		\mathfrak{B}_k\ ,
	\end{split}
\end{equation}
with $\mathfrak{p}_{k}(H^2,\w^2)$ being a homogeneous polynomial of degree $k$ in the variables $H^2$ and $\w^2$. An explicit expression is given by
\begin{equation}
	\begin{split}
		\mathfrak{p}_k
		&\equiv \frac{H^{2k}}{k!}	\sum_{m=0}^k (-)^m\binom{k}{m}\sum_{n=0}^m(-)^n\binom{m}{n} \sigma^{2k-2m}\frac{\Gamma(\alpha+m)\Gamma(1+\nu+m)}{\Gamma(\alpha+m-n)\Gamma(1+\nu+m-n)}\ \\
		&\quad \qquad\qquad \times\frac{\Gamma(\alpha+i\sigma+m-n) \Gamma(\alpha-i\sigma+m-n) }{\Gamma(\alpha+i\sigma) \Gamma(\alpha-i\sigma) }\ ,
	\end{split}
\end{equation}
where we have defined the variables
\begin{equation}
	\begin{split}
\alpha\equiv \frac{1}{2}(1+\nu-\mu)\ , \quad \sigma= \frac{\w}{2H}\ .
	\end{split}
\end{equation}
A useful fact about these polynomials is the leading $H$ scaling at small $H$
of these polynomials, given by 
\begin{equation}\label{eq:pHScaling}
	\begin{split}
 \mathfrak{p}_{3n-2}\ ,\ \mathfrak{p}_{3n-1}\ ,\ \mathfrak{p}_{3n}\ \propto H^{2n} \ .
	\end{split}
\end{equation}
Thus, to obtain an answer accurate up to $H^{2n}$, we only need the polynomials till $\mathfrak{p}_{3n}$.

We will now outline a derivation for the above expansion as follows: first, we use the Euler transformation on the hypergeometric functions to write
\begin{equation}\label{eq:EuXiN}
	\begin{split}
		\Xi_n &= \frac{1}{2\nu} r^{\nu-\frac{d}{2}+\frac{1}{2}(d-1-\nn)}(1-H^2r^2)^{-\alpha}\ {}_2F_1\left[\alpha+i\sigma,\alpha-i\sigma,1+\nu,-\frac{H^2r^2}{1-H^2 r^2}\right]\ ,
	\end{split}
\end{equation}
where the variables $\alpha$ and $\sigma$ are as defined above.
In the next step, we employ the Mellin-Barnes representation of the hypergeometric function, viz.\cite{NIST:DLMF},
\begin{equation}
	\begin{split}
{}_2F_1\left[ a,b,c,x\right]= \int_{-i\infty}^{i\infty}\frac{dz}{2\pi i}(-x)^z\Gamma(-z)\frac{\Gamma(a+z) \Gamma(b+z) \Gamma(c)}{\Gamma(a) \Gamma(b) \Gamma(c+z)}\ .
	\end{split}
\end{equation}
and expand the resultant integrand using
\begin{equation}
\begin{split}
(1-H^2r^2)^{-\alpha-z}\left(H^2r^2\right)^z
= \sum_{k=0}^\infty\frac{(H^2r^2)^{k+z}}{k!}
\frac{\Gamma\left[k+z+\frac{1}{2}(1+\nu-\mu)\right]}{\Gamma\left[\alpha+z+\frac{1}{2}(1+\nu-\mu)\right]}\ .
\end{split}
\end{equation}
Shifting the Mellin-Barnes integration variable, we get the following Mellin-integral representation for $\Xi_n$:
\begin{equation}
	\begin{split}
		\Xi_n&=\frac{1}{2\nu} r^{\nu-\frac{d}{2}+\frac{1}{2}(d-1-\nn)}
		\int_{-i\infty}^{i\infty}\frac{dz}{2\pi i}\left(\frac{\w r}{2}\right)^{2z}\Gamma(-z)\frac{\Gamma(c) }{\Gamma(c+z) } \widetilde{\Xi}(z)\  ,
	\end{split}
\end{equation}
where the Mellin-transform
\begin{equation}
	\begin{split}
	\widetilde{\Xi}(z)&\equiv	\left(\frac{2H}{\w}\right)^{2z}\sum_{k=0}^\infty(-)^k\binom{z}{k}\frac{\Gamma(a+z-k) \Gamma(b+z-k) \Gamma(c+z)}{\Gamma(a) \Gamma(b) \Gamma(c+z-k)}\frac{\Gamma(z+\alpha)}{\Gamma(z-k+\alpha)}\\
	&=\left(\frac{2H}{\w}\right)^{2z}\frac{\Gamma(a+z) \Gamma(b+z) }{\Gamma(a) \Gamma(b) }{}_3F_2\left[\begin{array}{ccc}1-c-z,& -z,& 1-\alpha-z\\ 1-a-z& , & 1-b-z\end{array}; 1\right]\ .
	\end{split}
\end{equation}
Here, $\alpha\equiv \frac{1}{2}(1+\nu-\mu)$ and $a,b,c$ denote the parameters of the hyper-geometric function appearing in Eq.\eqref{eq:EuXiN}. The Mellin-transform $\widetilde{\Xi}(z)$ evaluated at integer $z$ is, in fact,
a polynomial of degree $z$ in the variable $(H/\w)^2$: this can be gleaned from the fact that the series above truncates in this case with polynomial coefficients.

To determine the polynomials $\mathfrak{p}_n$, we should compare the polynomials $\widetilde{\Xi}(n)$ against the Mellin-transform of $\sum_k\mathfrak{p}_k\mathfrak{B}_k$. This can be done using the Mellin-Barnes representation of ${}_0F_1$, viz.\cite{NIST:DLMF}, 
\begin{equation}
	\begin{split}
		{}_0F_1\left[c,x\right]= \int_{-i\infty}^{i\infty}\frac{dz}{2\pi i}(-x)^z\Gamma(-z)\frac{\Gamma(c) }{\Gamma(c+z) }\ .
	\end{split}
\end{equation}
This, in turn, yields an expression of the form
\begin{equation}
	\begin{split}
		\widetilde{\Xi}(z)=\sum_{k=0}^\infty\left(\frac{2}{\w}\right)^{2k}\mathfrak{p}_k(\nu,H^2,\w^2)\ \frac{\Gamma(k-z)}{\Gamma(-z)}\ .
	\end{split}
\end{equation}
This series also truncates for integer $z$ and the above relation can then be inverted to give 
\begin{equation}
	\begin{split}
\mathfrak{p}_k&=	\frac{1}{k!}\left(\frac{\w}{2}\right)^{2k}	\sum_{m=0}^k (-)^m\binom{k}{m}\ \widetilde{\Xi}(m)\ .
	\end{split}
\end{equation}
The explicit expression quoted before follows from this equation. The first few polynomials  are given by
\begin{equation}
	\begin{split}
\mathfrak{p}_0 &=1\ ,\\
\mathfrak{p}_1 &=\frac{H^2}{2^2}(1+\nu+\mu)(1+\nu-\mu)\ ,\\
\mathfrak{p}_2 &=\frac{H^2}{2^3}\left\{-\w^2(2\nu+3)+\frac{H^2}{2^2}(1+\nu+\mu)(1+\nu-\mu)(3+\nu+\mu)(3+\nu-\mu)\right\}\ ,\\
\mathfrak{p}_3 &=\frac{H^2}{2^2\times 3!}\Biggl\{\w^4+\frac{H^2\w^2}{2^2}[
		3\mu^2(2\nu+5)
		-(103+132\nu)-3(17\nu^2+2\nu^3)]
		+H^4(\ldots)\Biggr\}\ ,
\end{split}
\end{equation}
The polynomials $\mathfrak{p}_4$ and higher are proportional to $H^4$ and, hence the above expressions are sufficient to obtain an answer accurate up to order $H^2$ terms. To get terms accurate up to order $H^4$, we also need the leading terms of the next three polynomials:
\begin{equation}
	\begin{split}
		\mathfrak{p}_4&=\frac{H^4}{2^5\times 4!}
		\Biggl\{\w^4[-4\mu^2+8\nu(2\nu+13)+157]+H^2(\ldots)
		\Biggr\}\ ,\\
		\mathfrak{p}_{5} &=\frac{H^4}{2^2\times 5!}
		\Biggl\{\w^6
		(5\nu+18)+H^2(\ldots)
		\Biggr\}\ ,\\
		\mathfrak{p}_{6}&=
		\frac{H^4}{2^7\times 3^2}
		\Biggl\{\w^8
		+H^2(\ldots)
		\Biggr\}\ .
	\end{split}
\end{equation}
The polynomials $\mathfrak{p}_7$ and higher are proportional to $H^6$, and hence can be ignored at this order. 

For odd values of $d$, the function $\Xi_{nn}$ is related to $\Xi_n$ simply by the transformation: $\nu\to -\nu$. This allows us to also obtain the flat space expansion for $\Xi_{nn}$ in odd $d$: 
\begin{equation}
	\begin{split}
		\Xi_{nn}|_\text{Odd $d$} =\sum_{k=0}^\infty \mathfrak{p}_{k}(-\nu,H^2,\w^2)
		\mathfrak{G}_k\ ,
	\end{split}
\end{equation}
where the functions $\mathfrak{G}_k$ are related to the $\mathfrak{B}_k$ by $\nu\to -\nu$:
\begin{equation}
	\begin{split}
		\mathfrak{G}_k&\equiv\frac{r^{-\nu-\frac{d}{2}+\frac{1}{2}(d-1-\nn)+2k}}{(-\nu+1)\ldots(-\nu+k)}\  {}_0F_1\left[1+k-\nu,-\frac{\w^2r^2}{4}\right]=-2\nu\frac{\Gamma(-\nu)\ r^{\frac{1}{2}(1-\nn)+k}}{2(\w/2)^{k-\nu}}J_{k-\nu}(\w r)\  .
	\end{split}
\end{equation}

We will conclude by giving the near-flat/high-frequency  expansion of $\kO$  in odd $d$. This can be achieved 
using Stirling approximation, i.e.,
\begin{equation}
\Gamma\left(z\right)\sim\exp\left\{\left(z-\tfrac{1}{2}\right)\ln z-z+%
\tfrac{1}{2}\ln\left(2\pi\right)+\sum_{k=1}^{\infty}\frac{B_{2k}}{2k(2k-1)z^{2%
k-1}}\right\}\ ,
\end{equation}
an approximation valid as long $z\to \infty$ away from  the negative real axis. We then obtain the following expansion for $\kO$ in odd $d$:
\begin{equation}\label{eq:Koutflat}
\begin{split}
\kO|_\text{Odd $d$} &= \frac{2\pi i}{\Gamma(\nu)^2}\left(\frac{\w}{2}\right)^{2\nu} \Bigg[1+(\nu^2+3\mu^2-1)\frac{\nu}{3!!}\frac{H^2}{\w^2}\Bigg.\\
&\quad+\frac{5\nu^4-4\nu^3+(30\mu^2-14)\nu^2-(60\mu^2-16)\nu+(45\mu^2-90\mu^2+21)}{2\times 3}\frac{\nu(\nu-1)}{5!!}\frac{H^4}{\w^4}\\
&\qquad+O\left(\frac{H^6}{\w^6}\right)\Bigg]\ .
\end{split}
\end{equation}
This expression describes how the radiation reaction kernel gets corrected due to the non-zero cosmological constant.

\begin{center}
\begin{landscape}
\begin{table}[h]
	\begin{tabular}{||c||c|c|c||}
		\hline\hline
		Symbol & ${}^0f^\mu_d$ & ${}^0f^\mu_{d-2}$ & ${}^0f^\mu_{d-4}$ \\
		\hline \hline & & & \\
		$\mathbb{D}_1$ & $\frac{\partial_t^d}{d!!}$ & $\frac{\partial_t^{d-2}}{(d-2)!!}$ & $\frac{\partial_t^{d-4}}{(d-4)!!}$\\ 	 & $\qquad$ & & \\ 	
	\hline & & & \\	
	$\mathbb{D}_2$ & $\frac{\partial_t^{d+2}}{(d+2)!!}-\frac{H^2}{3!}(d+1)\frac{\partial_t^{d}}{d!!}+\frac{H^4}{5!}\frac{7}{3}(d^2-1)\frac{\partial_t^{d-2}}{(d-2)!!}$ & $\frac{\partial_t^{d}}{d!!}-\frac{H^2}{3!}(d-1)\frac{\partial_t^{d-2}}{(d-2)!!}+\frac{H^4}{5!}\frac{7}{3}(d-1)(d-3)\frac{\partial_t^{d-4}}{(d-4)!!}$ & $\frac{\partial_t^{d-2}}{(d-2)!!}$\\ & $\qquad$ & & \\
		\hline & & & \\		  	
	$\mathbb{D}_1^X$ & $\frac{\partial_t^{d+2}}{(d+2)!!}+\frac{H^2}{3}(d-2)\frac{\partial_t^{d}}{d!!}-\frac{H^4}{45}(d^2-1)\frac{\partial_t^{d-2}}{(d-2)!!}$ & $\frac{\partial_t^{d}}{d!!}+\frac{H^2}{3}(d-4)\frac{\partial_t^{d-2}}{(d-2)!!}-\frac{H^4}{45}(d-1)(d-3)\frac{\partial_t^{d-4}}{(d-4)!!}$ & $\frac{\partial_t^{d-2}}{(d-2)!!}$\\		 & $\qquad$ & & \\ 	
	\hline & & & \\
	$\mathbb{D}_1^V$ & $\frac{\partial_t^d}{d!!}$ & $\frac{\partial_t^{d-2}}{(d-2)!!}$ & $\frac{\partial_t^{d-4}}{(d-4)!!}$\\ 	 & $\qquad$ & & \\ 	
	\hline & & & \\	
	$\mathbb{D}_0^{XX}$ & $\frac{d+2}{2}\frac{\partial_t^{d+2}}{(d+2)!!}+\frac{H^2}{4!}2(5d^2-15d-2)\frac{\partial_t^{d}}{d!!}$ & $\frac{d+2}{2}\frac{\partial_t^{d}}{d!!}+\frac{H^2}{4!}2(5d^2-25d+2)\frac{\partial_t^{d-2}}{(d-2)!!}$ & $\frac{d+2}{2}\frac{\partial_t^{d-2}}{(d-2)!!}$\\
	 &  $+\frac{H^4}{6!}(67d^3-526 d^2+833 d-14) $ & $+\frac{H^4}{6!}(67d^3-794 d^2+2125 d+42) $& \\
	 	 & $\qquad$ & & \\ 	
	 		\hline & & & \\	
	 $\mathbb{D}_0^{VV}$ & $\frac{\partial_t^{d-2}}{(d-2)!!}+\frac{H^2}{3!}(d-1)\frac{\partial_t^{d-4}}{(d-4)!!}+\frac{H^4}{5!}(d-1)(d-3)\frac{\partial_t^{d-6}}{(d-6)!!}$ & $\frac{\partial_t^{d-4}}{(d-4)!!}+\frac{H^2}{3!}(d-3)\frac{\partial_t^{d-6}}{(d-6)!!}+\frac{H^4}{5!}(d-3)(d-5)\frac{\partial_t^{d-6}}{(d-6)!!}$ & $\frac{\partial_t^{d-6}}{(d-6)!!}$\\	
	 & $\qquad$ & & \\ 	
	 \hline & & & \\	
	 $\mathbb{D}_0^{XV}$ & $\frac{\partial_t^{d}}{d!!}+\frac{H^2}{2!}(d-3)\frac{\partial_t^{d-2}}{(d-2)!!}+\frac{H^4}{4!}(d-1)(d-7)\frac{\partial_t^{d-4}}{(d-4)!!}$ & $\frac{\partial_t^{d-2}}{(d-2)!!}+\frac{H^2}{2!}(d-5)\frac{\partial_t^{d-4}}{(d-4)!!}+\frac{H^4}{4!}(d-3)(d-9)\frac{\partial_t^{d-6}}{(d-6)!!}$ & $\frac{\partial_t^{d-4}}{(d-4)!!}$\\	
	 & $\qquad$ & & \\ 
	 \hline \hline
	\end{tabular}
\caption{The differential operators that appear in dS radiation reaction (for $d$ odd). We have divided
up the sum into three columns where each column combines into  an
expression covariant under dS isometries. The entries in the second column must be multiplied by a relative factor of $-\frac{H^2}{4\times 3!}c_h$ 
with $c_h\equiv 12\mu^2+d^2-4$ and then added to the first column. Similarly, the third column should be multiplied by a relative factor of $\frac{H^4}{8\times 6!}[5c_h^2-40(d+2)c_h+32(d+2)(d^2-1)]$ and then added to the first two contributions. The sum of these contributions should be further multiplied by a factor of $\frac{(-1)^{\frac{d-1}{2}}}{|S^{d-1}|(d-2)!!}$.}
\label{tab:RRoperators}
\end{table}	
\end{landscape}
\end{center}

	\bibliographystyle{JHEP}
	\bibliography{biblio}
\end{document}